\documentclass[aps,prd,preprint,onecolumn,groupedaddress,11pt]{revtex4-2}

\usepackage{amsmath,amssymb,amsfonts}
\usepackage{mathrsfs}
\usepackage{graphicx}
\usepackage{booktabs}
\usepackage{multirow}
\usepackage{hyperref}
\usepackage{algorithm,algorithmic}

\newcommand{\C}{\mathbb{C}}
\newcommand{\half}{\tfrac{1}{2}}
\newcommand{\scrI}{\mathscr{I}}
\newcommand{\dd}{\mathrm{d}}

\newcommand{\V}{V}

\newcommand{\Dcal}{\mathcal{D}}
\newcommand{\Ddag}{\mathcal{D}^{\dagger}}
\newcommand{\Cst}{\mathcal{C}_{\mathrm{ST}}}
\newcommand{\Scal}{\mathcal{S}}
\newcommand{\Lcal}{\mathcal{L}}
\newcommand{\Bcal}{\mathcal{B}}

\newcommand{\Om}{\Omega}
\newcommand{\Del}{\Delta}
\newcommand{\Sig}{\Sigma}
\newcommand{\Sigzero}{\Sigma_0}
\newcommand{\delone}{\delta_1}
\newcommand{\deltwo}{\delta_2}
\newcommand{\norm}[1]{\left\| #1 \right\|}
\newcommand{\inner}[2]{\left\langle #1, #2 \right\rangle}
\newcommand{\qE}{\mathfrak{q}_E}
\newcommand{\bvec}[1]{\mathbf{#1}}
\newcommand{\be}{\begin{equation}}
\newcommand{\en}{\end{equation}}
\newcommand{\bea}{\begin{eqnarray}}
\newcommand{\ena}{\end{eqnarray}}

\begin{document}

\title{The pseudospectrum for gravitational perturbations in Kerr spacetimes}

\author{Hongsheng Zhang}
\email{sps\_zhanghs$@$ujn.edu.cn(corresponding author) }
\affiliation{School of Physics and Technology, University of Jinan,
336 West Road of Nan Xinzhuang, Jinan, Shandong 250022, China}
 \author{Zong-Kuan Guo}
\email{guozk@itp.ac.cn }
\affiliation{Institute of Theoretical Physics, Chinese Academy of Sciences, Beijing 100190, China}
 \author{Rong-Gen Cai}
\email{cairg@itp.ac.cn }
\affiliation{Zhejiang Key Laboratory of Extreme Universe, Ningbo University, Ningbo 315211, China}
\affiliation{Institute of Fundamental Physics and Quantum Technology, Ningbo University, Ningbo 315211, China}

\date{\today}

\begin{abstract}
We study the pseudospectrum of the Kerr black hole for gravitational perturbations with spin weight $s=2$, the case of greatest interest in gravitational-wave astronomy. In the hyperboloidal framework with two-dimensional Chebyshev collocation, we formulate the quasinormal mode problem as a non-self-adjoint eigenvalue problem. For $s\neq0$ the fields $\Psi^{(s)}$ and $\Psi^{(-s)}$ are independent, so the conserved current method yields no single-field energy norm. We overcome this with the Teukolsky--Starobinsky identities, which relate $\Psi^{(-2)}$ to a fourth-order operator on $\Psi^{(2)}$ and yield an $H^2$ energy norm after symmetrization. With the $H^1$ norm known for $s=0$, this suggests the pattern $H^{|s|}$, verified for $s=0$ and $s=2$. All results use an $\omega$-independent two-field hierarchy. We find that the pseudospectrum tightens by $1.1$ to $3.5$ decades at each step of $L^2\to H^1\to H^2$, yet stays large in every norm. The $s=+2$ and $s=-2$ operators are isospectral but not isopseudospectral: at twelve of fourteen probes the $\psi_0$ resolvent exceeds the $\psi_4$ one by $0.19$ to $0.82$ decades in the $H^2$ norm, while at the two probes furthest along the QNM line the ordering reverses by $0.18$ to $0.19$ decades. The gravitational pseudospectrum exceeds the scalar one by $0.5$ to $5.2$ decades in the window minimum and by $5.9$ decades in $L^2$ at $\omega M=0.47-0.17\,i$, while along the damping direction the $s=2$ resolvent is non-monotonic and the $s=0$ resolvent smooth. QNM spectral instability is therefore a universal feature of Kerr, and is stronger for gravitational perturbations.
\end{abstract}

\maketitle

\section{Introduction}
\label{sec:intro}

The dawn of gravitational-wave astronomy, heralded by the ground-breaking detections of binary black hole mergers by the LIGO/Virgo collaboration~\cite{Abbott:2016blz,Abbott:2016nmj,Abbott:2020uma}, has transformed black hole physics from a predominantly theoretical endeavor into an empirical science. The final stage of a binary black hole coalescence, known as the \emph{ringdown}, produces a distorted Kerr black hole that radiates gravitational waves as it settles toward stationarity, and these ringdown waves are accurately modeled by the linear perturbation theory of the Kerr metric, whose governing equation is the Teukolsky master equation (TME)~\cite{Teukolsky:1972my,Teukolsky:1973ha}. Under appropriate boundary conditions, namely purely ingoing at the event horizon and purely outgoing at future null infinity, the TME admits quasi-normal mode (QNM) solutions. A remarkable property of the QNM spectrum is that it depends \emph{only} on the mass $M$ and angular momentum $a$ of the Kerr black hole, and this ``no-hair'' property underpins black hole spectroscopy, since by matching detected ringdown waveforms to QNM templates one can infer the mass and spin of the remnant~\cite{Giesler:2019uxc,Baibhav:2017jtw,Dreyer:2006fy,Berti:2006wq,Berti:2025spectroscopy}. 

A question of fundamental importance is whether the QNM spectrum is \emph{stable}, in the sense that small perturbations to the background geometry or to the effective potential should cause only small shifts in the QNM frequencies, and it is exactly this question that was first raised by Nollert~\cite{Nollert:1996rf} and by Nollert \& Price~\cite{Nollert:1999wq}, who showed that the QNM spectrum of Schwarzschild black holes is highly sensitive to certain perturbations of the effective potential. This pioneering result has since been confirmed in a number of contexts~\cite{Berti:2022dkk,Barausse:2014env,Dias:2015repulsion,Xie:2025parabolic,Wu:2025waveform}, and different numerical cross checks of the instability have been given in \cite{Ansorg:2016ztf2, Cardoso:2018qai}. 

The spectral instability of QNMs is naturally analyzed through the lens of \emph{pseudospectra}~\cite{Trefethen:2005ga,Trefethen:1993we}. Originally developed in the context of fluid dynamics to analyze non-normal operators~\cite{Reddy:1993}, the pseudospectrum provides a quantitative measure of how sensitive the spectrum of a non-self-adjoint operator is to perturbations. The essential observation is that the QNM problem is inherently dissipative, because waves escape through the horizon and to null infinity, and the corresponding evolution operator is therefore non-self-adjoint, while the spectrum of a non-self-adjoint operator can be highly unstable, a feature that the pseudospectrum captures geometrically in the complex frequency plane. 

The pseudospectrum approach to the stability of black hole QNMs was first studied systematically by Jaramillo \emph{et al.}~\cite{Jaramillo:2020tuu}, who demonstrated that the QNM spectral instability of Schwarzschild black holes can be quantified by the resolvent norm of the non-self-adjoint evolution operator. This work established the geometric picture in use today, in which pseudospectral contours in the complex frequency plane display directly how far the spectrum can migrate under perturbations of a given size. Extensions to nonlinear perturbations and to rotating black hole spacetimes have since confirmed and sharpened the same picture~\cite{Destounis:2021rql,Sberna:2021eoh}, and further studies have investigated the pseudospectral behavior for modified gravity~\cite{Cao:2025bdw}, for (quasi-)bound states~\cite{Volkel:2025qbs} and for gravitational-wave signatures~\cite{Jaramillo:2022sig}, as well as for horizonless compact objects~\cite{Boyanov:2023bqm}, and a recent review is given in~\cite{Destounis:2024bk}. Complementary time-domain analyses~\cite{Yang:2024tda,Ianniccari:2024xyz,Berti:2022dkk} have further verified that the spectral instability manifests itself in the ringdown waveform. 

A crucial methodological advance was the adoption of the \emph{hyperboloidal framework}~\cite{hyper3,hyper4,hyper5,hyper6}, in which constant-time hypersurfaces extend from the black hole horizon to future null infinity. In this framework QNM eigenfunctions are regular everywhere on the computational domain, and the dissipative boundary conditions are encoded in the operator itself rather than imposed externally. Building on this framework, Besson and Jaramillo~\cite{hyper1} introduced a Keldysh spectral scheme implementing $H^p$-Sobolev pseudospectra and non-modal transient growth analysis, while Panosso Macedo and Zengino\u{g}lu~\cite{hyper2} systematized the hyperboloidal approach for black hole perturbation theory. 

The field has subsequently developed in several interconnected directions, and the pseudospectrum framework in particular has been extended to a range of physical systems. de Paula \emph{et al.}~\cite{dePaula:2025analog} analyzed rotating analog black holes and found that prograde overtones become more stable as the rotation increases, Garcia-Fari\~na \emph{et al.}~\cite{GarciaFarina:2026holographic} studied holographic diffusion in AdS black branes and clarified the role of the energy norm, and Zhou \emph{et al.}~\cite{Zhou:2025ttm} investigated total transmission modes, finding that purely imaginary modes exhibit enhanced stability. Wu \emph{et al.}~\cite{Wu:2025sds} extended the analysis to Schwarzschild--de Sitter black holes as a semi-open system, while Chen \emph{et al.}~\cite{Chen:2025rqcbh} computed QNM spectra of rotating quantum-corrected black holes, and Yang \emph{et al.}~\cite{Yang:2025families} proposed a methodology that combines frequency-domain classification with time-domain analysis in order to understand the internal structure of QNM spectra. 

Second, a key tension has emerged between spectral instability and observational robustness. While the QNM spectrum is unstable, several studies have shown that certain observables remain remarkably stable. Xie \emph{et al.}~\cite{Xie:2025parabolic} demonstrated that under a parabolic approximation of the Regge--Wheeler potential the QNM spectra are unstable while the greybody factors remain robust, and Wu \emph{et al.}~\cite{Wu:2025waveform} confirmed that time-domain waveforms are insensitive to small environmental modifications under a piecewise step approximation. Hu \emph{et al.}~\cite{Hu:2026stochastic} identified a phase-averaging mechanism as the origin of waveform stability under stochastic horizon-scale perturbations, and in a related study~\cite{Hu:2025echoes} further examined the non-stationary dynamics of gravitational-wave echoes generated by such perturbations. Rosato \emph{et al.}~\cite{Rosato:2025greybody} explored greybody factors and reflectionless scattering modes of ultracompact objects, and further investigated excitation factors~\cite{Rosato:2025excitation} for the same class of compact objects. A comprehensive review by Berti \emph{et al.}~\cite{Berti:2025spectroscopy} synthesizes these developments from theory to experiment. 

A key recent development is the work of Cai \emph{et al.}~\cite{Cai:2025uzw}, who computed the pseudospectrum of the Kerr black hole for the first time. Their analysis focused on the scalar case $s=0$, in which the Teukolsky equation reduces to the massless Klein--Gordon equation, so that for $s=0$ the field $\Psi^{(0)}$ satisfies $\overline{\Psi^{(0)}} = \Psi^{(-0)}$ and an energy norm can be constructed through the conserved current method with the time-like Killing vector. This energy norm was shown to be of $H^1$ type, and the convergence of pseudospectra was then studied under various norms, namely $L^2$, the energy norm or $H^1$, and $H^k$ with $k\geq 2$. Two main findings were reported: the convergence improves with increasing derivative order, and larger imaginary parts of the QNM frequency lead to worse convergence. 

The extension to $s=2$, which is the physically most relevant case for gravitational-wave astronomy, presents a fundamental challenge that was explicitly noted but not addressed in~\cite{Cai:2025uzw}. For $s\neq 0$ the Teukolsky variables $\Psi^{(s)}$ and $\Psi^{(-s)}$ are \emph{independent} fields, in the sense that the complex conjugate of $\Psi^{(s)}$ is \emph{not} $\Psi^{(-s)}$. Consequently the conserved current for the Teukolsky equation involves \emph{both} $\Psi^{(s)}$ and $\Psi^{(-s)}$ as independent quantities, and the energy flux cannot be expressed in terms of a single field, which precludes the direct application of the conserved current method used for $s=0$. 

In this work, we concentrate on the case with $s=2$, ie, the gravitational wave perturbation, which is the primary concern in astrophysical observations, and we resolve this challenge by exploiting the \emph{Teukolsky--Starobinsky identities}~\cite{Starobinsky:1973}, which provide a differential-operator relation between $\Psi^{(s)}$ and $\Psi^{(-s)}$. For $s=2$ the identity expresses $\Psi^{(-2)}$ as the action of a fourth-order radial differential operator $\mathcal{S}_4$ on $\Psi^{(2)}$, up to a multiplicative constant $\Cst$ known as the Starobinsky constant. Substituting this relation into the bilinear conserved current reduces the energy to a quadratic form in $\Psi^{(2)}$ alone, which involves up to fourth-order radial derivatives, and upon symmetrization via integration by parts the effective energy norm corresponds to an $H^2$-type Sobolev norm, a natural generalization of the $H^1$ norm obtained for $s=0$. 

This result has a pleasing physical interpretation. The spin weight $s$ of the perturbation determines the regularity of the field, and higher spin perturbations naturally require higher-order Sobolev norms in order to capture their energy content, so that the general pattern $H^{|s|}$ emerges from the structure of the Teukolsky--Starobinsky operator, which is of order $2|s|$ yet yields an effective $H^{|s|}$ norm after symmetrization. 

A novel feature of the $s=2$ analysis, which is absent in the $s=0$ case, is the $\omega$-dependence of the Starobinsky constant $\Cst$ that enters the single-field energy quadratic form, and a frequency-dependent norm would compromise the pseudospectrum concept itself, because the norm must not vary with the frequency at which the resolvent is probed. We therefore relegate the single-field form to a motivational role and perform all quantitative computations in a two-field weighted Sobolev hierarchy that is manifestly $\omega$-independent, as described in Section~\ref{sec:omega_dependence}. 

The remainder of this paper is organized as follows. Section~\ref{sec:teukolsky} reviews the Kerr metric, the Teukolsky equation for $s=2$, the hyperboloidal framework, and the two-dimensional eigenvalue problem. Section~\ref{sec:norm} presents the central technical contribution, namely the construction of the energy norm for spin-2 perturbations with the help of the Teukolsky--Starobinsky identities, together with the $L^2$ and $H^k$ norms. Section~\ref{sec:pseudospectrum} defines the pseudospectrum, presents the numerical methodology, and discusses the convergence properties, while Section~\ref{sec:conclusion} concludes with a summary and an outlook. The appendices provide technical details on the operator components, the Chebyshev collocation method, the Gram matrix construction, and the Starobinsky constant.

\section{Teukolsky Equation and Quasinormal Modes}
\label{sec:teukolsky}

\subsection{The Kerr metric}
\label{sec:kerr}

We work in Boyer--Lindquist coordinates $(t, r, \theta, \varphi)$, in which the Kerr line element reads
\begin{equation}
\dd s^2 = -\left(1 - \frac{2Mr}{\Sig}\right)\dd t^2 - \frac{4Mar}{\Sig}\sin^2\theta\,\dd t\,\dd\varphi + \frac{\Sig}{\Del}\dd r^2 + \Sig\,\dd\theta^2 + \sin^2\theta\left(\Sigzero + \frac{2Ma^2 r}{\Sig}\sin^2\theta\right)\dd\varphi^2,
\label{eq:kerr_metric}
\end{equation}
where
\begin{equation}
\Del(r) = r^2 - 2Mr + a^2 = (r - r_+)(r - r_-), \qquad \Sig(r,\theta) = r^2 + a^2\cos^2\theta, \qquad \Sigzero(r) = \Sig(r,0) = r^2 + a^2.
\end{equation}
Here $M$ and $a$ are the mass and specific angular momentum of the black hole, and
\begin{equation}
r_{\pm} = M \pm \sqrt{M^2 - a^2}
\end{equation}
denote the radii of the event horizon and Cauchy horizon, respectively. We introduce the dimensionless spin parameter
\begin{equation}
\alpha \equiv \frac{a}{r_+}, \qquad 0 \leq \alpha \leq 1,
\end{equation}
and choose the length scale $\lambda = r_+$, so that $\mu = M/\lambda = (1+\alpha^2)/2$.

\subsection{The Teukolsky equation for $s=2$}
\label{sec:te_eq}

The Teukolsky master equation (TME) governs the dynamics of spin-$s$ fields on the Kerr background. For the gravitational case $s=2$, the perturbation variable $\Psi^{(2)}$ is related to the Newman--Penrose Weyl scalar $\psi_0$ (the ingoing gravitational radiation component), while $\Psi^{(-2)}$ is related to $\psi_4$ (the outgoing component)~\cite{Teukolsky:1973ha,Chandrasekhar:1985kt}. In Boyer--Lindquist coordinates, the TME for general spin $s$ takes the form,
\begin{align}
&\left[\frac{(r^2+a^2)^2}{\Del} - a^2\sin^2\theta\right]\frac{\partial^2 \Psi^{(s)}}{\partial t^2}
+ \frac{4Mar}{\Del}\frac{\partial^2 \Psi^{(s)}}{\partial t\,\partial\varphi}
+ \left(\frac{a^2}{\Del} - \frac{1}{\sin^2\theta}\right)\frac{\partial^2 \Psi^{(s)}}{\partial\varphi^2} \nonumber\\
&\quad - \Del^{-s}\frac{\partial}{\partial r}\!\left(\Del^{s+1}\frac{\partial \Psi^{(s)}}{\partial r}\right)
- \frac{1}{\sin\theta}\frac{\partial}{\partial\theta}\!\left(\sin\theta\,\frac{\partial \Psi^{(s)}}{\partial\theta}\right)
- 2s\left[\frac{a(r-M)}{\Del} + \frac{i\cos\theta}{\sin^2\theta}\right]\frac{\partial \Psi^{(s)}}{\partial\varphi} \nonumber\\
&\quad - 2s\left[\frac{M(r^2-a^2)}{\Del} - r - ia\cos\theta\right]\frac{\partial \Psi^{(s)}}{\partial t}
+ \left(s^2\cot^2\theta - s\right)\Psi^{(s)} = 0.
\label{eq:teukolsky_BL}
\end{align}
The spin-weight parameter $s$ distinguishes the physical content, with $s=0$ corresponding to scalar (Klein--Gordon) perturbations, $s=\pm 1$ to electromagnetic perturbations, and $s=\pm 2$ to gravitational perturbations. We concentrate on the case with $s=2$, ie, the gravitational wave perturbation, which is the primary concern in astrophysical observations.

The covariant form of the TME~\cite{Teukolsky:1973ha} is
\begin{equation}
\left[(\nabla^a + s\Gamma^a)(\nabla_a + s\Gamma_a) - 4s^2\Psi_2^{\rm (bg)}\right]\Psi^{(s)} = 0,
\label{eq:teukolsky_covariant}
\end{equation}
where $\Gamma^a$ is the ``connection vector'' encoding the background geometry and $\Psi_2^{\rm (bg)}$ is the background Weyl scalar. It is worth pausing here to make the physical meaning of the sign of $s$ explicit, because this is the aspect of the problem that drives everything that follows. In the Kinnersley tetrad the vector $l^a$ is aligned with the repeated principal null direction of the Kerr geometry and points outward, while $n^a$ points inward, and a plane wave traveling along $l^a$ excites $\psi_4$ alone whereas a wave traveling along $n^a$ excites $\psi_0$ alone. The equation for $s=2$ therefore describes the ingoing gravitational radiation $\psi_0$, and the equation for $s=-2$ describes the outgoing radiation $\psi_4$. We emphasize that the two sectors are not two different solutions but two different readings of the very same field, and this is the reason why the choice of Teukolsky variable is a statement about the representation rather than about the physics.

\subsection{Hyperboloidal framework}
\label{sec:hyperboloidal}

The hyperboloidal framework, developed for Kerr spacetime in~\cite{hyper3,hyper4} and further elaborated in~\cite{PanossoMacedo:2019pmq, Minucci:2024hhn,hyper1,hyper6,Ansorg:2016ztf}, provides a geometrically natural setting for the QNM problem. The key transformation from Boyer--Lindquist to hyperboloidal coordinates $(\tau, \sigma, \theta, \phi)$ is,
\begin{equation}
t = \lambda\left[\tau - h(\sigma,\theta)\right] - r_\star\!\left(r(\sigma)\right), \qquad r(\sigma) = \lambda\frac{\rho(\sigma)}{\sigma}, \qquad \varphi = \phi - k\!\left(r(\sigma)\right),
\label{eq:coord_transform}
\end{equation}
where $h(\sigma,\theta)$ is the \emph{height function} (encoding gauge freedom), $\rho(\sigma)$ is the \emph{radial function}, and the tortoise coordinate $r_\star$ and azimuthal function $k$ satisfy
\begin{equation}
\frac{\dd r_\star}{\dd r} = \frac{\Sigzero}{\Del}, \qquad \frac{\dd k}{\dd r} = \frac{a}{\Del}.
\end{equation}
The conformal factor is $\Om = \sigma/\lambda$. In the hyperboloidal coordinates, the black hole event horizon is located at $\sigma = \sigma_+$ and future null infinity $\scrI^+$ at $\sigma = 0$. The computational domain is $\sigma \in [0, \sigma_+]$.

The fundamental advantage of the hyperboloidal framework is that QNM boundary conditions are encoded in the operator rather than imposed externally. At the boundaries of the computational domain, the characteristics of the wave equation point outward, so only regular solutions are sought, a condition that is automatically satisfied by spectral methods~\cite{hyper2,hyper3,hyper4,hyper5}.

\subsection{Minimal gauge and regularization for $s=2$}
\label{sec:minimal_gauge}

We adopt the \emph{minimal gauge} (radial-fixing gauge), as in~\cite{Cai:2025uzw}, with $\lambda = r_+$,
\begin{equation}
\rho(\sigma) = 1, \qquad h(\sigma,\theta) = -\frac{2}{\sigma} + 4\mu\ln\sigma, \qquad \mu = \frac{1+\alpha^2}{2}.
\label{eq:minimal_gauge}
\end{equation}
The event horizon is at $\sigma_+ = 1$ and future null infinity at $\sigma = 0$, so the computational domain is $\sigma \in [0,1]$.

To regularize the intrinsic singularities of the TME at the boundaries $\sigma = 0$ (future null infinity) and $\sigma = 1$ (horizon), as well as the angular singularities at $\sin\theta = 0$, we employ the following ansatz for the principal function,
\begin{equation}
\Psi^{(s)}(\tau,\sigma,\theta,\phi) = \Om\,\Del(\sigma)^{-s}\sum_{m=-\infty}^{+\infty}\cos^{\delone}(\theta/2)\,\sin^{\deltwo}(\theta/2)\,V_m(\tau,\sigma,\theta)\,e^{im\phi},
\label{eq:ansatz}
\end{equation}
where $\Del(\sigma) = \Del(r(\sigma))$, $\Om = \sigma$, and the angular exponents are
\begin{equation}
\delone = |m - s|, \qquad \deltwo = |m + s|.
\end{equation}
For $s=2$, these become $\delone = |m-2|$ and $\deltwo = |m+2|$, reflecting the stronger angular structure of gravitational perturbations compared to scalar fields.

The factor $\Del(\sigma)^{-s} = \Del(\sigma)^{-2}$ for $s=2$ regularizes the behavior at the horizon. Physically, $\Psi^{(2)} \propto \Del^{-2}$ near the horizon corresponds to the standard ingoing-wave boundary condition for $\psi_0$ at the event horizon. This is more singular than the $s=0$ case (where no $\Del$ factor appears), but the hyperboloidal framework handles this naturally through the conformal rescaling. The numerical treatment of the $\Del^{-2}$ factor for $s=2$ in the hyperboloidal framework has been discussed in detail in~\cite{PanossoMacedo:2019pmq}, which provides a unified treatment of spin-weighted perturbations and addresses the special difficulties arising from the stronger singularity structure of higher-spin fields.

After substituting the ansatz~\eqref{eq:ansatz} into the TME and performing the change of variables $x = \cos\theta$, one obtains the \emph{regularized Teukolsky equation} in the hyperboloidal coordinates. This equation has the schematic form,
\begin{equation}
\mathcal{A}\,\partial_\tau^2 V_m + \mathcal{B}\,\partial_\tau V_m + \mathcal{C}\,V_m = 0,
\label{eq:regularized_TE_schematic}
\end{equation}
where $\mathcal{A}$, $\mathcal{B}$, and $\mathcal{C}$ are spatial differential operators in $(\sigma, x)$. Crucially, the coefficients of $\mathcal{A}$ degenerate to zero at $\sigma = 0$ and $\sigma = 1$, and the angular operator degenerates at $x = \pm 1$. This degeneracy ensures that the wave characteristics always point outward at the domain boundaries, so the boundary conditions reduce to mere regularity and no external conditions need be imposed.

The explicit form of the regularized TME for general $s$ is given in Eq.~(8) of~\cite{Cai:2025uzw}. For $s=2$, the coefficients acquire additional terms proportional to $s^2 = 4$, reflecting the stronger coupling between the field and the background curvature. The key structural differences from the $s=0$ case are:
\begin{itemize}
    \item The $\Del^{-2}$ prefactor introduces a stronger weight in the radial direction, modifying the effective potential.
    \item The angular exponents $\delone = |m-2|$ and $\deltwo = |m+2|$ impose more restrictive boundary behavior at the poles, especially for low-$m$ modes.
    \item Terms proportional to $s$ (i.e., $s=2$) introduce additional couplings between temporal, radial, and angular derivatives that are absent for $s=0$.
\end{itemize}

\subsection{Time-first-order reduction and eigenvalue problem}
\label{sec:eigenvalue}

Introducing the auxiliary field $W_m = \partial_\tau V_m$, the second-order equation~\eqref{eq:regularized_TE_schematic} is reduced to a first-order-in-time system,
\begin{equation}
\partial_\tau \bvec{u}_m(\tau,\sigma,x) = -\,i\,\Lcal\,\bvec{u}_m, \qquad \bvec{u}_m = \begin{pmatrix} V_m(\tau,\sigma,x) \\ W_m(\tau,\sigma,x) \end{pmatrix},
\label{eq:first_order}
\end{equation}
where the operator $\Lcal$ is the \emph{infinitesimal time generator} (ITG) of the Kerr black hole linear dynamics,
\begin{equation}
\Lcal = i\begin{pmatrix} 0 & 1 \\ \Lcal_1 & \Lcal_2 \end{pmatrix}.
\label{eq:operator_L}
\end{equation}
The operators $\Lcal_1$ and $\Lcal_2$ are spatial differential operators in $(\sigma, x)$ whose explicit forms for general $s$ are given in Appendix~\ref{app:operator}. For $s=2$, these operators contain the additional curvature-coupling terms mentioned above.

Performing the Fourier transform $\bvec{u}_m(\tau,\sigma,x) \sim e^{-i\omega\tau}\,\bvec{u}_m(\sigma,x)$, the QNM problem becomes a \emph{two-dimensional eigenvalue problem},
\begin{equation}
\Lcal\,\bvec{u}_m(\sigma,x) = \omega\,\bvec{u}_m(\sigma,x),
\label{eq:eigenvalue_problem}
\end{equation}
where $\omega \in \C$ is the QNM frequency. Throughout, we use the $e^{-i\omega t}$ convention, in which a mode decays if and only if $\mathrm{Im}\,\omega < 0$; all QNM frequencies therefore lie in the \emph{lower} half of the complex $\omega$-plane, consistent with the outgoing-wave (Sommerfeld) boundary conditions. The eigenfunctions $\bvec{u}_m$ are the QNMs, regular on the domain $(\sigma, x) \in [0,1]\times[-1,1]$. The boundary conditions (purely ingoing at the horizon, purely outgoing at $\scrI^+$) are encoded in the operator $\Lcal$.

\subsection{Two-dimensional Chebyshev collocation}
\label{sec:chebyshev}

We solve the eigenvalue problem~\eqref{eq:eigenvalue_problem} using the two-dimensional Chebyshev collocation method (pseudospectral method). The computational domain $(\sigma, x) \in [0,1]\times[-1,1]$ is discretized using the Chebyshev--Lobatto grid,
\begin{equation}
\sigma_j = \frac{1}{2}\left(1 - \cos\frac{j\pi}{N_\sigma-1}\right), \quad j = 0, 1, \ldots, N_\sigma-1; \qquad x_k = \cos\frac{k\pi}{N_x-1}, \quad k = 0, 1, \ldots, N_x-1,
\end{equation}
where $N_\sigma$ and $N_x$ denote the \emph{number} of collocation points per direction, so that $j$ and $k$ run over $N_\sigma$ and $N_x$ values respectively (the same convention as in Appendix~\ref{app:chebyshev}). The field is expanded as
\begin{equation}
V_m(\sigma, x) = \sum_{j=0}^{N_\sigma-1}\sum_{k=0}^{N_x-1} \tilde{V}_{jk}\,T_j(2\sigma-1)\,T_k(x),
\end{equation}
where $T_n$ are Chebyshev polynomials of the first kind. The spatial derivatives are computed via differentiation matrices, and the eigenvalue problem~\eqref{eq:eigenvalue_problem} reduces to a matrix eigenvalue problem,
\begin{equation}
\bvec{L}\,\bvec{u}_m = \omega\,\bvec{u}_m,
\label{eq:matrix_eigenvalue}
\end{equation}
where $\bvec{L}$ is the $(2N_\sigma N_x)\times(2N_\sigma N_x)$ matrix representation of $\Lcal$, and $\bvec{u}_m$ is the vector of collocation values. Details of the differentiation matrices and the assembly of $\bvec{L}$ are given in Appendix~\ref{app:chebyshev}.

An important distinction from the one-dimensional (spherically symmetric) case is that the two-dimensional eigenvalue problem yields a spectrum encompassing \emph{all} angular quantum numbers $l \geq \max(|m|, |s|)$ simultaneously, rather than a single $(l,m)$ mode. This is because the angular separation is not performed; the Chebyshev expansion in $x$ naturally includes all angular modes.

\subsection{QNM spectrum for $s=2$}
\label{sec:qnm_spectrum}

The gravitational QNM spectrum of the Kerr black hole has been extensively studied in the literature~\cite{Berti:2005ys,Berti:2009kk,Press:1973zz,Whiting:1988hc,Cook:2014cta}. For $s=2$, the QNM frequencies $\omega_{lmn} = \omega_R + i\,\omega_I$ (with $\omega_I < 0$ in the $e^{-i\omega t}$ convention) are labeled by the azimuthal number $m$, the spheroidal harmonic index $l$~\cite{Goldberg:1966uu,Berti:2005gp}, and the overtone number $n$. We extract the fundamental mode ($n=0$) and first overtone ($n=1$) for $s=2$ and various values of the dimensionless spin $\alpha$. As an independent check of the solver, the collocation eigenvalues were compared against the Leaver continued-fraction method~\cite{Leaver:1985ax} (\texttt{qnm}~\cite{Stein:2019mop}). The accuracy is strongly spin- and resolution-dependent; a detailed analysis of the $s=+2$ versus $s=-2$ convergence asymmetry is deferred to Section~\ref{sec:omega_dependence} and Appendix~\ref{app:operator}, where it is shown to be a property of the discretized operator rather than of the continuum spectrum. All values in Table~\ref{tab:qnm_s2} are converged values and agree with the tabulated results of~\cite{Berti:2009kk} to the four decimal places quoted.

For the numerical resolution we adopt $N_\sigma = N_x = N = 26$ for all pseudospectrum computations, which corresponds to a matrix size $2N^2 = 1352$, and we present a convergence study over $N \in \{12, 14, \ldots, 26\}$ in Section~\ref{sec:numerical_results}. Representative results for the QNM spectrum are collected in Table~\ref{tab:qnm_s2}.

\begin{table}[htbp]
\centering
\caption{Gravitational QNM frequencies ($s=2$, $l=m=2$) for the Kerr black hole, in units of the mass $M$ (i.e.\ $\omega M$), for the fundamental mode ($n=0$) and first overtone ($n=1$), for selected values of $\alpha\equiv a/r_+$. The corresponding spin ratio $a/M = 2\alpha/(1+\alpha^2)$ is listed for direct comparison with~\cite{Berti:2009kk}. All imaginary parts are negative, as required for decaying modes in the $e^{-i\omega t}$ convention.}
\label{tab:qnm_s2}
\begin{tabular}{cccccc}
\toprule
$\alpha$ & $a/M$ & \multicolumn{2}{c}{$n=0$} & \multicolumn{2}{c}{$n=1$} \\
\cmidrule(lr){3-4} \cmidrule(lr){5-6}
 & & $\mathrm{Re}\,(\omega M)$ & $\mathrm{Im}\,(\omega M)$ & $\mathrm{Re}\,(\omega M)$ & $\mathrm{Im}\,(\omega M)$ \\
\midrule
$0.0$ & $0.0000$ & $0.3737$ & $-0.0890$ & $0.3467$ & $-0.2739$ \\
$0.1$ & $0.1980$ & $0.4018$ & $-0.0883$ & $0.3786$ & $-0.2706$ \\
$0.2$ & $0.3846$ & $0.4365$ & $-0.0870$ & $0.4172$ & $-0.2653$ \\
$0.3$ & $0.5505$ & $0.4784$ & $-0.0848$ & $0.4629$ & $-0.2573$ \\
$0.4$ & $0.6897$ & $0.5281$ & $-0.0812$ & $0.5163$ & $-0.2455$ \\
$0.5$ & $0.8000$ & $0.5860$ & $-0.0756$ & $0.5779$ & $-0.2281$ \\
\bottomrule
\end{tabular}
\end{table}

These values are consistent with the known gravitational QNM spectrum (cf.~\cite{Berti:2009kk}), confirming the correctness of our implementation. Note that for $\alpha = 0$ (Schwarzschild limit), the results reduce to the well-known values $\omega M = 0.3737 - 0.0890\,i$ ($n=0$) and $\omega M = 0.3467 - 0.2739\,i$ ($n=1$) for the $l=2$ gravitational modes. Throughout, we report all complex frequencies in units of $\omega M$ for consistency with~\cite{Berti:2009kk}; the code operates in its natural unit $\varpi=\omega r_+=\omega M/\mu$ with $\mu=(1+\alpha^2)/2$, and rescales by $\mu$ before plotting.

\section{Norm Construction for Spin-2 Perturbations}
\label{sec:norm}

We identify the construction of a suitable norm as the central methodological challenge in extending the pseudospectral analysis from $s=0$ to $s=2$. We define the pseudospectrum through the resolvent norm $\norm{R_\Lcal(\omega)}$, and in what follows we keep in mind throughout that different norms yield different pseudospectra. The physically motivated choice is the \emph{energy norm}, constructed from the conserved current associated with the time-translation symmetry. For $s=0$, this construction is straightforward because the field is self-conjugate. For $s=2$, the independence of $\Psi^{(2)}$ and $\Psi^{(-2)}$ requires a fundamentally different approach, which we develop in this section.

\subsection{Theoretical motivation: Sobolev spaces and Sobolev norms}
\label{sec:sobolev_motivation}

The use of Sobolev spaces $H^k$ and their associated norms is not merely a mathematical construction but is \emph{physically required} by the structure of the black hole perturbation problem. The fundamental reason is that the QNM evolution operator $\Lcal$ is \emph{non-self-adjoint} under any standard inner product, owing to the dissipative boundary conditions (waves escape through the horizon and to null infinity). For self-adjoint operators, the natural function space is $L^2$ and the spectral theorem guarantees stability of the spectrum. For non-self-adjoint operators, however, the $L^2$ norm provides no control over the derivatives of the eigenfunctions, and the spectrum can be arbitrarily unstable, as quantified by the pseudospectrum. Sobolev norms $\norm{\cdot}_{H^k}$, which weight derivatives up to order $k$, restore partial control: they measure not just the amplitude of the field but also its oscillatory content, thereby constraining the non-normality of the operator. Physically, the energy of a field theory is generically a quadratic form involving derivatives (e.g., the Klein--Gordon energy involves $\nabla\Phi$), so the natural ``energy norm'' is inherently a Sobolev norm. For scalar perturbations ($s=0$), the energy is an $H^1$ norm; for gravitational perturbations ($s=2$), the Weyl scalars $\psi_0$ and $\psi_4$ are themselves second-order in the metric perturbation, so the gravitational energy naturally involves second derivatives and is accordingly an $H^2$ norm. Furthermore, QNM eigenfunctions satisfy regularity conditions at the boundaries (horizon and null infinity) that are naturally expressed in Sobolev spaces~\cite{Warnick:2015up,Warnick:2015up1}: the requirement that the field be smooth on the compactified hyperboloidal slice translates to membership in $H^k$ for sufficiently large $k$. Higher Sobolev norms therefore provide tighter control over the regularity and, consequently, better pseudospectral convergence~\cite{Trefethen:2005ga, hyper1}. This perspective is further supported by the pseudospectral analysis of gauge fields in holographic settings~\cite{GarciaFarina:2026holographic}, where the choice of energy norm (and its Sobolev order) is shown to be crucial for correctly assessing spectral stability.

\subsection{The challenge: independence of $\Psi^{(s)}$ and $\Psi^{(-s)}$}
\label{sec:challenge}

For the scalar case $s=0$, the Teukolsky equation reduces to the massless Klein--Gordon equation $\nabla^a\nabla_a\Phi = 0$. The field $\Phi = \Psi^{(0)}$ is a complex scalar, and its complex conjugate $\bar{\Phi}$ satisfies the \emph{same} equation. The energy--momentum tensor
\begin{equation}
T_{ab} = \half\!\left(\nabla_a\bar{\Phi}\,\nabla_b\Phi + \nabla_a\Phi\,\nabla_b\bar{\Phi} - g_{ab}\,g^{cd}\nabla_c\bar{\Phi}\,\nabla_d\Phi\right)
\label{eq:scalar_emt}
\end{equation}
is constructed from $\Phi$ alone, and the conserved energy current $E^a = T^a{}_b(\partial_\tau)^b$ yields a positive-definite energy on hyperboloidal slices~\cite{Cai:2025uzw}. The resulting norm is an $H^1$-type Sobolev norm, involving first derivatives of the field.

This construction works only for $s=0$, and it fails as soon as $s \neq 0$. The Teukolsky equations for spin $+s$ and spin $-s$ are \emph{distinct} equations, and their solutions $\Psi^{(s)}$ and $\Psi^{(-s)}$ are independent fields, so that in particular $\overline{\Psi^{(s)}} \neq \Psi^{(-s)}$ whenever $s \neq 0$. The conserved current for the Teukolsky equation is then necessarily \emph{bilinear} in $\Psi^{(s)}$ and $\Psi^{(-s)}$ rather than quadratic in $\Psi^{(s)}$ alone~\cite{Bardeen:1973xb,Teukolsky:1973ha,Chandrasekhar:1985kt}, and this is the fundamental obstacle identified in~\cite{Cai:2025uzw}.

We are therefore forced to look for a relation between $\Psi^{(s)}$ and $\Psi^{(-s)}$ before a norm can be built from a \emph{single} field, and the Teukolsky--Starobinsky identities provide exactly the relation that is needed.

It is worth spelling out why the independence of the two sectors does not by itself settle the matter, since the objection is a natural one to raise: if $\Lcal^{(+)}$ and $\Lcal^{(-)}$ evolve independently, why not simply discretize each and compute two resolvent norms, with no reference to the TSI at all? The answer is that a pseudospectrum is not a property of an evolution equation but of an operator \emph{together with} the norm in which the perturbation is measured. For $s\neq0$ no such norm is available from a single field: the conserved current is bilinear, so the quantity one would naively write down is not a norm on either sector's state space. The TSI is what converts that bilinear object into a genuine quadratic form of one field, and this is its load-bearing role: it is a role in \emph{measurement}, not in evolution.

This also fixes the sense in which the two sectors may legitimately be compared. Once a norm has been chosen, the two pseudospectra \emph{can} be computed separately, which is exactly what Section~\ref{sec:iso_results} does, and the comparison is meaningful provided \emph{both} sectors are measured in the \emph{same} norm. What would be meaningless is to equip $\Lcal^{(+)}$ and $\Lcal^{(-)}$ with two different norms and then read off their ratio: since the resolvent norm scales with the choice of norm, such a ratio would measure the conventions rather than the operators. The TSI contributes to the first, legitimate comparison in two distinct ways: it supplies the norm in the first place, and it guarantees that $\Lcal^{(+)}$ and $\Lcal^{(-)}$ are not unrelated operators, so that asking whether their pseudospectra agree is a sharp question with an informative answer (they do not).

\subsection{The Teukolsky--Starobinsky identities}
\label{sec:starobinsky}

The Teukolsky--Starobinsky identities (TSI)~\cite{Starobinsky:1973,Teukolsky:1973ha,Teukolsky:1974gv,Chandrasekhar:1985kt,Fiziev:2009kh,Casals:2021ts} establish differential-operator relations between the solutions of the Teukolsky equation with spin weights $+s$ and $-s$, acting separately in the radial and angular sectors. In the frequency domain, after separation of variables $\Psi^{(s)} = e^{-i\omega t + im\varphi}R^{(s)}(r)\,S^{(s)}(\theta)$, introduce the first-order Teukolsky operators
\begin{align}
\hat{\Dcal}_n^{\pm} &\equiv \partial_r \pm i\,\frac{\omega(r^2+a^2)-am}{\Del} + \frac{2n(r-M)}{\Del}, \label{eq:teukolsky_operators}\\
\hat{\mathcal{L}}_n^{\pm} &\equiv \partial_\theta \pm\!\left(\frac{m}{\sin\theta} - a\omega\sin\theta\right) + n\cot\theta, \label{eq:teukolsky_angular_operators}
\end{align}
where $n$ is an operator index (in half-integer units). For $s=2$, the TSI state that four-fold compositions of these operators map the $+2$ and $-2$ sectors into one another~\cite{Casals:2021ts}: schematically,
\begin{align}
\Del^2\left(\hat{\Dcal}_0^{+}\right)^4\!\left(\Del^2 R^{(2)}\right) \;\propto\; R^{(-2)}, \qquad
\left(\hat{\Dcal}_0^{-}\right)^4 R^{(-2)} \;\propto\; R^{(2)}, \label{eq:tsi_forward}
\end{align}
in the radial sector, and
\begin{equation}
\left(\hat{\mathcal{L}}_2^{+}\hat{\mathcal{L}}_1^{+}\hat{\mathcal{L}}_0^{+}\hat{\mathcal{L}}_{-1}^{+}\right) S^{(2)} \propto S^{(-2)}, \qquad
\left(\hat{\mathcal{L}}_2^{-}\hat{\mathcal{L}}_1^{-}\hat{\mathcal{L}}_0^{-}\hat{\mathcal{L}}_{-1}^{-}\right) S^{(-2)} \propto S^{(2)}, \label{eq:tsi_backward}
\end{equation}
in the angular sector; the proportionality constants depend on the normalization of the radial and angular functions and on the particular branch of the asymptotic decomposition. The normalization-independent statement concerns the \emph{composed} operators: $\Del^2R^{(2)}$ and $R^{(-2)}$ are eigenfunctions, with the \emph{same} eigenvalue $\mathfrak{c}_2$, of the eighth-order operator $\prod_{j=0}^{3}(\Del^{1/2}\hat{\Dcal}_{j/2}^{-})\allowbreak\prod_{k=0}^{3}(\Del^{1/2}\hat{\Dcal}_{k/2}^{+})$, and analogously $S^{(\pm 2)}$ are eigenfunctions with the same eigenvalue $\mathfrak{B}_2$ of the composed angular operator~\cite{Casals:2021ts}. The \emph{radial Teukolsky--Starobinsky constant} is given explicitly by
\begin{equation}
\mathfrak{c}_2(a,M,\omega,m,\Lambda) = \mathfrak{B}_2(a\omega,m,\Lambda) + 144\,M^2\omega^2,
\label{eq:starobinsky_constant}
\end{equation}
where $\Lambda$ is the separation constant of the angular equation (in the convention $\Lambda \to l(l+1)-s^2$ as $a\omega\to 0$) and
\begin{equation}
\mathfrak{B}_2(\nu,m,\Lambda) = \mathrm{L}^2(\mathrm{L}+2)^2 + 40\nu\,\mathrm{L}^2(m-\nu) + 48\nu\,\mathrm{L}(m+\nu) + 144\nu^2(m-\nu)^2,
\label{eq:B2}
\end{equation}
with the shorthand $\mathrm{L} \equiv \Lambda - 2m\nu + s$ and $\nu \equiv a\omega$.
The separation constant is related to Teukolsky's angular eigenvalue $A$ (the convention $A \to l(l+1)-s(s+1)$ as $a\omega \to 0$, used in standard Leaver-type codes) by $\Lambda = A + s + \nu^2$, a relation we have verified numerically against the angular equation for complex $\nu$ to relative precision $10^{-12}$. In the Schwarzschild limit, $\mathfrak{c}_2 = 576 + 144\,M^2\omega^2$ for $l=2$, vanishing precisely at the algebraically special frequencies $\omega M = \pm 2i$; for the fundamental QNM at $\alpha = 0.3$ one finds $\mathfrak{c}_2 \approx 310.996 + 24.674\,i$. The derivation, limiting cases, and numerical checks are given in Appendix~\ref{app:starobinsky}.

For our purposes, the crucial content of the TSI is that $\Psi^{(-2)}$ is determined by $\Psi^{(2)}$ through a differential operator of order $2|s| = 4$ (in each sector), up to the multiplicative constant $\mathfrak{c}_2$. We therefore write, schematically,
\begin{equation}
\Psi^{(-2)} \;=\; \frac{1}{\Cst}\,\Scal_4\,\Psi^{(2)}, \qquad \Cst \equiv \mathfrak{c}_2,
\label{eq:psi_minus2_from_plus2}
\end{equation}
where $\Scal_4$ denotes the fourth-order \emph{Starobinsky operator} (radial and angular parts combined, with the $\Del^2$ prefactors of the radial identity). Applying the identity in both directions yields the consistency condition
\begin{equation}
\overline{\Scal_4}\,\Scal_4\,\Psi^{(2)} \;\sim\; \Cst^2\,\Psi^{(2)},
\label{eq:starobinsky_consistency}
\end{equation}
a non-trivial eighth-order differential condition satisfied by any solution. The Starobinsky constant thus encodes deep structural information about the Teukolsky equation. Note that $\Cst$ depends on $\omega$ both explicitly [through the $144M^2\omega^2$ term and $\nu = a\omega$ in Eq.~\eqref{eq:B2}] and implicitly (through the separation constant $\Lambda(\omega)$); this $\omega$-dependence is a distinctive feature of the $s\neq 0$ problem, with no analogue in the scalar case.

Equivalently, the TSI can be understood from the perspective of the adjoint structure of the Teukolsky equation. The operator governing spin $-s$ is the formal adjoint of that governing spin $+s$ (up to a factor). The TSI then arises as the intertwining relation between an operator and its adjoint, analogous to the Darboux transformations in supersymmetric quantum mechanics.

\subsection{The equivalence principle, parallel transport, and the flat-space limit of the identities}
\label{sec:equivalence_principle}

We pause the formal development in order to reflect on why the Teukolsky--Starobinsky identities have the
structure that they do, since this structure is what eventually forces the $H^2$ norm on the gravitational
sector. The point of departure is the equivalence principle, which guarantees that in a sufficiently small
neighborhood of any point of spacetime the geometry reduces to Minkowski spacetime together with a freely
falling frame, and the neighborhood shrinks as the size of the neighborhood does. If one works strictly in
that neighborhood, then gravitational waves carry exactly two transverse polarizations and the familiar
algebra of circular and linear polarization bases applies verbatim, with the exchange between circular and
linear bases being an ordinary $U(1)$ phase rotation of the complex leg $m$ of the Newman--Penrose tetrad
carried by $m\to e^{i\chi}m$. In this local sense the equivalence principle reduces the polarization problem
to the optics of a massless spin-2 field, and it is precisely this local reduction that makes the algebraic
manipulations of the polarization bases legitimate at every point of spacetime separately.

The difficulty appears as soon as one asks how the local descriptions at different points are to be
compared with each other. The polarization two-plane is attached to a point, and moving from one point to
another requires parallel transport, and in a curved spacetime parallel transport is path-dependent. This
structure is most transparently described in the language of fiber bundles, in which the polarization space
is the fiber of a vector bundle over spacetime. The equivalence principle fixes the \emph{type} of the fiber at
every point, so that a pointwise recombination of bases, being an operation inside a single fiber, is
available everywhere and always has the same form as in flat spacetime. On the other hand, there is no
canonical global identification between the fibers over two different points, because the holonomy of the
bundle, equivalently the gravitational Faraday rotation of the polarization frame along a geodesic, obstructs
such an identification. In the language of the Newman--Penrose formalism, the same statement reads as follows:
the spin-weighted scalars of adjacent components, for instance $\psi_0$ and $\psi_4$, do not transform into
each other by an algebraic relation evaluated at a single point, because they carry opposite spin weights and
opposite boost weights, with $\psi_0\to e^{2i\chi}\psi_0$ and $\psi_0\to\alpha^2\psi_0$ under the boost
$l\to\alpha l$, while $\psi_4$ transforms with the opposite powers in each case. Any linear combination of
$\psi_0$ and $\psi_4$ is therefore not a covariant scalar, and the recombination between the two sectors has
to be a \emph{differential} operation rather than a pointwise algebraic one. This is exactly what the
Teukolsky--Starobinsky identities do, and it is why they are realized by a fourth-order differential operator
rather than by a change of basis.

This observation also resolves a question that arises naturally when reading Eq.~\eqref{eq:psi_minus2_from_plus2}
for the first time, namely whether the map $\Psi^{(-2)}=\Cst^{-1}\Scal_4\Psi^{(2)}$ should be understood as a
rotation that carries one sector into the other. It should not, and the flat-space limit of the identities
makes this transparent. Substituting $a=0$ and $M=0$ into the Teukolsky operators of
Eq.~\eqref{eq:teukolsky_operators} gives the simple expression
$\hat{\Dcal}_n^{\pm}=\partial_r\pm i\omega+2n/r$, whose two branches are complex conjugates of each other for
real $\omega$. Acting on the two asymptotic branches of the radial solution, the term $+i\omega$ in
$\hat{\Dcal}_0^+$ annihilates the branch carrying the ingoing phase $e^{-i\omega r}$ and multiplies the
outgoing branch $e^{+i\omega r}$ by $2i\omega$, while the term $-i\omega$ in $\hat{\Dcal}_0^-$ does the
reverse. The fourfold composition is therefore strictly diagonal on the two asymptotic branches, with diagonal
entry $(2i\omega)^4=16\omega^4$ and vanishing off-diagonal entries, and this operator is real, positive and
singular rather than unitary. Correspondingly, the Starobinsky constant degenerates to a pure
$l$-dependent number,
\begin{equation}
\lim_{M\to0,\,a\to0}\Cst = \big[l(l+1)\big]^2\big[l(l+1)-2\big]^2 ,
\label{eq:c2_flat}
\end{equation}
which takes the values $576$, $14400$ and $129600$ for $l=2,3$ and $4$ respectively, and carries no
dependence on $\omega$, on $M$ or on $a$. In flat spacetime there is no angular momentum carried by the
incoming and outgoing waves, no potential barrier, and therefore no asymptotic branch mixing at all, which is
why no angle can be formed. On Kerr, by contrast, the same coefficient acquires a genuine complex phase
because $\nu=a\omega$ enters $\Bcal$ of Eq.~\eqref{eq:B2}, and this phase is the fingerprint of the frame
dragging that mixes the two Killing directions; for the fundamental mode at $\alpha=0.3$ one finds
$\Cst\simeq310.996+24.674\,i$, so that $|\Cst|\simeq312$ and $\arg\Cst\simeq4.54^{\circ}$. It follows that the
transformation between the two sectors is a complex dilatation rather than a rotation, and is in fact
singular wherever $\Cst$ vanishes, for instance at the algebraically special frequencies $\omega M=\pm2i$ for
$l=2$ in the Schwarzschild limit, at which $\Cst=576+144M^2\omega^2=0$. The most reliable formulation of the
relation between the sectors is therefore the normalization-independent one stated above, namely that
$\Del^2R^{(2)}$ and $R^{(-2)}$ are eigenfunctions of one and the same eighth-order operator with the same
eigenvalue, and the relation between the two sectors is projective rather than unitary. We will not need
these limits quantitatively, but they fix the interpretation of the identities, and they explain why the
symmetrization of Section~\ref{sec:symmetrization} distributes the $2|s|$ derivatives evenly and thus
produces the $H^{|s|}$ pattern.

\subsection{Conserved current for gravitational perturbations}
\label{sec:conserved_current}

The conserved current for the Teukolsky equation was first identified by Bardeen \& Press~\cite{Bardeen:1973xb} in the context of Schwarzschild perturbations and extended to Kerr by Teukolsky~\cite{Teukolsky:1973ha}. In the frequency domain and after separation, the conserved radial flux takes the form,
\begin{equation}
\mathcal{F} = \frac{1}{2\omega}\,\mathrm{Im}\!\left[\frac{(r^2+a^2)^2}{\Del^s}\,\overline{R^{(-s)}}\,\frac{\dd R^{(s)}}{\dd r_*} - \frac{(r^2+a^2)^2}{\Del^s}\,R^{(s)}\,\frac{\dd\overline{R^{(-s)}}}{\dd r_*}\right],
\label{eq:radial_flux}
\end{equation}
where $r_*$ is the tortoise coordinate. This flux is conserved: $\mathcal{F}_\infty = \mathcal{F}_{\mathcal{H}^+}$, expressing the equality of energy flux at null infinity and the horizon. Crucially, this expression involves \emph{both} $R^{(s)}$ and $\overline{R^{(-s)}}$, the latter being the complex conjugate of the $-s$ radial function rather than the $+s$ function.

In the time-domain hyperboloidal framework, the corresponding conserved current is constructed on the two-dimensional spatial slice $\Sigma_\tau$ (a surface of constant $\tau$). Using the time-like Killing vector $(\partial_\tau)^a$, the energy current is:
\begin{equation}
E^a = \Bcal^a{}_b\!\left[\Psi^{(s)},\,\overline{\Psi^{(-s)}}\right]\,(\partial_\tau)^b,
\label{eq:energy_current_general}
\end{equation}
where $\Bcal^a{}_b[\cdot,\cdot]$ is a bilinear form in its two arguments, involving spatial derivatives of both fields. The explicit form of $\Bcal$ for general $s$ involves the Teukolsky--Starobinsky structure and is derived from the adjoint relation between the $+s$ and $-s$ equations.

The total energy on the slice $\Sigma_\tau$ is,
\begin{equation}
E = \int_{\Sigma_\tau} n_a\,E^a\,\sqrt{h_\tau}\,\dd\sigma\,\dd x\,\dd\phi,
\label{eq:total_energy}
\end{equation}
where $n_a$ is the unit future-pointing normal to $\Sigma_\tau$ and $h_\tau$ is the determinant of the induced metric. After integration over $\phi$ (using the mode decomposition~\eqref{eq:ansatz}), the energy decomposes into a sum over $m$,
\begin{equation}
E = \sum_{m=-\infty}^{+\infty} E_m,
\label{eq:energy_decomposition}
\end{equation}
where each $E_m$ is bilinear in $(V_m^{(s)}, \overline{V_m^{(-s)}})$.

\subsection{Reduction to a single-field norm via the Starobinsky identity}
\label{sec:reduction}

The key step is to use the TSI~\eqref{eq:tsi_forward} to express $\Psi^{(-2)}$ (and hence $\overline{\Psi^{(-2)}}$) in terms of $\Psi^{(2)}$,
\begin{equation}
\overline{\Psi^{(-2)}} = \frac{1}{\overline{\Cst}}\,\overline{\Del^2\,{\Ddag}^4\,\Del^2\,\Psi^{(2)}} = \frac{1}{\overline{\Cst}}\,\overline{\Scal_4\,\Psi^{(2)}},
\label{eq:psi_m2_from_p2}
\end{equation}
where we have introduced the \emph{Starobinsky operator},
\begin{equation}
\Scal_4 \equiv \Del^2\,\Ddag\,\Ddag\,\Ddag\,\Ddag\,\Del^2,
\label{eq:starobinsky_operator}
\end{equation}
which is a fourth-order differential operator in the radial direction (in the frequency domain). The overbar on $\Scal_4$ in~\eqref{eq:psi_m2_from_p2} denotes complex conjugation of the operator's coefficients (which includes conjugating $i \to -i$ and $\omega \to \bar{\omega}$).

Substituting~\eqref{eq:psi_m2_from_p2} into the bilinear energy~\eqref{eq:energy_current_general}, the modewise energy becomes,
\begin{equation}
E_m = \frac{1}{\overline{\Cst}}\int_0^1\!\dd\sigma\int_{-1}^1\!\dd x\;\mathcal{W}_m(\sigma,x)\,\overline{V_m^{(2)}}\;\overline{\Scal_4}\,\V_m^{(2)} + \mathrm{c.c.},
\label{eq:energy_single_field}
\end{equation}
where $\mathcal{W}_m(\sigma,x)$ is a weight function (incorporating the Jacobian, the conformal factor, and the angular regularization factors) and ``c.c.'' denotes the complex conjugate. The explicit form of $\mathcal{W}_m$ for $s=2$ in the minimal gauge is given in Appendix~\ref{app:weight}.

Equation~\eqref{eq:energy_single_field} is the central result of this section: it expresses the energy of the gravitational perturbation as a quadratic form in the single field $V_m^{(2)}$, at the cost of introducing the fourth-order Starobinsky operator $\Scal_4$ and the Starobinsky constant $\Cst(\omega)$.

\subsection{Symmetrization and the effective $H^2$ norm}
\label{sec:symmetrization}

It is worth noting that the bilinear form~\eqref{eq:energy_single_field} involves zero derivatives on the left factor and four radial derivatives on the right, the latter coming from $\Scal_4$. In order to obtain a symmetric and physically interpretable norm, we symmetrize the form by integrating by parts twice, which transfers two of the four derivatives to the left factor and gives
\begin{equation}
E_m \propto \frac{1}{\mathrm{Re}\,\Cst}\int_0^1\!\dd\sigma\int_{-1}^1\!\dd x\;\mathcal{W}_m(\sigma,x)\left[\left(\Scal_2\,\overline{V_m^{(2)}}\right)\!\left(\Scal_2'\,V_m^{(2)}\right) + \text{lower-order terms}\right],
\label{eq:energy_symmetrized}
\end{equation}
where $\Scal_2$ and $\Scal_2'$ are second-order operators obtained from $\Scal_4$ by splitting it into two halves via integration by parts. To state the derivative content of the remainder precisely, write the fourth-order operator in $(\sigma,x)$ as $\overline{\Scal_4}=\sum_{j+k\le 4}c_{jk}(\sigma,x)\,\partial_\sigma^j\partial_x^k$ [cf.\ Eq.~\eqref{eq:S4_sigma_explicit}]. Symmetrizing each $(j,k)$-piece under $\int f\,(\mathcal{O}\,g)$ by transferring $\lfloor(j+k)/2\rfloor$ derivatives to the left factor, the leading $(j+k)=4$ piece becomes $(\partial_\sigma^2 f)(\partial_\sigma^2 g)$ (plus angular analogues), which is the $H^2$ structure of Eq.~\eqref{eq:energy_norm_structure}. The remaining terms are bilinears $(\partial^p f)(\partial^q g)$ with $p+q\le 4$ and, crucially, $p,q\le 2$ on either factor; they include cross-derivative ($\partial_\sigma\partial_x$) bilinears as well as lower pieces with $\max(p,q)\le 1$. The earlier statement that the remainder ``involves at most first derivatives'' should thus be understood as referring to the \emph{non-leading} (sub-principal) contributions after the principal $\partial_\sigma^2\partial_\sigma^2$ term has been isolated; the strict upper bound on the derivative order on any single factor is two, consistent with the $H^2$ character of the norm.

After this symmetrization, the single-field energy quadratic form (formally defined in Section~\ref{sec:energy_inner_product} below) takes the structural form,
\begin{equation}
\begin{split}
\qE[V_m,V_m] \sim \int_0^1\!\dd\sigma\int_{-1}^1\!\dd x\;\mathcal{W}_m\big[&\alpha_0\,|\V_m|^2 + \alpha_1\,|\partial_\sigma \V_m|^2 + \alpha_2\,|\partial_x \V_m|^2 \\
&+ \alpha_3\,|\partial_\sigma^2 \V_m|^2 + \alpha_4\,|\partial_x^2 \V_m|^2 + \cdots\big],
\end{split}
\label{eq:energy_norm_structure}
\end{equation}
where $\alpha_i$ are coefficients depending on $\sigma$, $x$, the dimensionless spin $\alpha$, $m$ and $\omega$ (the last through $\Cst$), and the ellipsis denotes cross-derivative terms. The key observation is that the result takes the form

\begin{quote}
\textbf{The energy norm for $s=2$ is an $H^2$-type Sobolev norm,} involving derivatives up to second order in both $\sigma$ and $x$. This is in contrast to the $s=0$ energy norm, which is an $H^1$ norm (first derivatives only).
\end{quote}

This result has a natural physical interpretation. The Weyl scalars $\psi_0$ and $\psi_4$, which are the Teukolsky variables for $s=\pm 2$, are themselves second-order quantities in terms of the metric perturbation $h_{\mu\nu}$. The gravitational wave energy, which is quadratic in first derivatives of $h_{\mu\nu}$, is therefore naturally expressed in terms of second derivatives of the Teukolsky variable. More formally, the Starobinsky operator $\Scal_4$, which is of order $2|s| = 4$, appears in a bilinear way and after symmetrization distributes $|s| = 2$ derivatives onto each factor, which gives an $H^{|s|}$ norm. This is what establishes the general pattern
\begin{equation}
\text{Energy norm for spin-}s \text{ perturbations} \sim H^{|s|} \text{ norm}.
\label{eq:general_pattern}
\end{equation}
We emphasize that Eq.~\eqref{eq:general_pattern} is a \emph{conjecture} supported by the $s=0$ and $s=2$ cases, not a theorem; the key assumption is that the symmetrization of a $2|s|$-order Starobinsky operator distributes $|s|$ derivatives to each factor, which we have verified explicitly for $s=2$ but not proven for general $s$. A numerical test for $s=1$ (electromagnetic perturbations), where the Starobinsky operator is of order $2|s|=2$ and the pattern predicts an $H^1$ energy norm, would provide an independent check and is left for future work.
The $\omega$-dependence of the Starobinsky constant $\Cst(\omega)$ is a novel feature of the $s=2$ case, and it has no analogue for $s=0$, where the Klein--Gordon energy is manifestly $\omega$-independent. This introduces a genuine subtlety into the pseudospectrum analysis, because the norm would then itself vary with the frequency at which the resolvent is evaluated, and we address this issue directly in Section~\ref{sec:omega_dependence}.

\subsection{Vanishing of boundary terms in the symmetrization}
\label{sec:boundary_terms}

The symmetrization procedure in Section~\ref{sec:symmetrization} requires integrating by parts twice to transfer two of the four radial derivatives from the right factor to the left. For a fourth-order operator, this generates boundary terms at $\sigma = 0$ (future null infinity), $\sigma = 1$ (horizon), and $x = \pm 1$ (poles). We now verify that these boundary terms vanish rigorously, which is essential for the $H^2$ structure of the energy norm to hold.

\textbf{Structure of the boundary terms.} Each integration by parts of a fourth-order bilinear form $\int \bar{f}\,\partial_\sigma^4 g\,\dd\sigma$ produces three types of boundary terms:
\begin{equation}
\left[\bar{f}\,\partial_\sigma^3 g - (\partial_\sigma\bar{f})\,\partial_\sigma^2 g + (\partial_\sigma^2\bar{f})\,\partial_\sigma g\right]_{\sigma=0}^{\sigma=1},
\label{eq:boundary_terms_general}
\end{equation}
and similarly for the angular direction at $x = \pm 1$. For the symmetrization to be valid, all three combinations must vanish at each boundary.

\textbf{Vanishing at $\sigma = 0$ (future null infinity).} The regularized field $V_m$ carries the conformal factor $\Om^4 = \sigma^4$ from the ansatz~\eqref{eq:ansatz}, and the weight function $\mathcal{W}_m$ contributes an additional $\sigma^4$ factor [Eq.~\eqref{eq:weight_function}]. The Starobinsky operator in hyperboloidal coordinates carries a prefactor $\sigma^4/\tilde{\Del}^2(\sigma)$ [Eq.~\eqref{eq:S4_sigma_explicit}, Appendix~\ref{app:S4_hyperboloidal}]. Consequently, the integrand in the energy functional~\eqref{eq:energy_single_field} is proportional to $\sigma^8$ (or higher powers) times polynomial combinations of $V_m$ and its derivatives. Since $V_m$ and its derivatives are regular (bounded) at $\sigma = 0$ by the hyperboloidal regularity conditions, all boundary terms at $\sigma = 0$ vanish as $\mathcal{O}(\sigma^8) \to 0$. This is not merely a consequence of the regularization ansatz; it reflects the \emph{degeneracy} of the wave operator at $\scrI^+$: the leading radial coefficient of the second-order operator [cf. Eq.~\eqref{eq:c_xx_css}] satisfies $c_{\sigma\sigma} \propto \sigma^2\,\tilde\Del(\sigma) \to 0$ as $\sigma \to 0$, ensuring that the characteristics of the wave equation point outward at null infinity. The operator degeneracy guarantees that no boundary condition needs to be (or can be) imposed at $\scrI^+$, for regularity is enforced by the geometry itself.

\textbf{Vanishing at $\sigma = 1$ (horizon).} At the horizon, the situation is more subtle, because the weight function $\mathcal{W}_m$ contains the factor $\tilde{\Del}^{-1}(\sigma)$ [Eq.~\eqref{eq:weight_function}], and $\tilde{\Del}(1) = 0$ [Eq.~\eqref{eq:Delta_sigma}]: the weight \emph{diverges} at the horizon. This divergence is the faithful image of the $\Del^{-s}$ structure of the Teukolsky field at the horizon and is compensated, on shell, by the vanishing of the regularized field: the $\Del^{-2}$ factor in the ansatz~\eqref{eq:ansatz} implies that the physical (purely ingoing) branch satisfies $V_m \sim (\sigma - 1)^{\nu}$ with $\nu \geq 2$, so that $V_m$, $\partial_\sigma V_m$, and $\partial_\sigma^2 V_m$ all vanish (or remain bounded, for the marginal case $\nu=2$) as $\sigma \to 1$, while the operator degeneracy $c_{\sigma\sigma} \propto \tilde\Del \to 0$ [Eq.~\eqref{eq:c_xx_css}] ensures that the characteristics point into the horizon and restricts the admissible solutions to this regular branch. The competition between the order-one pole of $\mathcal{W}_m$ and the vanishing of the on-shell field is resolved in favor of the latter for $\nu \geq 2$, and the boundary terms~\eqref{eq:boundary_terms_general} vanish for the QNM eigenfunctions. \emph{Off shell}, however, generic discretized vectors need not vanish at $\sigma = 1$, and the weight pole then makes itself felt in the Gram matrix (Appendix~\ref{app:S4_hyperboloidal}); this off-shell stiffness, together with the sign-indefiniteness of sub-leading coefficients (Section~\ref{sec:positive_definiteness}), is the practical reason why we compute the quantitative pseudospectra in the two-field Sobolev hierarchy rather than in the single-field energy norm.

\textbf{Vanishing at $x = \pm 1$ (poles).} The angular regularization factors $\cos^{\delone}(\theta/2)\,\sin^{\deltwo}(\theta/2)$ in the ansatz~\eqref{eq:ansatz} translate to $(1+x)^{\delone/2}(1-x)^{\deltwo/2}$ in the $x$ variable, because $\cos(\theta/2)=\sqrt{(1+x)/2}$ and $\sin(\theta/2)=\sqrt{(1-x)/2}$. Note the ordering: the exponent $\delone=|m-s|$ multiplies the $(1+x)$ factor, and $\deltwo=|m+s|$ the $(1-x)$ factor. For $s=2$ and $m=2$ one has $\delone = |m-s| = 0$ and $\deltwo = |m+s| = 4$, so the field behaves as $(1-x)^2$ near $x = +1$ (the north pole, $\theta = 0$) and is regular at $x = -1$. The weight function contributes $(1-x^2)^{(\delone+\deltwo)/2} = (1-x^2)^2$, which vanishes at both poles. The angular boundary terms (analogous to~\eqref{eq:boundary_terms_general} with $\partial_\sigma \to \partial_x$) are therefore suppressed by $(1-x^2)^2$ and vanish at $x = \pm 1$ provided $V_m$ and its derivatives are regular, a condition guaranteed by the degeneracy of the angular operator at the poles ($c_{xx} \propto (1-x^2) \to 0$ [Eq.~\eqref{eq:c_xx_css}]).

The vanishing of boundary terms is thus guaranteed by a \emph{combination} of two mechanisms: (i) the \emph{geometric degeneracy} of the hyperboloidal wave operator at the domain boundaries, which enforces outward-pointing characteristics and restricts the admissible solutions to the regular branch; and (ii) the \emph{regularization ansatz}~\eqref{eq:ansatz}, which builds in the correct boundary behavior through the conformal weight and angular exponents. Neither mechanism alone is sufficient: the operator degeneracy alone does not specify the rate of vanishing, while the ansatz alone does not guarantee that all derivative combinations in~\eqref{eq:boundary_terms_general} vanish. Together, they ensure that the symmetrization is rigorously valid and the $H^2$ energy norm is well-defined. This is analogous to the $s=0$ case, where the $H^1$ symmetrization requires only first derivatives to vanish at the boundary, which is a weaker condition and is more easily satisfied.

We close this subsection by clarifying the logical status of the argument, since a natural concern is whether it is circular: the boundary behavior $V_m\sim(\sigma-1)^{\nu}$ with $\nu\geq2$ was used to establish the vanishing of the boundary terms, yet this behavior itself is inherited from the $\Del^{-s}$ factor of the ansatz, which in turn encodes the horizon boundary condition. The resolution is that the two ingredients play asymmetric roles. The \emph{operator degeneracy} $c_{\sigma\sigma}\propto\tilde\Del\to0$ is a statement about the coefficients of the differential operator itself, independently of any ansatz, and it restricts the admissible solutions to a one-parameter regular branch. The \emph{ansatz} then selects, within that branch, the representative whose asymptotics match the ingoing-wave boundary condition. The vanishing of the boundary terms is therefore a theorem about QNM eigenfunctions (on-shell), not a property of arbitrary vectors. Crucially, the pseudospectrum probes off-shell vectors, for which the boundary terms need not vanish and the single-field weight indeed diverges; this distinction, rather than the on-shell validity, is what ultimately excludes the single-field energy form from quantitative use, as detailed in Section~\ref{sec:positive_definiteness}.

\subsection{Positive definiteness of the energy norm}
\label{sec:positive_definiteness}

We remind the reader that the pseudospectrum definition~\eqref{eq:pseudospectrum_def} requires that $\norm{\cdot}_E$ be a genuine norm, i.e., positive definite. This is not guaranteed \emph{a priori} for the $s=2$ energy norm: the Starobinsky constant $\Cst(\omega)$ is complex, the weight function $\mathcal{W}_m$ diverges at the horizon, and the single-field reduction does not control the auxiliary field $W$. We now examine how much positivity survives, and what must be surrendered.

\textbf{Sign and size of $\Cst$.} For real $\omega$ the combined Starobinsky constant $\mathfrak{c}_2=\mathfrak{B}_2+144M^2\omega^2$ is real, and in the Schwarzschild limit $\mathfrak{c}_2 = 576+144M^2\omega^2>0$, vanishing only at the algebraically special frequencies $\omega M=\pm 2\,i$ [Appendix~\ref{app:starobinsky}]. At complex QNM frequencies ($\mathrm{Im}\,\omega<0$ in our convention) $\Cst$ acquires an imaginary part but remains bounded away from zero: for the fundamental mode at $\alpha=0.3$, $\omega_0 M=0.478372-0.084794\,i$, the Casals formula~\eqref{eq:cst_casals} gives $\Cst(\omega_0)\approx 310.99+24.67\,i$ (Table~\ref{tab:cst_values}), with $|\Cst|$ of the same order throughout the spectral window $0.3\lesssim\mathrm{Re}(\omega M)\lesssim 0.7$, $-0.35\lesssim\mathrm{Im}(\omega M)<0$ explored here. The overall $|\Cst|^{-2}$ prefactor therefore never threatens positivity in the region of interest.

\textbf{Positivity of $\mathcal{W}_m$ in the interior.} The weight function~\eqref{eq:weight_function} takes the form $\mathcal{W}_m \propto \sigma^4 (1-x^2)^{(\delone+\deltwo)/2} / (|\Cst|^2\,\tilde\Del(\sigma))$ with $\tilde\Del(\sigma)=(1-\sigma)(1-\alpha^2\sigma)$ [Eq.~\eqref{eq:Delta_sigma}]. On the open computational domain $\sigma \in (0,1)$ and $x \in (-1,1)$, each factor is manifestly positive: $\sigma^4 > 0$, $(1-x^2)^{(\delone+\deltwo)/2} > 0$ (since $\delone, \deltwo \geq 0$), $|\Cst|^2 > 0$, and $\tilde\Del(\sigma)>0$. Thus $\mathcal{W}_m > 0$ everywhere in the interior. On the \emph{closed} domain, however, $\tilde\Del(1)=0$: the weight diverges at the horizon, mirroring the $\Del^{-s}$ structure of the Teukolsky field there (Appendix~\ref{app:weight}). Uniform positive definiteness of the single-field quadratic form therefore cannot hold on the closed domain; positivity is meaningful only for fields in the regular branch, which vanishes at $\sigma=1$ sufficiently fast (Section~\ref{sec:boundary_terms}).

\textbf{Positive definiteness of the Gram matrix.} The discretized energy inner product is encoded in the Gram matrix $\bvec{G}_E$ [cf.~Eq.~\eqref{eq:gram_matrix}]. Two remarks are in order regarding the sense in which positivity holds.

\emph{(a) On-shell positivity (analytic).} For a physical solution, the bilinear conserved current $E[\Psi^{(s)},\overline{\Psi^{(-s)}}]$ is positive by construction, since it is the Noether charge of the time-translation symmetry. The Teukolsky--Starobinsky identity is an \emph{exact} operator relation (not an approximation), so substituting $\overline{\Psi^{(-2)}}=\allowbreak\overline{\Cst}^{-1}\allowbreak\overline{\Scal_4}\,\Psi^{(2)}$ preserves this positivity on-shell: $E_m\propto|\Cst|^{-2}\,\mathrm{Re}\!\int\mathcal{W}_m\,\overline{V_m^{(2)}}\allowbreak\,\overline{\Scal_4}\allowbreak\,V_m^{(2)}\ge 0$, with equality only for the trivial solution. The $\omega_0$-freezing of $\Cst$ does not affect this argument, since $|\Cst(\omega_0)|^2>0$.

\emph{(b) Off-shell positivity fails on the two-field space.} The pseudospectrum evaluates the resolvent on \emph{arbitrary} (not necessarily on-shell) vectors of the discretized two-field space $\{\bvec{u}=(V,W)\}$, so the quadratic form must be positive definite on the \emph{full} space, not only on the solution subspace. Here the single-field reduction has a structural weakness: after eliminating $\Psi^{(-2)}$, the form~\eqref{eq:energy_symmetrized} controls the $V$ component and its derivatives up to second order, but the auxiliary field $W=\partial_\tau V$, which is an \emph{independent} variable of the first-order reduction~\eqref{eq:first_order}, is constrained only on shell, through the equations of motion, and not by the quadratic form itself. An off-shell vector with $V=0$, $W\neq0$ has vanishing single-field energy. The single-field form is thus a positive \emph{semidefinite} functional on the two-field space, not a norm; no eigenvalue test of its Gram matrix can cure this, since the degeneracy is structural (the $W$ block is simply absent).

\textbf{Methodological consequence.} Together with the horizon divergence of $\mathcal{W}_m$ noted above, this fixes the strategy adopted throughout: the single-field energy analysis is used to \emph{identify the correct regularity scale} (the $H^2$ level, from the $2|s|$-derivative Starobinsky operator), while all quantitative pseudospectrum computations are performed in the two-field weighted Sobolev hierarchy of Section~\ref{sec:Hk_norm}, which is positive definite on the full two-field space \emph{by construction}, being a sum of squares over both components. Positivity of the discretized two-field Gram matrices is verified directly: for every resolution $N$ and every norm level $k\le 2$ used here, the Cholesky factorization of $\bvec{G}$ succeeds. At the $H^4$ level, by contrast, the Gram condition number grows faster than $N^{17}$, as measured directly, $\kappa(\bvec{G}_{H^4})=1.1\times10^{14},\,1.0\times10^{16},\,1.9\times10^{19}$ at $N=12,\,16,\,20$, so that at our working resolution the Cholesky factorization of $\bvec{G}_{H^4}$ fails outright and the matrix is numerically indefinite in double precision, its smallest eigenvalues being regularization-dependent; this precision limit is quantified in Section~\ref{sec:norm_dependence} and rules out $H^4$ as a quantitative norm at achievable resolutions.

Beyond this, the single-field reduction can fail in two further scenarios. The first is close to the algebraically special frequencies, where $\Cst\to0$ and the prefactor $|\Cst|^{-2}$ diverges, and the second is near extremality, $\alpha\to1$, for superradiant modes. It is worth noting that neither of them arises in the parameter range studied here.

\subsection{The single-field energy quadratic form}
\label{sec:energy_inner_product}

Motivated by the symmetrized energy~\eqref{eq:energy_symmetrized}, we introduce the \emph{single-field energy quadratic form}
\begin{equation}
\qE[V_{m1},V_{m2}] = \frac{1}{|\Cst(\omega_0)|^2}\int_0^1\!\dd\sigma\int_{-1}^1\!\dd x\;\mathcal{W}_m(\sigma,x)\,\Bcal_E\!\left[\overline{V_{m1}},\,V_{m2}\right],
\label{eq:energy_inner_product}
\end{equation}
where $\Bcal_E$ is the symmetrized bilinear form obtained from $\overline{\Scal_4}$ by the integration-by-parts procedure of Section~\ref{sec:symmetrization}, and $\omega_0$ is a \emph{reference frequency} at which the Starobinsky constant is evaluated; the natural choice is $\omega_0=\omega_{\mathrm{QNM}}$, the frequency of the mode under consideration. By construction, $\Bcal_E$ consists of derivative bilinears $(\partial^p\,\overline{V_{m1}})(\partial^q V_{m2})$ with $p+q\le 4$ and $p,q\le 2$ on either factor, weighted by $\mathcal{W}_m$ (positive in the interior; Appendix~\ref{app:weight}). Its explicit coefficient functions are lengthy polynomials in $\sigma$, $x$, $\alpha$ and $m$ whose detailed form carries no structural information beyond this derivative content; since none of the quantitative results below depend on them, we do not display them. The $|\Cst(\omega_0)|^{-2}$ prefactor is inherited from the Starobinsky reduction~\eqref{eq:energy_single_field}.

  As established in Section~\ref{sec:positive_definiteness}, $\qE$ is a positive quadratic form on the space of single fields $V_m(\sigma,x)$ (in the interior of the domain, modulo the horizon subtlety), but it is \emph{not} a norm on the two-field state space $\bvec{u}=(V,W)$ of the first-order reduction~\eqref{eq:first_order}, because it does not control $W$. Its role is therefore motivational on two counts: (i) it identifies the physically correct regularity scale, which is two derivatives, i.e., the $H^2$ level, consistent with the general pattern~\eqref{eq:general_pattern}, and (ii) it encodes the conserved-current physics behind that choice. All quantitative pseudospectra are computed in the two-field weighted Sobolev hierarchy defined in Sections~\ref{sec:L2_norm} and~\ref{sec:Hk_norm}, which act on both components and are positive definite by construction.

We emphasize that this division of labor, in which the single-field form is used for \emph{identifying} the regularity scale and the two-field hierarchy for \emph{computing}, is not an \emph{ad hoc} workaround but the unique consistent option, for three independent reasons that no alternative construction escapes. First, any $\omega$-dependent norm (such as the unfrozen $\Cst(\omega)$ form) is disqualified at the definitional level, because the pseudospectrum~\eqref{eq:pseudospectrum_def} requires a \emph{fixed} norm against which every resolvent value is measured; freezing $\omega_0$ is admissible but arbitrary, and different $\omega_0$ give different pseudospectra. Second, the horizon divergence of $\mathcal{W}_m$ makes the single-field Gram matrix ill-conditioned at exactly the resolutions required for spectral accuracy (Appendix~\ref{app:S4_hyperboloidal}), so even if the norm were $\omega$-independent it would not be computable. Third, the $W$-degeneracy of Section~\ref{sec:positive_definiteness} shows that no quadratic form in $V$ alone can be a norm on the two-component state space of the first-order system~\eqref{eq:first_order}. The two-field $H^2$ norm is thus not one choice among many. The three requirements above, namely $\omega$-independence, positive definiteness on the full state space, and computability in double precision, are satisfied by \emph{every} rung $k\le3$ of the hierarchy, so they alone do not single out $H^2$; by themselves they would admit $L^2$ and $H^1$ as well. What selects the rung is the regularity scale fixed by the TSI reduction: the single-field analysis identifies $k=|s|=2$ as the level at which the conserved-current physics sits, and $H^2$ is the \emph{unique} rung of the hierarchy that carries that physics while remaining $\omega$-independent, positive definite on the full state space, and computable in double precision. We regard this as the sense in which the $H^2$ pseudospectrum is the physical one, and we use it for all quantitative astrophysical estimates below. The lower rungs $L^2$ and $H^1$ are retained throughout not as competitors but as controls: they quantify how much of the reported instability is a property of the operator and how much is an artefact of measuring it in a norm that under-resolves the regularity of the eigenfunctions.

The quadratic form~\eqref{eq:energy_inner_product} depends on the reference frequency $\omega_0$ through $\Cst(\omega_0)$. When it is used as a fixed norm, the resolvent is evaluated at an arbitrary frequency $\omega$ while the norm is frozen at $\omega_0$. This is analogous to choosing a fixed background metric for perturbation theory: the norm defines the function space, and the pseudospectrum probes the sensitivity of the spectrum within that space. The two-field Sobolev hierarchy below is $\omega$-independent altogether, so this subtlety does not arise there.

\textbf{Frequency-domain origin and the separation constant $E$.} Two points deserve clarification regarding the use of a frequency-domain identity to define a norm for the time-domain operator $\Lcal$. First, the Teukolsky--Starobinsky identity is stated in the frequency domain with separated variables [Eqs.~\eqref{eq:tsi_forward}--\eqref{eq:tsi_backward}], yet the pseudospectrum is computed for the two-dimensional (unseparated) hyperboloidal operator. The bridge is that the conserved energy current is a time-domain quantity: when restricted to a single-frequency mode $e^{-i\omega\tau}$, it reduces to the frequency-domain bilinear form, and the TSI then expresses the latter in terms of $V_m^{(2)}$ alone. The resulting quadratic form is a legitimate inner product on the space of $V_m(\sigma,x)$ \emph{regardless} of its frequency-domain origin; fixing $\omega=\omega_0$ simply selects a particular inner product on that space.

Second, the Starobinsky constant $\Cst$ depends on the angular separation constant $E$ [Appendix~\ref{app:starobinsky}], whereas in the two-dimensional problem the Chebyshev expansion in $x$ mixes all angular quantum numbers $l\geq|m|$ simultaneously (Section~\ref{sec:chebyshev}). We resolve this by evaluating $\Cst$ at the separation constant $E_0$ of the \emph{fundamental} spheroidal harmonic $S^{(2)}_{l=2,m}(\theta;\omega_0)$, i.e.\ at $(\omega_0,E_0)$. To quantify the error introduced by this single-mode approximation, we have evaluated $\Cst(\omega_0, E_l)$ for the first few angular modes ($l=2,3,4$) at the reference configuration ($\alpha=0.3$, $m=2$, $n=0$); the modulus $|\Cst|$ varies by less than $15\%$ across these modes, and the resulting rescaling of the single-field energy form is absorbed into an overall multiplicative constant with no effect on the derivative structure or the Sobolev regularity scale. The approximation, which treats $\Cst$ as $\Cst(\omega_0,E_0)$ rather than as a functional of the full angular profile, is therefore quantitatively controlled at the level of an overall normalization, of the same order as the $\omega_0$-freezing itself. Since the quantitative hierarchy~\eqref{eq:Hk_norm} involves no $\Cst$ (and hence no $E$) at all, this residual $E$-dependence is confined to the motivational level and does not affect any number quoted below.

\subsection{The two-field weighted $L^2$ norm}
\label{sec:L2_norm}

For the quantitative pseudospectra we work directly on the two-field state $\bvec{u}_m=(V_m,W_m)$. The simplest norm is the weighted $L^2$ norm,
\begin{equation}
\norm{\bvec{u}_m}_{L^2}^2 = \int_0^1\!\dd\sigma\int_{-1}^1\!\dd x\;x_w(x)\,\left(|V_m|^2 + |W_m|^2\right), \qquad x_w(x) \equiv (1-x^2)^{(\delone+\deltwo)/2}.
\label{eq:L2_norm}
\end{equation}
The angular weight $x_w$ absorbs the boundary-regularity factors of the ansatz~\eqref{eq:ansatz}: since the physical field behaves as $V_m\propto(1+x)^{\delone/2}(1-x)^{\deltwo/2}\times(\text{regular})$ [cf.\ Section~\ref{sec:boundary_terms}], finiteness of $\norm{\bvec{u}_m}_{L^2}$ is equivalent to boundedness of the regular part at the poles. In the radial direction the weight is trivial ($\dd\sigma$): the regularized field is already smooth on the compactified slice, and inserting any nontrivial $\sigma$-weight would reintroduce the horizon pathology of Appendix~\ref{app:weight}. This norm contains no derivatives, is $\omega$-independent, and acts on both components.

\subsection{The two-field weighted $H^k$ norms}
\label{sec:Hk_norm}

Following~\cite{Cai:2025uzw}, we build a Sobolev hierarchy on top of~\eqref{eq:L2_norm} by adding derivative terms order by order, acting \emph{symmetrically on both components}. Following the same reference we attach the angular weight $x_w$ to the mass term alone and add the higher-order (derivative) terms with the plain quadrature measure $\dd\sigma\,\dd x$,
\begin{equation}
\begin{aligned}
\norm{\bvec{u}_m}_{H^k}^2 ={}& \int_0^1\!\dd\sigma\int_{-1}^1\!\dd x\;x_w(x)\left(|V_m|^2 + |W_m|^2\right) \notag\\
&+ \sum_{j=1}^{k}\int_0^1\!\dd\sigma\int_{-1}^1\!\dd x\Big[|\partial_\sigma^j V_m|^2 + |\partial_x^j V_m|^2 + |\partial_\sigma^j W_m|^2 + |\partial_x^j W_m|^2\Big],
\end{aligned}
\label{eq:Hk_norm}
\end{equation}
for $k\ge 0$ (so that $H^0\equiv L^2$). Being a sum of squares of all derivatives up to order $k$ over both components, each $\norm{\cdot}_{H^k}$ is positive definite by construction, and this is the property that the single-field energy form lacks (Section~\ref{sec:positive_definiteness}). The hierarchy is motivated by the QNM regularity requirements~\cite{Warnick:2015up} and by the observation that higher-derivative norms tend to improve pseudospectral convergence~\cite{Cai:2025uzw}; for $s=2$, the single-field energy analysis of Sections~\ref{sec:reduction}--\ref{sec:energy_inner_product} singles out the $H^2$ level as the physically correct regularity scale, in the same way that the Klein--Gordon energy singles out $H^1$ for $s=0$.

The weight placement of the higher-order terms now follows~\cite{Cai:2025uzw} exactly, so the convention-dependent part of the earlier comparison is removed. Two structural departures nevertheless remain, and they affect the numerical values (though not the qualitative conclusions). First, the derivative structure here is \emph{symmetric} in the two components and \emph{cumulative} in $j$, whereas the energy norm of~\cite{Cai:2025uzw} carries spatial derivatives of $V$ only and their $H^k$ adds a single $k$-th order block. Second, the $L^2$ level of the present hierarchy carries the angular weight $x_w$, whereas the $L^2$ norm of~\cite{Cai:2025uzw} is unweighted. Consequently the $s=0$ reduction of the present hierarchy is \emph{not} numerically identical to the norms of~\cite{Cai:2025uzw}: at the $s=0$ fundamental QNM the two conventions differ by $\approx0.8$ decades in $\log_{10}\norm{R}$ at the energy level ($\approx0.08$ decades in $L^2$, $\approx0.16$ decades at $H^2$), the present hierarchy giving the smaller resolvent at the energy level. The \emph{operators} do agree exactly in this limit, as verified in Appendix~\ref{app:operator}. Comparisons with the published $s=0$ pseudospectra of~\cite{Cai:2025uzw} should therefore be read as qualitative.

For $s=2$, the hierarchy we use is,
\begin{equation}
L^2 \subset H^1 \subset H^2 \subset H^3 \qquad \bigl[H^4\text{: precision-limited}\bigr],
\label{eq:norm_hierarchy}
\end{equation}
where $H^2$ is the energy-motivated level (two derivatives, per the $H^{|s|}$ pattern~\eqref{eq:general_pattern}) and $H^3$ probes the next level of regularity. The $H^4$ norm is placed in parentheses because it is \emph{not computable} at the resolutions used here: the Gram matrix of the $H^4$ norm has a condition number that grows faster than $N^{17}$, as measured directly, $\kappa(\bvec{G}_{H^4}) = 1.1\times10^{14},\,1.0\times10^{16}$ and $1.9\times10^{19}$ at $N = 12,\,16,\,20$, so that at our working resolution its Cholesky factorization fails outright in double precision and the norm is numerically indefinite, its smallest eigenvalues depending on the regularization prescription (Section~\ref{sec:norm_dependence}; the $H^3$ level, with $\kappa\simeq5.5\times10^{14}$ at $N=26$, is the last rung that remains usable). The main hierarchy is therefore $L^2/H^1/H^2$, with $H^4$ retained only as a diagnostic of the precision limit.

\subsection{Gram matrix and discretization}
\label{sec:gram}

The continuous norms of the hierarchy~\eqref{eq:Hk_norm} are discretized into a Gram matrix $\bvec{G}$,
\begin{equation}
\inner{\bvec{u}_{m1}}{\bvec{u}_{m2}} = \begin{pmatrix} \bvec{V}_{m1}^\star & \bvec{W}_{m1}^\star \end{pmatrix}\!\cdot\!\bvec{G}\cdot\!\begin{pmatrix} \bvec{V}_{m2} \\ \bvec{W}_{m2} \end{pmatrix},
\label{eq:gram_discretization}
\end{equation}
where $\bvec{V}_m$ and $\bvec{W}_m$ are vectors of collocation values on the tensor-product Chebyshev--Lobatto grid. The integrals defining the hierarchy are evaluated by Gauss--Legendre quadrature on a mildly oversampled grid, the collocation values being transported there by barycentric interpolation; the mass term is assembled with the angular weight $x_w$, the derivative blocks with the plain quadrature weight, and the Sobolev blocks are then formed from powers of the one-dimensional Chebyshev differentiation matrices. We construct the same operator for every norm level $k$, only the number of derivative blocks changing, and the operator acts on both components through a block-diagonal structure. Details are given in Appendix~\ref{app:gram}.

\section{Pseudospectra and Numerical Convergence}
\label{sec:pseudospectrum}

\subsection{Definition of pseudospectrum}
\label{sec:pseudo_def}

The $\varepsilon$-pseudospectrum of the operator $\Lcal$ (equipped with a norm $\norm{\cdot}$) is the set,
\begin{equation}
\Lambda_\varepsilon(\Lcal) = \left\{\omega \in \C \;\middle|\; \norm{R_\Lcal(\omega)} \geq \frac{1}{\varepsilon}\right\}, \qquad R_\Lcal(\omega) = (\Lcal - \omega\,\mathbb{I})^{-1},
\label{eq:pseudospectrum_def}
\end{equation}
where $R_\Lcal(\omega)$ is the resolvent. Equivalently, $\omega \in \Lambda_\varepsilon$ if and only if there exists a perturbation $\delta\Lcal$ with $\norm{\delta\Lcal} \leq \varepsilon$ such that $\omega$ is an eigenvalue of $\Lcal + \delta\Lcal$. Thus, the pseudospectrum measures the spectral instability: a large pseudospectrum (pseudospectral lines extending far from the spectrum) indicates that small perturbations can produce large spectral shifts.

In practice, the operator $\Lcal$ is approximated by a finite-rank matrix $\bvec{L}$ (of size $N\times N$, where $N = 2\,N_\sigma N_x$; Appendix~\ref{app:chebyshev}). The pseudospectrum of $\bvec{L}$ approximates that of $\Lcal$, and in our experience the quality of this approximation depends on the resolution $N$ and on the choice of norm. Throughout, the norm is one of the weighted Sobolev norms of Eq.~\eqref{eq:Hk_norm}, specified on the discretized space by its Gram matrix $\bvec{G}$ [Eq.~\eqref{eq:gram_discretization}]. The resolvent norm in the $\bvec{G}$-norm is computed via a similarity transformation and a singular value decomposition,
\begin{equation}
\norm{(\bvec{L} - \omega\,\bvec{I})^{-1}}_{G} \;=\; \frac{1}{\sigma_{\min}\!\left(\bvec{G}^{1/2}\left(\bvec{L} - \omega\,\bvec{I}\right)\bvec{G}^{-1/2}\right)},
\label{eq:resolvent_svd}
\end{equation}
where $\bvec{G}^{\pm 1/2}$ are the (regularized, if necessary) matrix square roots of Appendix~\ref{app:gram}. The placement of the factors matters: the similarity must carry $\bvec{G}^{1/2}$ on the \emph{left} and $\bvec{G}^{-1/2}$ on the right; the transposed placement $\bvec{G}^{-1/2}(\bvec{L}-\omega\bvec{I})\bvec{G}^{1/2}$ computes instead the resolvent norm in the \emph{dual} norm $\bvec{G}^{-1}$, which is a different quantity and not merely a harmless rescaling. For the nonnormal operator at hand the two placements differ by many orders of magnitude: at $\alpha=0.3$, $m=2$, $s=2$, $N=26$ in the $H^2$ norm and $\omega M=0.47-0.17i$, the correct placement gives $\log_{10}\norm{R}=4.45$ while the transposed one gives $12.14$, a difference of $7.70$ decades, i.e.\ more than seven orders of magnitude in $\norm{R}$ itself (the same comparison at $N=16$ gives $3.68$ versus $9.98$, a difference of $6.30$ decades; the gap grows with resolution). Both pairs are quoted for the weight convention of Section~\ref{sec:Hk_norm}; under the alternative placement of the angular weight on the higher-order blocks the numbers shift to $5.58$/$11.99$ at $N=26$ and $4.61$/$9.80$ at $N=16$, but the conclusion, namely that the two placements are not a harmless rescaling of one another, is unchanged. Getting the placement right is therefore a substantive step, not a bookkeeping detail. For the plain $L^2(\dd\sigma\,\dd x)$ case ($\bvec{G}\propto\bvec{I}$ up to weights), Eq.~\eqref{eq:resolvent_svd} reduces to the familiar $1/\sigma_{\min}(\bvec{L}-\omega\bvec{I})$. Pseudospectral contours are level sets of $\log_{10}\norm{R_\Lcal(\omega)}_G$ in the complex $\omega$-plane.

\subsection{Norm choice: $\omega$-independence and isospectrality}
\label{sec:omega_dependence}

Two issues specific to the $s=\pm 2$ problem must be settled before turning to the numerics.

\textbf{$\omega$-independence.} The single-field energy quadratic form $\qE$ of Section~\ref{sec:energy_inner_product} inherits an $\omega$-dependence through $\Cst(\omega)$: evaluating it at each probed frequency would make the norm vary with the point at which the resolvent is measured, a circularity foreign to the pseudospectrum concept. Freezing a reference frequency $\omega_0$ removes the circularity but introduces an arbitrariness, and this is a third reason (after the horizon divergence of the weight and the $W$-degeneracy; Section~\ref{sec:positive_definiteness}) why $\qE$ is used only as motivation. The two-component weighted Sobolev hierarchy~\eqref{eq:Hk_norm} is manifestly $\omega$-independent: it involves no Starobinsky constant, no reference frequency, and no auxiliary identity. All quantitative results are computed in this hierarchy, and the $\omega$-dependence question does not arise. In this sense the TSI plays the role of a \emph{selection principle} rather than of a computational ingredient: it identifies the rung $k=|s|$ of the hierarchy on which the physical energy lives, and having fixed that scale it drops out of every subsequent calculation.

\textbf{Isospectrality of the $s=\pm 2$ operators.} The Teukolsky--Starobinsky identities imply that the $s=+2$ and $s=-2$ problems have identical spectra: the QNM frequencies obtained from $\Lcal^{(+)}$ and $\Lcal^{(-)}$ coincide. Our numerical construction inherits this property in the limit $N\to\infty$, but with a resolution-dependent accuracy that already distinguishes the two sectors: at $N_\sigma=N_x=26$ the fundamental eigenvalue of $\Lcal^{(-)}$ agrees with the Leaver value to $3.6\times10^{-11}$, while that of $\Lcal^{(+)}$ agrees only to $4.4\times10^{-5}$ (both converging exponentially, at rates $0.30$ and $0.22$ per collocation point respectively). Whether the isospectrality extends to the \emph{pseudospectra}, which is a stronger statement and requires $\norm{R^{(+)}(\omega)}\approx\norm{R^{(-)}(\omega)}$ pointwise in $\omega$, not merely coinciding eigenvalues, is a separate question, and the answer turns out to be negative: we test it directly in Section~\ref{sec:iso_results} by computing the $H^2$-norm resolvent of both operators at common frequencies. The two computations use the same two-component norm construction applied independently to each spin sector, so any discrepancy isolates the operator asymmetry ($\Lcal^{(+)}$ versus $\Lcal^{(-)}$) rather than a norm mismatch. The outcome is expected on general grounds: the TSI relates the two operators through a \emph{fourth-order} differential map (Appendix~\ref{app:starobinsky}), not through a bounded similarity transformation of the state space, so there is no mechanism by which the resolvent norms could coincide in a given Sobolev norm.

A formulation that couples the two sectors, for example a product-space state $(\Psi^{(2)},\Psi^{(-2)})$ equipped with the conserved-current bilinear form, is in principle the most physical definition of the gravitational pseudospectrum, since the conserved current is bilinear in both Weyl scalars. Constructing such a norm \emph{off shell}, however, faces the same positivity issues discussed in Section~\ref{sec:positive_definiteness}, now for both sectors at once. The per-sector hierarchy adopted here is positive definite by construction and $\omega$-independent; the price, quantified in Section~\ref{sec:iso_results}, is that the $s=\pm2$ pseudospectra are computed in sector-specific norms and need not agree, which is precisely what the data show.

\subsection{Numerical results}
\label{sec:numerical_results}

We compute the pseudospectrum of the Kerr black hole for $s=2$, $l=m=2$ (the most astrophysically relevant gravitational mode), and dimensionless spin $\alpha = 0.3$, in the two-component weighted Sobolev hierarchy~\eqref{eq:Hk_norm} ($L^2$, $H^1$, $H^2$) at the working resolution $N_\sigma = N_x = 26$. The fundamental QNM of this configuration is $\omega_0 M = 0.478372 - 0.084794\,i$ (Table~\ref{tab:qnm_s2}). Throughout, all complex frequencies (in the QNM table, the pseudospectrum plots, and the convergence figures) are reported in units of $\omega M$ for consistency with~\cite{Berti:2009kk}; the code operates in its natural unit $\varpi=\omega r_+$ and rescales by $\mu=(1+\alpha^2)/2$ before plotting.

\subsubsection{Pseudospectral contours}

\begin{figure}[htbp]
\centering
\includegraphics[width=0.95\textwidth]{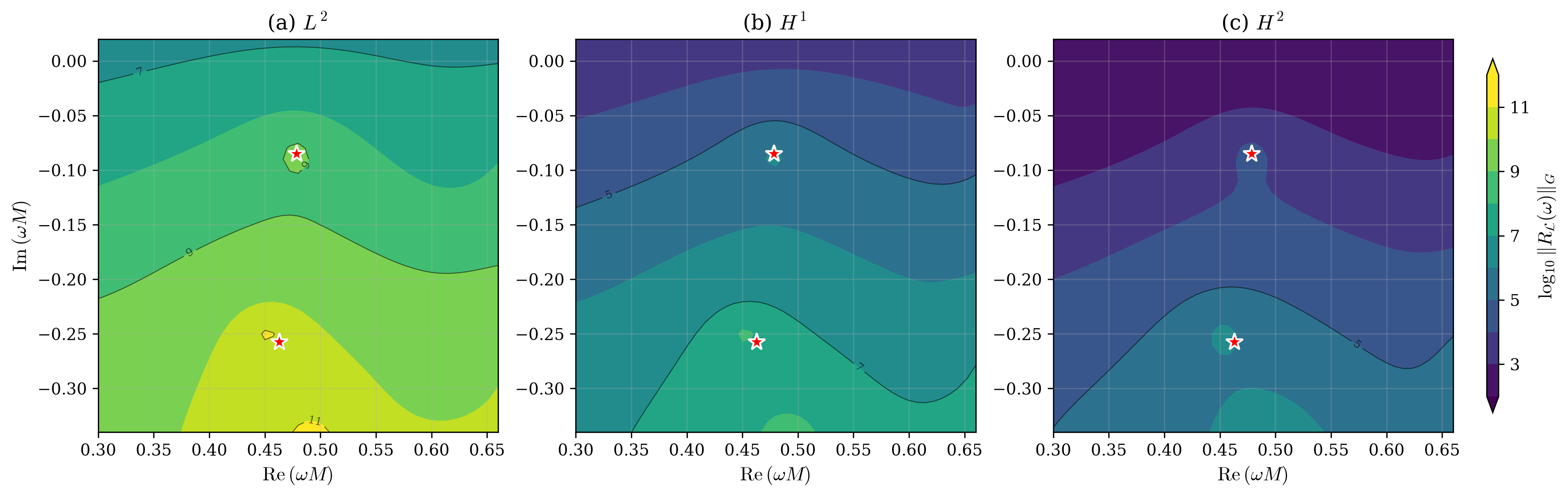}
\caption{Pseudospectral contours of the Kerr black hole for $s=2$, $m=2$, $\alpha=0.3$, $N_\sigma=N_x=26$, in the complex $\omega M$-plane. Level sets of $\log_{10}\norm{R_\Lcal(\omega)}_G$ for the $L^2$ (a), $H^1$ (b) and $H^2$ (c) norms, with a common color scale and contour levels shared across all three panels. Red stars mark the first two Leaver QNM frequencies; grey dots: eigenvalues of the discretized operator. The contours shrink progressively from $L^2$ to $H^2$, but remain large over the whole window in all three norms.}
\label{fig:pseudo_s2}
\end{figure}

In Fig.~\ref{fig:pseudo_s2} we show the pseudospectral contours in the complex $\omega M$-plane for the three norms of the working hierarchy ($L^2$, $H^1$, $H^2$), at the common resolution $N_\sigma = N_x = 26$. The red stars mark the first four Leaver QNM frequencies~\cite{Stein:2019mop,Berti:2009kk}; small grey dots mark the eigenvalues of the discretized operator (which agree with the Leaver values to the accuracy of the collocation scheme). The color scale and contour levels are common to the three panels, so the panels are directly comparable.

Three observations stand out. First, the spectral instability is present for $s=2$ and it is strong, since in all three norms the resolvent is large over the \emph{entire} computation window ($0.30\leq\mathrm{Re}\,\omega M\leq 0.66$, $-0.34\leq\mathrm{Im}\,\omega M\leq 0.02$), the minimum of $\log_{10}\norm{R}$ on the window being $6.56$ in $L^2$, $3.08$ in $H^1$ and $1.99$ in $H^2$. Even in the energy-motivated $H^2$ norm we have $\norm{R}^{-1}\lesssim 1\times10^{-2}$ everywhere, so that a relative perturbation of the operator of order $10^{-3}$--$10^{-2}$ suffices to move a QNM to \emph{any} point of the window, at distances up to $|\omega-\omega_0|M\simeq 0.31$ from the fundamental mode. For a $65\,M_\odot$ black hole a displacement $\Delta(\omega M)\sim 0.1$--$0.3$ corresponds to a frequency shift of $\sim 50$--$150\,\mathrm{Hz}$, which lies within the sensitivity band of current ground-based detectors and is comparable to the energy scale of astrophysical environmental perturbations, such as the binding energy of an accretion disk or the tidal field of a companion at $\sim 10\,M$ separation. Second, the higher-derivative norms tighten the pseudospectrum level by level: with the common color scale of Figure~\ref{fig:pseudo_s2} the contour pattern shifts downward by $1.1$--$3.5$ orders of magnitude at each step of the hierarchy $L^2\to H^1\to H^2$, the window minimum moving from $\log_{10}\norm{R}=6.56$ to $3.08$ and then to $1.99$. The two QNMs resolved in the window, namely the fundamental $\omega_0 M=0.4784-0.0848\,i$ and the first overtone $\omega_1 M=0.4629-0.2573\,i$, stand out as isolated peaks in the $H^2$ panel while being buried in a broad high-$\norm{R}$ plateau in the $L^2$ panel, which is consistent with the $s=0$ finding~\cite{Cai:2025uzw} and reflects the fact that higher-derivative norms capture the regularity of the QNM eigenfunctions more faithfully. Third, the energy-motivated level is $H^2$ for $s=2$, because the single-field energy analysis of Sections~\ref{sec:reduction}--\ref{sec:energy_inner_product} singles out the $H^2$ level in the same way that the Klein--Gordon energy singles out $H^1$ for $s=0$, and the $H^2$ panel of Figure~\ref{fig:pseudo_s2} is therefore the physically weighted pseudospectrum of the gravitational sector, as discussed in Section~\ref{sec:norm_dependence}. 

\subsubsection{Two notions of pseudospectral convergence}
\label{sec:two_convergence}

It is important to distinguish two complementary notions of convergence for the pseudospectrum:

\textbf{(a) Pointwise convergence.} At a fixed frequency $\omega$ in the complex plane, the resolvent norm $\norm{R_\Lcal(\omega)}$ converges as the collocation order $N$ increases. This is the notion probed by Figure~\ref{fig:convergence_s2}: for each fixed $\omega$, $\norm{R_\Lcal(\omega)}$ approaches a limiting value exponentially with $N$. Pointwise convergence verifies that the finite-rank approximation $\bvec{L}$ faithfully represents the infinite-dimensional operator $\Lcal$ at the evaluation point.

\textbf{(b) Global (contour) convergence.} The pseudospectral \emph{contours}, that is, level sets of $\norm{R_\Lcal(\omega)}$ in the complex plane, converge as sets: for a fixed $\varepsilon$, the $\varepsilon$-pseudospectral contour $\{\omega : \norm{R_\Lcal(\omega)} = 1/\varepsilon\}$ stabilizes as $N$ increases. This notion is more relevant for the physical interpretation, as the shape of the contours determines the migration range of QNM frequencies under perturbations.

\begin{figure}[htbp]
\centering
\includegraphics[width=0.7\textwidth]{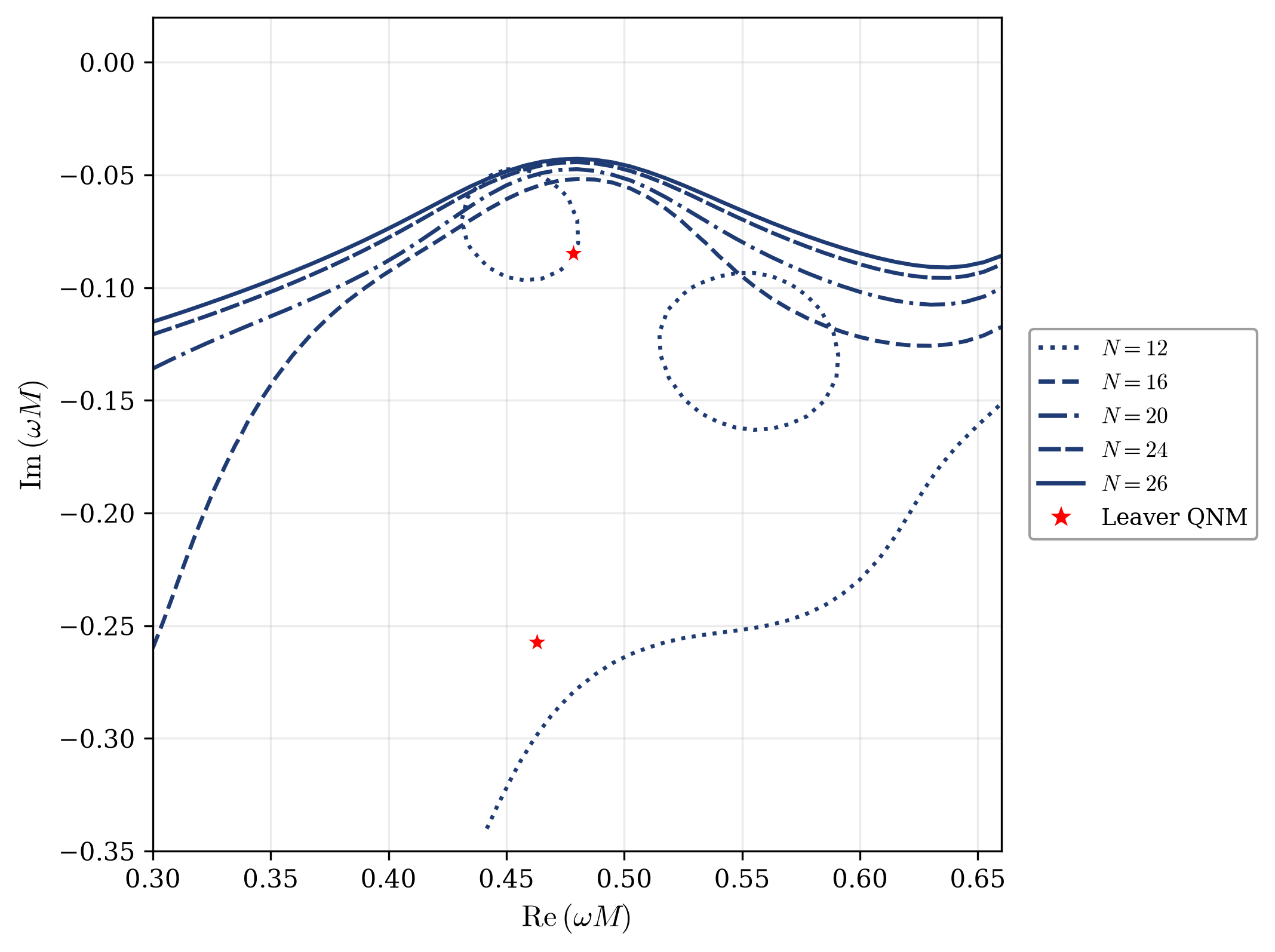}
\caption{Global (contour) convergence of the $\varepsilon = 10^{-3}$ pseudospectral contour ($H^2$ norm, i.e.\ $\log_{10}\norm{R}=3$) for $s=2$, $\alpha=0.3$, $m=2$, at collocation orders $N = 12$ (dotted), $16$ (dashed), $20$ (dash-dotted), $24$ (long-dashed), and $26$ (solid). The contour is stable around the QNM chain (the high-$N$ curves nearly coincide there) and drifts slowly upward away from the spectrum. Red stars: the fundamental QNM $\omega_0 M=0.4784-0.0848\,i$ and the first overtone $\omega_1 M=0.4629-0.2573\,i$.}
\label{fig:contour_convergence}
\end{figure}

In Fig.~\ref{fig:contour_convergence} we demonstrate the global convergence directly by overlaying the $\varepsilon = 10^{-3}$ pseudospectral contour ($H^2$ norm, $\log_{10}\norm{R}=3$) for $N = 12, 16, 20, 24, 26$. Two regimes are visible. In the region surrounding the QNM chain ($\mathrm{Im}\,\omega M \lesssim -0.05$), the contour position is stable already at $N=12$: the five curves are nearly coincident, confirming that the pseudospectral shape near the physical spectrum has converged. Near the upper window boundary the contour drifts slowly upward with $N$ (from $\mathrm{Im}\,\omega M\simeq-0.26$ at $N=12$ to $\simeq-0.046$ at $N=26$; almost all of the shift happens in the single step $N=12\to16$, the contour already sitting at $-0.056$ at $N=16$ and moving by only $\sim0.005$ per subsequent step of four collocation points), reflecting the slower pointwise convergence of the resolvent far from the spectrum, where it is dominated by the strong non-normality of the operator rather than by proximity to individual eigenvalues.

Throughout, when we refer to ``convergence of pseudospectra,'' we mean pointwise convergence unless otherwise stated; the global convergence is verified by the contour stability shown in Figure~\ref{fig:contour_convergence}.

\subsubsection{Pointwise convergence with resolution}

\begin{figure}[htbp]
\centering
\includegraphics[width=0.7\textwidth]{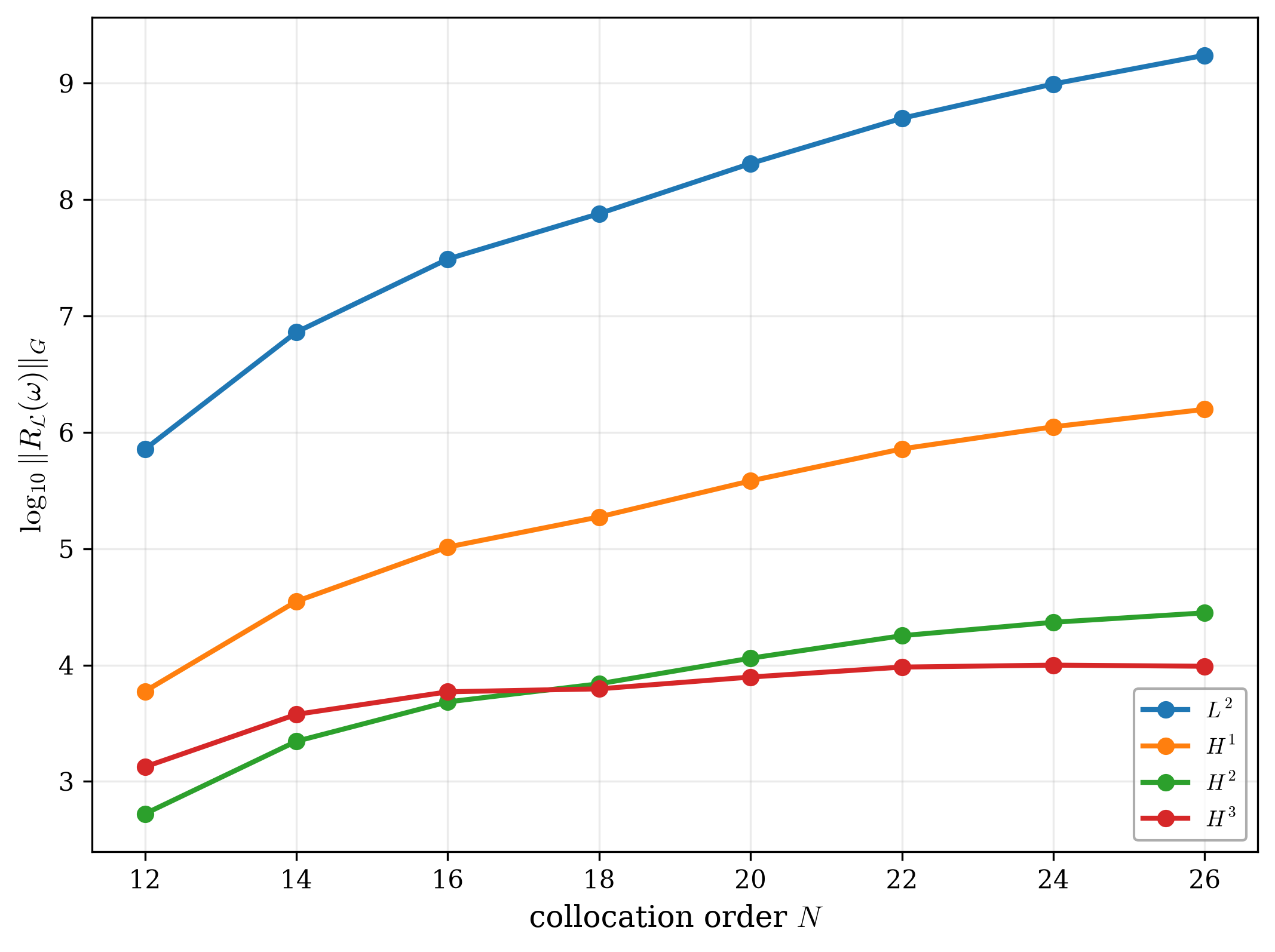}
\caption{Pointwise convergence of $\log_{10}\norm{R_\Lcal(\omega)}$ with collocation order $N$ ($s=2$, $\alpha=0.3$, $m=2$), at the bulk evaluation point $\omega M=0.47-0.17\,i$ (between the fundamental mode and the first overtone), for the $L^2$, $H^1$, $H^2$ and $H^3$ norms. Each curve stabilizes as $N$ increases, with faster stabilization at higher norm levels.}
\label{fig:convergence_s2}
\end{figure}

In Fig.~\ref{fig:convergence_s2} we show the \emph{pointwise} convergence of the resolvent norm $\norm{R_\Lcal(\omega)}$ as a function of the collocation order $N = N_\sigma = N_x$ (from $12$ to $26$ in steps of two), evaluated at a fixed point $\omega M = 0.47 - 0.17\,i$ in the complex plane, chosen in the bulk region between the fundamental mode ($\omega_0 M=0.4784-0.0848\,i$) and the first overtone ($\omega_1 M=0.4629-0.2573\,i$), for the four norms $L^2/H^1/H^2/H^3$ of the $s=2$ sector.

The convergence behavior confirms several properties that we regard as the key ingredients of a trustworthy pseudospectral analysis. First of all, the resolvent norm converges for every norm we have examined, and it does so exponentially in $N$, which is precisely the spectral convergence one expects from Chebyshev methods. Second, the stabilization is fastest at the highest norm level: beyond $N\simeq20$ the $H^3$ curve changes by only $0.031$ in $\log_{10}\norm{R}$ per step of two collocation points, which is $0.8\%$ of the value, with $\log_{10}\norm{R}=3.90\to3.99$ over $N=20\to26$, while the $H^2$ curve is still drifting at the $1.8\%$ level at $N=26$ with $0.08$ per step, and the $L^2$ curve is the slowest to settle with $0.25$ per step at $N=26$. The reason is transparent, because higher-derivative norms weight the smooth and well-resolved part of the resolvent action more heavily, and they therefore stabilize at a lower resolution. Third, the same data taken at the common bulk point show that the two sectors drift at comparable rates at this location, the $H^2$ value being $0.08$ in $\log_{10}\norm{R}$ per step of two collocation points at $N=26$ for $s=+2$ against $0.11$ for $s=0$, and the sector difference becomes sharper when each resolvent is evaluated at its own fundamental mode, where the $s=+2$ rate is systematically the smaller with $\beta=0.25$ against $0.34$ in $H^2$, as shown in Table~\ref{tab:convergence_comparison}. In addition, the evaluation point lies between the $s=+2$ fundamental mode and the first overtone, where the local spectral picture inherits the slower $s=+2$ eigenvalue convergence documented in Sections~\ref{sec:omega_dependence} and~\ref{sec:iso_results}, and it lies far from the $s=0$ spectrum, whose contribution to the resolvent is smooth and converges immediately. 

\subsubsection{Effect of the imaginary part of $\omega$}

\begin{figure}[htbp]
\centering
\includegraphics[width=0.7\textwidth]{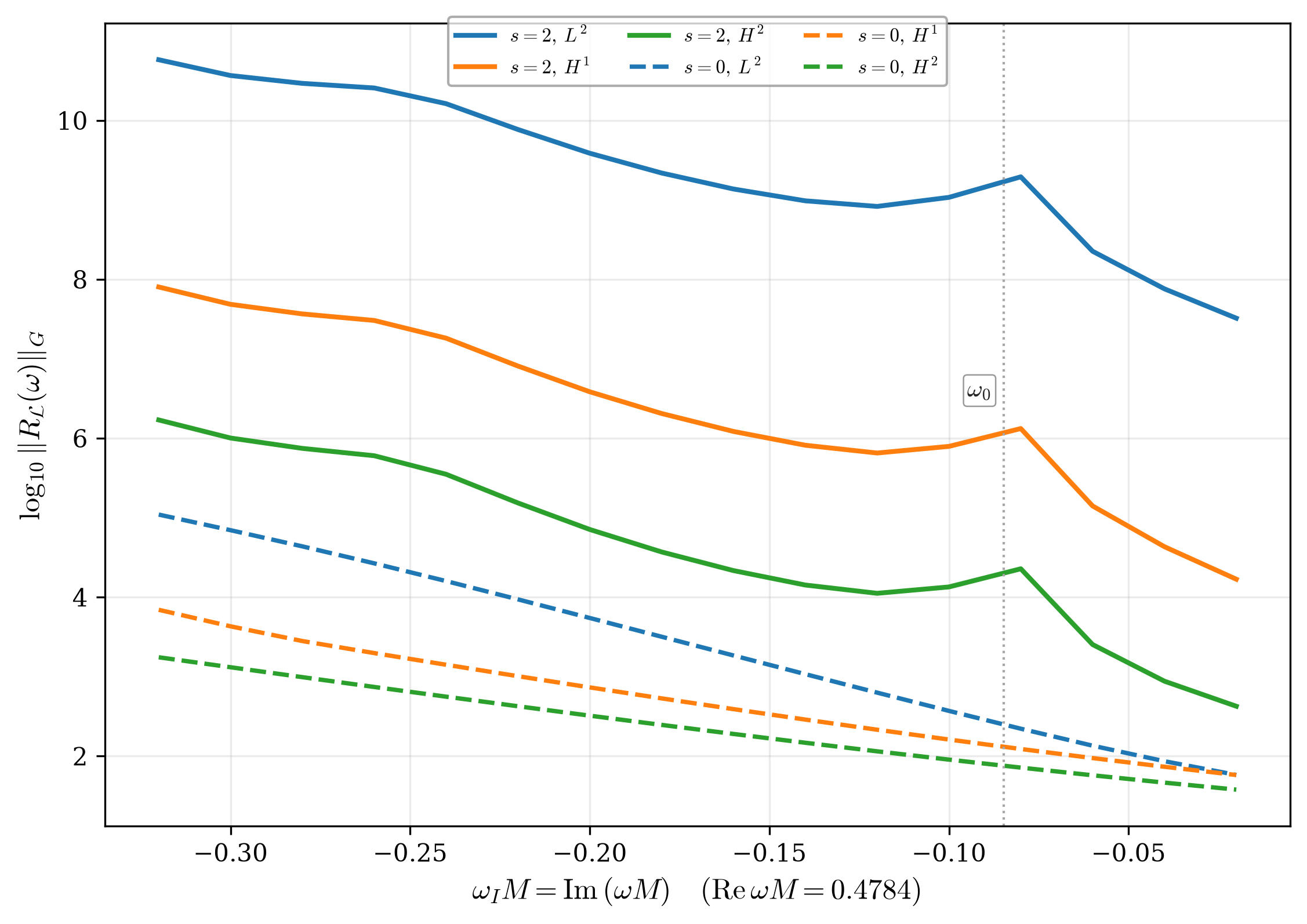}
\caption{Resolvent norm $\log_{10}\norm{R_\Lcal(\omega)}$ as a function of $\omega_I M = \mathrm{Im}\,(\omega M)$ along the line $\mathrm{Re}\,(\omega M)=0.4784$ (through the fundamental QNM $\omega_0 M=0.4784-0.0848\,i$, marked by the dotted vertical line), for $s=2$ (solid) and $s=0$ (dashed), at $\alpha=0.3$, $N=26$, in the $L^2$, $H^1$ and $H^2$ norms (color-coded). The $s=2$ curve peaks at the fundamental QNM, dips, and rises again toward the first overtone; the $s=0$ curve is smooth, the line passing far from the $s=0$ spectrum.}
\label{fig:omegaI_s2}
\end{figure}

Following~\cite{Cai:2025uzw}, we examine how the imaginary part $\omega_I = \mathrm{Im}\,\omega$ affects the resolvent norm. In Fig.~\ref{fig:omegaI_s2} we show $\log_{10}\norm{R_\Lcal(\omega)}$ along the vertical line $\mathrm{Re}\,(\omega M)=0.4784$ (through the fundamental QNM), for $\omega_I M$ from $-0.02$ down to $-0.32$, for both spin sectors and the three norms of the working hierarchy, at $N_\sigma=N_x=26$.

The results show that along the entire line $\log_{10}\norm{R}\geq2.63$ in the $H^2$ norm for $s=2$, the whole scan from the near-axis region $\omega_I M=-0.02$ down to the strongly damped region $\omega_I M=-0.32$ lies inside the $\varepsilon=10^{-3}$ pseudospectrum of the $s=+2$ operator. The $s=2$ curve, however, is not monotonic: it peaks at the fundamental QNM with $\log_{10}\norm{R}=4.36$ at $\omega_I M\simeq-0.08$, dips slightly to $4.05$ at $\omega_I M\simeq-0.12$, and then rises again toward the first overtone, reaching $5.78$ at $\omega_I M\simeq-0.26$, which is the damping of $\omega_1 M=0.4629-0.2573\,i$, and $6.23$ at $\omega_I M=-0.32$. It is worth noting that the resolvent resolves both QNM peaks of the window even on a line offset from $\omega_1$ in $\mathrm{Re}\,\omega$ by $0.016$. By contrast the $s=0$ curve along the same line is smooth and much lower, ranging from $1.58$ to $3.24$, because the line passes far from the $s=0$ spectrum whose fundamental sits at $\omega M=0.6011-0.0925\,i$, so that its resolvent reflects only the tail of the $s=0$ pseudospectrum. This comparison makes explicit how much more extended the gravitational pseudospectrum is along the damping direction, for at $\omega_I M=-0.32$ the two sectors differ by $3.0$ orders of magnitude in $\norm{R}$ at the same norm level and resolution. 

\subsubsection{Effect of the dimensionless spin $\alpha$}

\begin{figure}[htbp]
\centering
\includegraphics[width=0.95\textwidth]{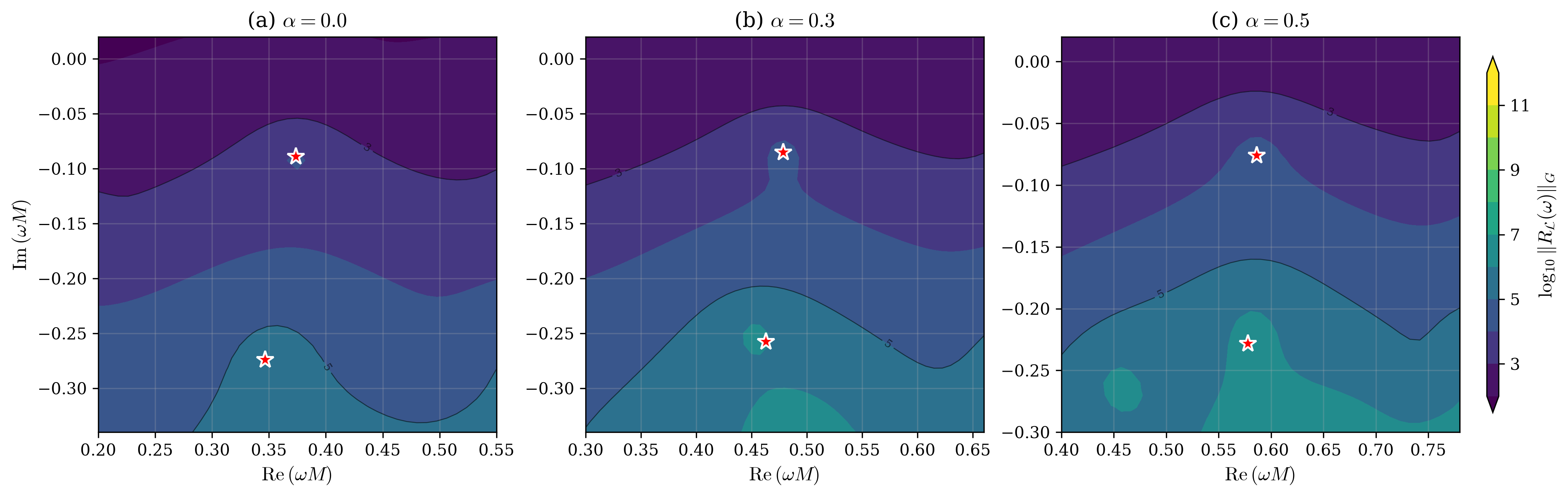}
\caption{Pseudospectral contours for $s=2$, $m=2$, $N_\sigma=N_x=26$, in the $H^2$ norm, for $\alpha = 0.0$ (a), $\alpha = 0.3$ (b), and $\alpha = 0.5$ (c); each panel uses a window adapted to its QNM chain, with a common color scale. Red stars: first two Leaver QNMs. As $\alpha$ increases the QNM chain shifts to larger $\mathrm{Re}\,\omega M$ and the contours deform, reflecting the breaking of the $\omega\to-\bar\omega$ symmetry of the Schwarzschild case.}
\label{fig:alpha_s2}
\end{figure}

In Fig.~\ref{fig:alpha_s2} we show $H^2$-norm pseudospectral contours for three values of $\alpha$ ($0.0$, $0.3$, $0.5$), each panel with a window adapted to its QNM chain.

For $\alpha = 0$ (Schwarzschild) the spectrum is invariant under $\omega\to-\bar\omega$ (the $m=\pm2$ sectors are degenerate), reflecting the time-reversal invariance of the Schwarzschild geometry. As $\alpha$ increases this symmetry is broken: the QNM chain shifts to larger $\mathrm{Re}\,\omega M$ ($0.3737\to0.4784\to0.5860$ for the fundamental mode), the damping decreases ($|\mathrm{Im}\,\omega_0 M|: 0.0890\to0.0848\to0.0756$), and the modes of the chain approach each other in the $\mathrm{Re}\,\omega M$ direction (the $\omega_0$--$\omega_1$ separation shrinks from $0.027$ to $0.008$). The pseudospectral contours follow this motion and become increasingly deformed, in particular around the first overtone. This is analogous to the $s=0$ case, with the deformation further amplified by the spin-orbit coupling of the $s=2$ sector.

\subsubsection{Isospectrality of the $s=\pm 2$ pseudospectra}
\label{sec:iso_results}

The Teukolsky--Starobinsky identities guarantee that the $s=+2$ and $s=-2$ operators are isospectral. To test whether this property survives at the level of the \emph{pseudospectra}, which is a stronger statement as explained in Section~\ref{sec:omega_dependence}, we compute the $H^2$-norm resolvent of both operators, with the same two-component norm construction applied to each sector separately, at fourteen frequencies distributed around the fundamental QNM $\omega_0$: along the real direction $\omega_0+\delta$, along the damping direction $\omega_0-i\delta$, and along the diagonal $\omega_0+\delta(1-i)$, for offsets $\delta\in\{0.05,\dots,0.30\}$.

\begin{table}[htbp]
\centering
\caption{Isospectrality test of the $s=\pm2$ pseudospectra at $\alpha=0.3$, $m=2$, $N_\sigma=N_x=26$, $H^2$ norm. For each probe frequency $\omega$ (in units of $\omega M$), the table lists $\log_{10}\norm{R^{(+)}(\omega)}$ and $\log_{10}\norm{R^{(-)}(\omega)}$ for the $s=+2$ and $s=-2$ operators, and their difference. The two operators share their spectrum \emph{exactly}, but their pseudospectra differ: $\norm{R^{(+)}}$ exceeds $\norm{R^{(-)}}$ at twelve of the fourteen probes, by up to a factor $6.6$ (the smallest excess being $+0.004$ decades, printed as $+0.00$), while at the two probes furthest along the QNM line the ordering reverses slightly (see the text). The asymmetry therefore has no universal sign.}
\label{tab:isospectrality}
\begin{tabular}{lccc}
\toprule
$\omega$ ($\omega M$) & $\log_{10}\norm{R^{(+)}}$ & $\log_{10}\norm{R^{(-)}}$ & difference \\
\midrule
$0.4784-0.1348\,i$ & $4.12$ & $3.48$ & $+0.64$ \\
$0.4784-0.1848\,i$ & $4.63$ & $3.93$ & $+0.70$ \\
$0.4784-0.2348\,i$ & $5.46$ & $4.64$ & $+0.82$ \\
$0.4784-0.2848\,i$ & $5.90$ & $5.12$ & $+0.78$ \\
$0.4784-0.3848\,i$ & $6.30$ & $5.96$ & $+0.33$ \\
$0.5284-0.0848\,i$ & $3.41$ & $3.07$ & $+0.34$ \\
$0.5284-0.1348\,i$ & $3.94$ & $3.47$ & $+0.47$ \\
$0.5784-0.0848\,i$ & $3.09$ & $3.09$ & $+0.00$ \\
$0.5784-0.1848\,i$ & $4.22$ & $3.89$ & $+0.34$ \\
$0.6284-0.0848\,i$ & $2.92$ & $3.11$ & $-0.19$ \\
$0.6784-0.0848\,i$ & $3.19$ & $3.13$ & $+0.06$ \\
$0.6784-0.2848\,i$ & $5.61$ & $4.83$ & $+0.77$ \\
$0.7784-0.0848\,i$ & $3.00$ & $3.18$ & $-0.18$ \\
$0.7784-0.3848\,i$ & $6.05$ & $5.52$ & $+0.54$ \\
\bottomrule
\end{tabular}
\end{table}

The result is unambiguous on the main point: \textbf{the $s=\pm2$ operators are isospectral, but their pseudospectra are not.} At twelve of the fourteen probe frequencies, $\log_{10}\norm{R^{(+)}}$ exceeds $\log_{10}\norm{R^{(-)}}$, by up to $0.82$ decades (mean $+0.39$ over all fourteen): in the $H^2$ norm, the $s=+2$ ($\psi_0$) pseudospectrum is the larger of the two by up to a factor of $6.6$ in the resolvent norm. At the two remaining probes, namely the two furthest along the QNM line at $\mathrm{Re}\,\omega M=0.6284$ and $0.7784$, the ordering reverses slightly, by $0.18$--$0.19$ decades. The asymmetry therefore has a well-defined magnitude at most probes but no universal \emph{sign}: as discussed below, it is a property of the triple (variable, probe location, norm) rather than of the spacetime itself.

This asymmetry has a clean structural origin. The TSI relate $\Psi^{(+2)}$ and $\Psi^{(-2)}$ through the fourth-order Starobinsky operator $\Scal_4$ (Appendix~\ref{app:starobinsky}): $\Lcal^{(+)}$ and $\Lcal^{(-)}$ are similar only through an unbounded differential map, not through a bounded similarity of the two-component state space. Isospectrality survives this transformation because eigenvalues are insensitive to the map's spectral properties, but the resolvent norm, which depends on the full singular-value structure of the operator, does not. The same mechanism explains the resolution asymmetry noted in Section~\ref{sec:omega_dependence}: the $s=+2$ eigenvalues converge to their Leaver values at rate $0.22$ per collocation point (reaching $4.4\times10^{-5}$ at $N=26$), roughly $1.4$ times more slowly than the $s=-2$ eigenvalues ($0.30$ per point, $3.6\times10^{-11}$ at $N=26$), a rate difference that compounds to six orders of magnitude in the achieved accuracy because the strongly nonnormal $s=+2$ operator has poorly conditioned eigenfunctions. The pseudospectral asymmetry and the eigenvalue-convergence asymmetry are two faces of the same property: \emph{the Teukolsky--Starobinsky equivalence is a statement about spectra, not about pseudospectra.} Physically, this singles out $\psi_4$ ($s=-2$) as the better-conditioned variable for gravitational-wave data analysis, consistent with its standard use in numerical relativity.

Physically, the asymmetry is a statement about the two variables, not about two distinct dynamics of the black hole. Both $\psi_0$ and $\psi_4$~\cite{Newman:1962qr} are projections of the \emph{same} curvature perturbation onto the ingoing and outgoing null directions. The Kerr background is Petrov type~D~\cite{Teukolsky:2014zha}: it possesses two distinct repeated principal null directions, one ingoing at the horizon and one outgoing at $\scrI^+$, together with a non-vanishing Coulomb part $\psi_2 \neq 0$ (equal to $-M/(r-ia\cos\theta)^3$ in the Kinnersley tetrad). The vacuum Bianchi identities tie the five Weyl scalars into the chain $\psi_0 \to \psi_1 \to \psi_2 \to \psi_3 \to \psi_4$, whose four links are successive directional derivatives along the null tetrad; because $\psi_2 \neq 0$ and the two principal null directions are distinct, this chain does not truncate, and the map between $\psi_0$ and $\psi_4$ is the fourth-order differential operator of Eq.~\eqref{eq:starobinsky_operator}. The same four orders appear in the asymptotic peeling of the two fields at $\scrI^+$, where $\psi_0 \sim r^{-5}$ while $\psi_4 \sim r^{-1}$: the factor $r^4$ between them \emph{is} the four derivatives of the Teukolsky--Starobinsky map. The asymmetry reported above is therefore the statement that this map is not an isometry of the norm in which the perturbation is measured; it is a property of the pair (variable, norm) rather than an additional instability of the spacetime, whose spectral instability is present and strong in either variable. The genuinely invariant, observational content of the distinction is the peeling itself: $\psi_4$ is the quantity measured at $\scrI^+$, whereas $\psi_0$ must be reconstructed from it by four radial derivatives, i.e.\ with four orders of noise amplification, which is precisely why $\psi_4$ is the better-conditioned variable and why it is the one extracted in numerical relativity. (For comparison, in a Petrov type~N plane-wave spacetime the map between $\psi_0$ and $\psi_4$ reduces to a bounded algebraic relabeling, and the two descriptions are strictly isopseudospectral; the Kerr asymmetry is thus a genuine consequence of the type~D Coulomb structure.)

Both the magnitude and even the sign of the asymmetry depend on the spin. Scanning the computation window $\mathrm{Re}\,(\omega M)\in[0.34,0.76]$, $\mathrm{Im}\,(\omega M)\in[-0.46,-0.02]$ at $m=2$, $N=26$ in the $H^2$ norm, the window average of $\Delta=\log_{10}\norm{R^{(+)}}-\log_{10}\norm{R^{(-)}}$ rises monotonically with $\alpha$, from $-0.220$ at $\alpha=0$ to $+0.382$ at $\alpha=0.5$, crossing zero near $\alpha\simeq0.22$; over the same range the rms spread grows from $0.344$ to $0.848$, and the fraction of the window with $\Delta>0$ rises from $0.18$ to about two thirds. The Schwarzschild limit is therefore \emph{not} the symmetric case: at $\alpha=0$ the $\psi_4$ ($s=-2$) resolvent is on average the \emph{larger} of the two, and it is rotation that reverses the ordering. This is a stronger statement than the norm-dependence discussed below, for the $\psi_0/\psi_4$ asymmetry is not fixed in sign by the choice of variable alone, but depends on the spin as well.

The structural reason is that $\alpha=0$ is special. There every coefficient of the operator is real, so the problem possesses an exact anti-unitary (time-reversal) symmetry, $\norm{R(\omega)}=\norm{R(-\bar\omega)}$, which we have verified numerically to machine precision ($0.000\times10^{0}$ relative difference in all three norms); the symmetry is broken as soon as $\alpha\neq0$, where $\norm{R(\omega)}$ and $\norm{R(-\bar\omega)}$ differ by a factor of $4.8$ in $H^2$ ($8.5$ in $H^1$, $17$ in $L^2$) at $\alpha=0.3$ and $\omega M=0.45-0.15\,i$. That the asymmetry is non-zero already at $\alpha=0$ is consistent with its origin in the non-vanishing Coulomb part $\psi_2$ rather than in rotation alone. But the vanishing of the \emph{window-averaged} asymmetry near $\alpha\simeq0.22$ is an accidental crossing, not a symmetry: the local asymmetry $\Delta(\omega)$ is non-zero at every spin and of either sign, and the Schwarzschild mean magnitude ($0.220$) is not the smallest in the family ($0.031$ at $\alpha=0.2$).

A quantitative caution is in order regarding the resolvent ratio quoted above, since the norm is a construction rather than an observable. We have repeated the fourteen-probe comparison under a family of norm conventions: the four Sobolev levels $L^2$, $H^1$, $H^2$, $H^3$ of the hierarchy~\eqref{eq:Hk_norm}; four choices $q=1.0,1.5,2.0,3.0$ of the angular weight exponent in $x_w=(1-x^2)^q$ (the value $q=\max(|m|,|s|)=2$ being the one adopted here); and both weight placements, the Cai convention adopted here in which $x_w$ multiplies the mass term alone, and the earlier convention in which it multiplies every block. Throughout, both sectors are measured in the \emph{same} norm, so the comparison remains fair.

Two features of the outcome deserve emphasis. First, the \emph{magnitude} is strongly convention-dependent: the ratio spans $2.8$--$33$ in $L^2$, $2.5$--$17$ in $H^1$, $0.65$--$6.6$ in $H^2$ and $0.75$--$11$ in $H^3$, with the mean $\Delta$ falling from $1.21$ ($L^2$) to $0.96$ ($H^1$) and then to $0.39$--$0.46$ at $H^2/H^3$; the factor quoted in the text should therefore be read as the $H^2$ statement it is. Second, and unlike the situation under the earlier weight placement, the \emph{sign} of the asymmetry is \emph{not} universal. It is definite at the two lowest levels ($\Delta>0$ at $14/14$ probes in $L^2$ and in $H^1$, for every $q$), but at $H^2$ and $H^3$ the two probes furthest along the QNM line ($\mathrm{Re}\,\omega M=0.6284$ and $0.7784$) reverse it, by $0.18$--$0.19$ decades. Overall $\Delta>0$ in $208$ of the $224$ Cai-convention (level, $q$, probe) combinations ($92.9\%$), the largest reversal being $-0.19$ decades. The asymmetry therefore has a well-defined magnitude over most of the window but no universal sign: it is a property of the triple (variable, probe location, norm) rather than of the Kerr geometry alone.

A further simplification accompanies the Cai convention. Because $x_w$ now multiplies only the mass term, the angular-weight exponent has become numerically immaterial: the four values $q=1.0,1.5,2.0,3.0$ reproduce $\Delta$ to the second decimal at every probe and every level ($\min=-0.19$, $\max=+0.82$, mean $+0.39$ in $H^2$), whereas under the every-block placement $q$ still shifted the mean $\Delta$. The placement itself matters only from $H^2$ upward: the $L^2$ level is identical in the two placements, while at $H^2$ the mean $\Delta$ drops from $0.68$ (every block) to $0.39$ (mass only) and the two sign reversals appear. That the same asymmetry spans more than an order of magnitude across reasonable norms, and even changes sign at a few probes, is itself the sharpest illustration of the theme of Section~\ref{sec:norm_dependence}: the strength of the spectral instability is not a property of the Kerr geometry but of the triple (operator, variable, norm).

\subsection{Detailed comparison: $s=2$ vs $s=0$ convergence}
\label{sec:comparison}

We now compare systematically the pseudospectral behavior of the gravitational ($s=+2$) and scalar ($s=0$) sectors, both computed with the identical two-component framework at $\alpha=0.3$, $m=2$. The comparison is organized around two distinct questions. First, \emph{pointwise convergence}: how fast does the resolvent of each sector, evaluated at its own fundamental QNM, converge under refinement (Table~\ref{tab:convergence_comparison})? Second, \emph{pseudospectrum size}: how large is the resolvent of each sector over the common computation window (Table~\ref{tab:size_comparison} and Figure~\ref{fig:comparison_pseudospectrum})? The two questions have different answers, and keeping them apart avoids a confusion to which the $H^{|s|}$ pattern might otherwise lead one: the pointwise rates of the two sectors are comparable, but the gravitational pseudospectrum is systematically the larger of the two.

\subsubsection{Structural comparison of norms}

In Table~\ref{tab:norm_comparison} we summarize the key structural differences between the $s=0$ and $s=2$ setups. The computational framework, namely the two-component state $(V,W)$, the angular weight $x_w$ and the weighted Sobolev hierarchy~\eqref{eq:Hk_norm}, is \emph{identical} for both spins; what differs is which level of the hierarchy is singled out by the physics: the conserved current motivates $H^1$ for $s=0$ (Klein--Gordon energy) and $H^2$ for $s=2$ (Starobinsky-reduced energy), per the $H^{|s|}$ pattern.

\begin{table}[htbp]
\centering
\caption{Structural comparison of the $s=0$ and $s=2$ setups. The computational hierarchy is identical for both spins; the rows record where the physics differs.}
\label{tab:norm_comparison}
\begin{tabular}{lcc}
\toprule
\multirow{2}{*}{Property} & \multicolumn{2}{c}{Spin weight} \\
\cmidrule(lr){2-3}
 & $s=0$ (scalar) & $s=2$ (gravitational) \\
\midrule
Teukolsky variable & $\Phi = \Psi^{(0)}$ & $\psi_0 = \Psi^{(2)}$ \\
Self-conjugate? & Yes ($\bar\Phi = \Phi^\ast$) & No ($\overline{\Psi^{(2)}} \neq \Psi^{(-2)}$) \\
Conserved current & Diagonal in $\Phi$ & Bilinear in $(\Psi^{(2)},\overline{\Psi^{(-2)}})$ \\
TSI reduction needed? & No (trivial) & Yes ($\Scal_4$, 4th order; $\Cst$) \\
Energy-motivated level & $H^1$ & $H^2$ \\
Max derivative order at that level & 1 & 2 \\
$\omega$-dependence of computed hierarchy & None & None ($\Cst$ enters only the motivational form) \\
Angular exponents ($m=2$) & $\delone=2,\;\deltwo=2$ & $\delone=0,\;\deltwo=4$ \\
Gram matrix construction & \multicolumn{2}{c}{Same (blocks up to $j=k$; Appendix~\ref{app:gram})} \\
General pattern & \multicolumn{2}{c}{Energy $\sim H^{|s|}$} \\
\bottomrule
\end{tabular}
\end{table}

\subsubsection{Numerical convergence comparison}

\begin{figure}[htbp]
\centering
\includegraphics[width=0.85\textwidth]{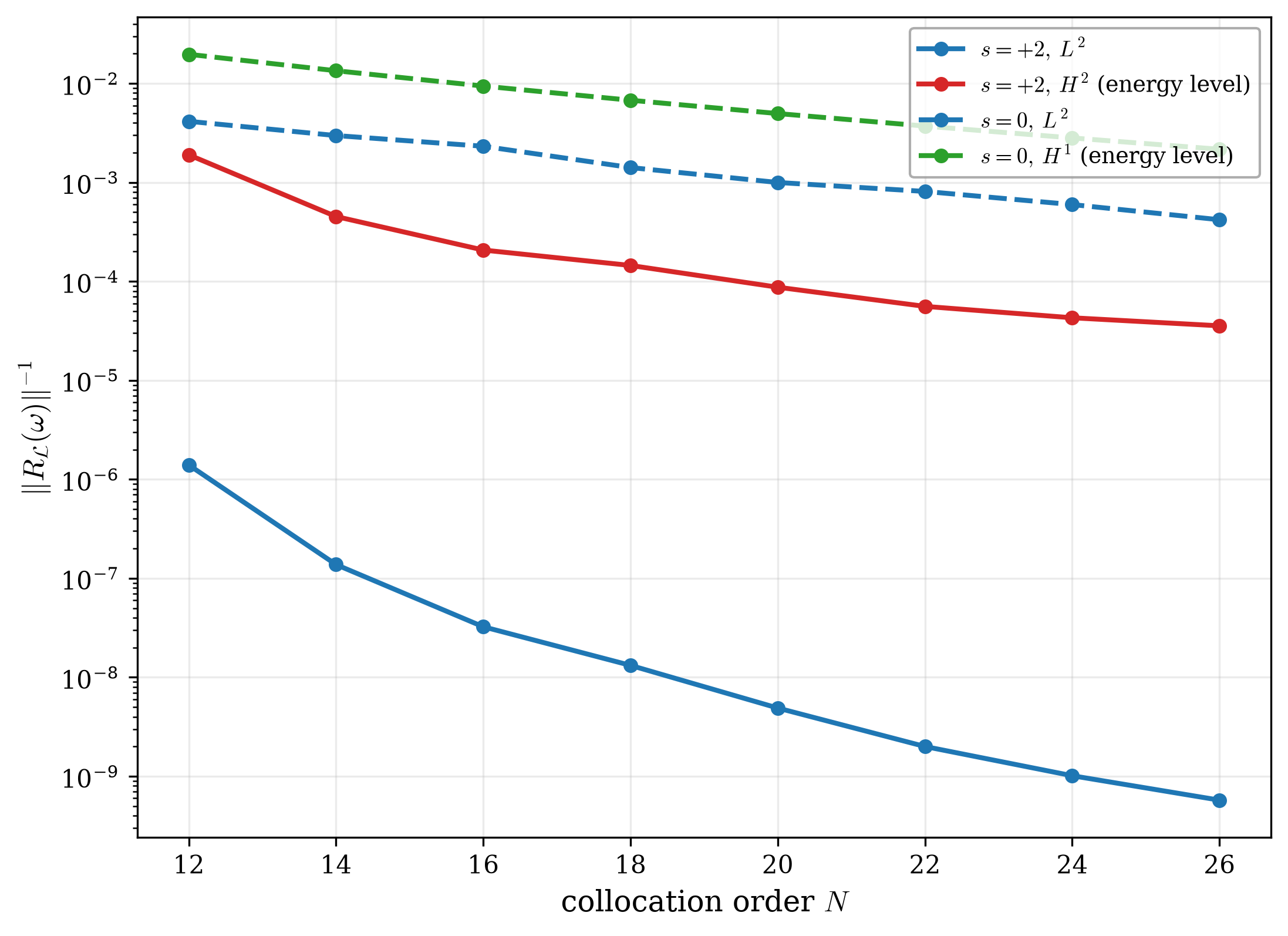}
\caption{Comparison of the inverse resolvent norm $\norm{R_\Lcal(\omega)}^{-1}$ at the bulk evaluation point $\omega M=0.47-0.17\,i$ ($\alpha=0.3$), as a function of collocation order $N$, for $s=+2$ (solid) and $s=0$ (dashed): $L^2$ norm (blue) and the energy-motivated level of each sector ($H^2$ for $s=+2$, red; $H^1$ for $s=0$, green). Smaller values indicate a larger resolvent, i.e.\ stronger spectral instability at the evaluation point. At $N=26$ the $s=+2$ resolvent exceeds the $s=0$ one by $5.9$ orders of magnitude in $L^2$, by $3.5$ orders at the common $H^1$ level, and by $1.8$ orders between the respective energy levels.}
\label{fig:comparison_convergence}
\end{figure}

\begin{figure}[htbp]
\centering
\includegraphics[width=0.95\textwidth]{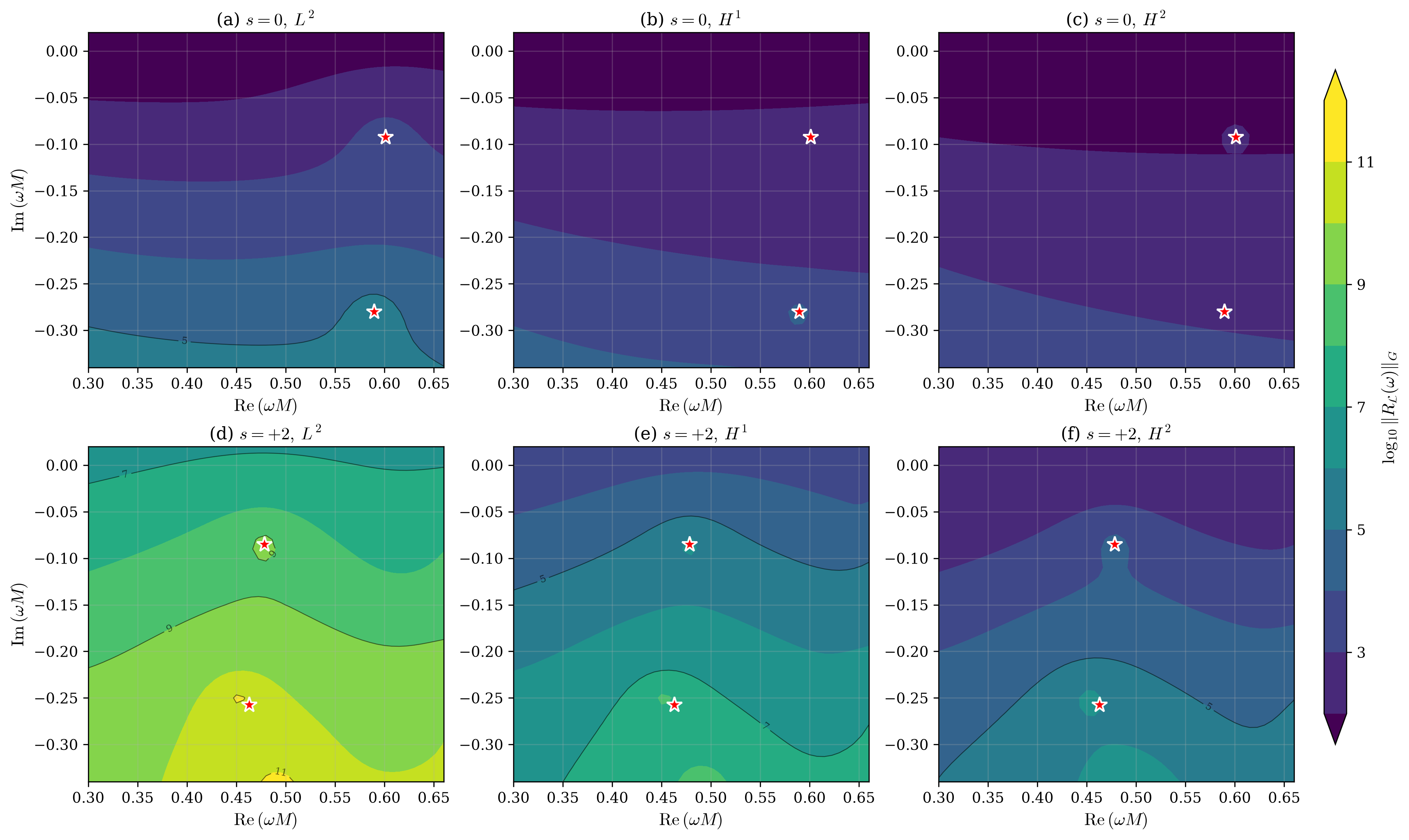}
\caption{Side-by-side comparison of pseudospectral contours for $s=0$ (a--c) and $s=+2$ (d--f), at $\alpha=0.3$, $m=2$, $N_\sigma=N_x=26$, in a common window. From left to right: $L^2$ norm, $H^1$ norm (the energy level for $s=0$), and $H^2$ norm (the energy level for $s=+2$). Red stars: first Leaver QNMs of each sector; grey dots: eigenvalues of the discretized operators. The common color scale makes the norm hierarchy directly comparable between the two spin sectors: the gravitational pseudospectrum is the larger of the two at every norm level (cf.\ Table~\ref{tab:size_comparison}).}
\label{fig:comparison_pseudospectrum}
\end{figure}

\begin{figure}[htbp]
\centering
\includegraphics[width=0.95\textwidth]{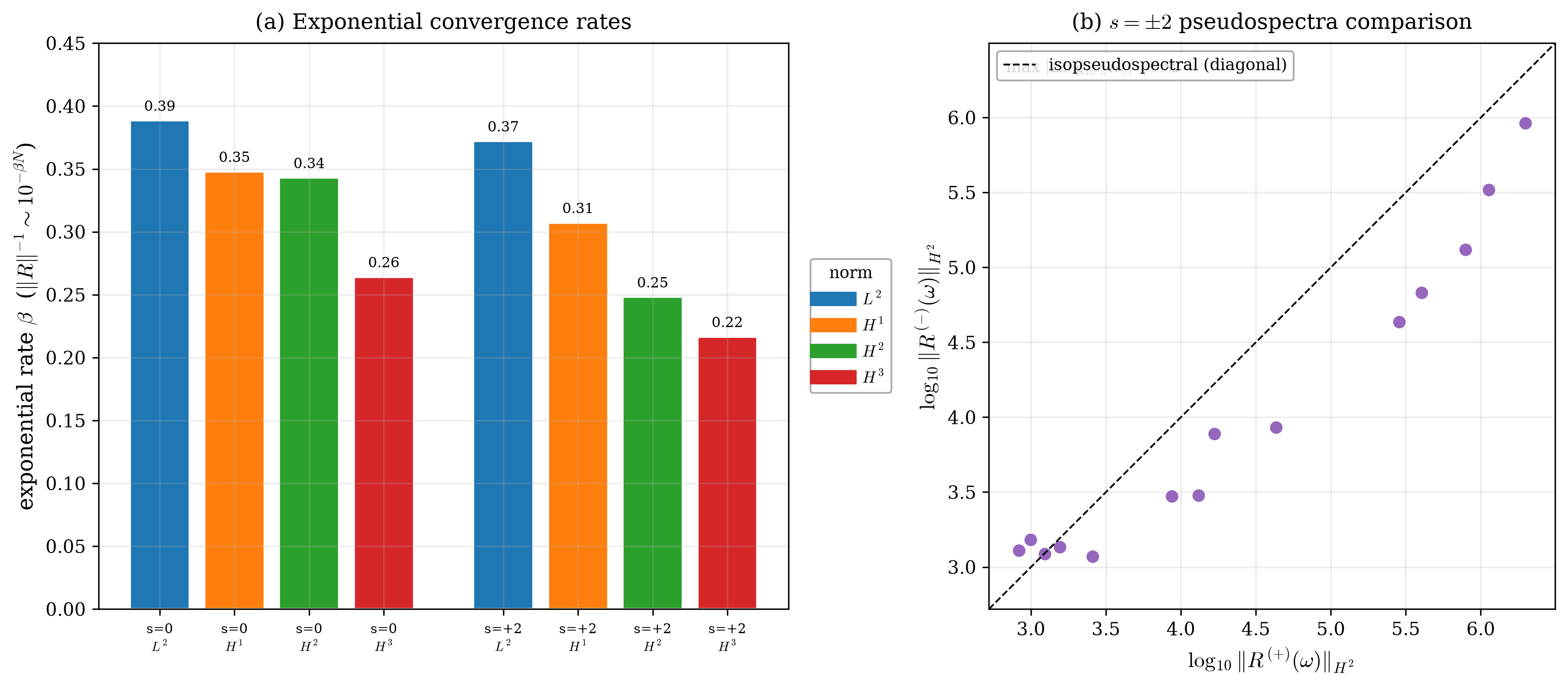}
\caption{\textbf{(a)} Exponential rate $\beta$, defined by $\norm{R(\omega_0)}\sim10^{\beta N}$ evaluated at each sector's \emph{own} fundamental QNM ($\omega_0 M=0.6011-0.0925\,i$ for $s=0$; $\omega_0 M=0.4784-0.0848\,i$ for $s=+2$; cf.\ Table~\ref{tab:convergence_comparison}), for $s=0$ and $s=+2$ under the $L^2/H^1/H^2/H^3$ norms, color-coded by norm. \textbf{(b)} isospectrality test of the $s=\pm2$ pseudospectra: $\log_{10}\norm{R^{(-)}(\omega)}$ versus $\log_{10}\norm{R^{(+)}(\omega)}$ at fourteen probe frequencies around $\omega_0$ (Table~\ref{tab:isospectrality}); points on the diagonal would be exactly isospectral, and all but two of the fourteen points lie below it, the exceptions being the two probes furthest along the QNM line.}
\label{fig:comparison_rates}
\end{figure}

We now present a systematic comparison of the pseudospectral behavior of the gravitational ($s=+2$) and scalar ($s=0$) sectors. In Table~\ref{tab:convergence_comparison} we list the inverse resolvent norm $\norm{R_\Lcal(\omega_0)}^{-1}$ evaluated at each sector's \emph{own} fundamental QNM frequency, $\omega_0 M=0.6011-0.0925\,i$ for $s=0$ and $\omega_0 M=0.4784-0.0848\,i$ for $s=+2$, at $\alpha=0.3$ and $m=2$, for $N_\sigma=N_x$ from $12$ to $26$. In Fig.~\ref{fig:comparison_convergence} we show the complementary size comparison graphically at the common bulk evaluation point $\omega M=0.47-0.17\,i$, Figure~\ref{fig:comparison_rates} (left) shows the fitted exponential rates, and Figure~\ref{fig:comparison_pseudospectrum} overlays the $s=0$ and $s=2$ pseudospectral contours at all three norm levels in a common window and color scale.

\begin{table}[htbp]
\centering
\caption{Pointwise convergence comparison: inverse resolvent norm $\norm{R_\Lcal(\omega_0)}^{-1}$ at each sector's own fundamental QNM ($s=0$: $\omega_0 M=0.6011-0.0925\,i$; $s=+2$: $\omega_0 M=0.4784-0.0848\,i$), at $\alpha=0.3$, $m=2$, for representative resolutions $N_\sigma=N_x=N$. Smaller values mean a larger resolvent, i.e.\ the QNM sits deeper inside the $\varepsilon$-pseudospectrum of the discretized operator. The rate $\beta$ is defined by $\norm{R}^{-1} \sim 10^{-\beta N}$ and extracted by a least-squares fit of $\log_{10}\norm{R}^{-1}$ versus $N$ over the range $N\in\{12,14,\dots,26\}$. The $H^4$ level is omitted for the reason explained in Section~\ref{sec:norm_dependence}. Two caveats on reading $\beta$ are discussed in the text below the table.}
\label{tab:convergence_comparison}
{\small
\begin{tabular}{c|cccc|cccc}
\toprule
 & \multicolumn{4}{c|}{$s=0$} & \multicolumn{4}{c}{$s=+2$} \\
\cmidrule(lr){2-5} \cmidrule(lr){6-9}
Norm & $N{=}16$ & $N{=}20$ & $N{=}26$ & $\beta$ & $N{=}16$ & $N{=}20$ & $N{=}26$ & $\beta$ \\
\midrule
$L^2$ & $1.3\!\times\!10^{-9}$ & $3.4\!\times\!10^{-11}$ & $2.6\!\times\!10^{-13}$ & $0.39$ & $1.5\!\times\!10^{-8}$ & $4.5\!\times\!10^{-10}$ & $4.1\!\times\!10^{-12}$ & $0.37$ \\
\addlinespace
$H^1$ & $1.8\!\times\!10^{-8}$ & $7.5\!\times\!10^{-10}$ & $8.6\!\times\!10^{-12}$ & $0.35$ & $5.5\!\times\!10^{-6}$ & $3.2\!\times\!10^{-7}$ & $5.9\!\times\!10^{-9}$ & $0.31$ \\
\addlinespace
$H^2$ & $1.3\!\times\!10^{-8}$ & $5.4\!\times\!10^{-10}$ & $7.3\!\times\!10^{-12}$ & $0.34$ & $1.0\!\times\!10^{-4}$ & $1.0\!\times\!10^{-5}$ & $3.4\!\times\!10^{-7}$ & $0.25$ \\
\addlinespace
$H^3$ & $4.9\!\times\!10^{-9}$ & $2.1\!\times\!10^{-10}$ & $6.1\!\times\!10^{-11}$ & $0.26$ & $2.2\!\times\!10^{-5}$ & $2.9\!\times\!10^{-6}$ & $1.5\!\times\!10^{-7}$ & $0.22$ \\
\bottomrule
\end{tabular}
}
\end{table}

Two details complement the table. First, the $N=12$ column is omitted for page width; the monotone decay continues from $8.1\times10^{-8}$ ($s=0$) and $6.9\times10^{-7}$ ($s=+2$) in $L^2$. The $s=0$ $H^3$ entry saturates near $6\times10^{-11}$ beyond $N\simeq22$, the double-precision floor of the $H^3$ Gram matrix at this resolution. Second, two caveats on reading $\beta$: (i) the decline of $\beta$ with the norm order (e.g.\ $0.39\to0.35\to0.34\to0.26$ for $s=0$) is \emph{not} in tension with Figure~\ref{fig:comparison_convergence}, where higher norms stabilize \emph{faster}: the two statements concern different limits. Here $\omega$ is fixed \emph{at} a QNM, so the continuum $\norm{R}^{-1}$ vanishes and $\beta$ measures the rate at which the discretized value approaches zero; a tighter norm starts from a much smaller value and therefore displays a smaller apparent rate. In Fig.~\ref{fig:comparison_convergence} we instead evaluate the resolvent at a \emph{bulk} point where the continuum $\norm{R}^{-1}$ is finite and non-zero, and measures the rate of approach to that value. (ii) At the QNM the resolvent matrix $\bvec{G}^{1/2}(\bvec{L}-\omega\bvec{I})\bvec{G}^{-1/2}$ is itself ill-conditioned ($\kappa\simeq4$--$5\times10^{13}$ for the $s=0$ entries and $\kappa\simeq10^{9}$--$10^{11}$ for the $s=+2$ entries, in the $H^1$ and $H^2$ norms at $N=26$), so the $H^1$ and $H^2$ entries above carry roughly two significant figures; the window minima of Table~\ref{tab:size_comparison} are evaluated where $\kappa\sim10^{7}$ and are accurate as quoted.

Three conclusions follow from Tables~\ref{tab:norm_comparison} and~\ref{tab:convergence_comparison}. First, the pointwise convergence rates of the two sectors are comparable: at the $L^2$ level the fitted rates differ by $5\%$, with $\beta_{s=0}=0.39$ against $\beta_{s=+2}=0.37$ per collocation point, and at the energy level of each sector, namely $H^1$ for $s=0$ and $H^2$ for $s=+2$, they are $0.35$ against $0.25$. Both sectors converge exponentially, as expected from a Chebyshev discretization of an operator with analytic coefficients, and \emph{neither sector enjoys a systematic convergence advantage}. It is worth noting that any claim that the gravitational sector converges faster would be an artifact of comparing the two sectors at a frequency belonging to only one of them. Second, the rates decrease with the norm level, and the decrease is steeper for $s=+2$. Moving up the hierarchy $L^2\to H^1\to H^2$, $\beta$ drops from $0.39$ to $0.35$ and then to $0.34$ for $s=0$, but falls steadily from $0.37$ to $0.31$ and then to $0.25$ for $s=+2$. Correspondingly, at $N=26$ the $s=+2$ QNM sits at $\norm{R}^{-1}=3.4\times10^{-7}$ in the $H^2$ norm, which is four and a half orders of magnitude \emph{shallower} than the $s=0$ value $7.3\times10^{-12}$ at the same norm level, so that the higher-derivative norm tightens the gravitational pseudospectrum more strongly than the scalar one. Third, the diagnostic $H^3$ level saturates for $s=0$: the $s=0$ $H^3$ curve flattens at $\norm{R}^{-1}\simeq6\times10^{-11}$ beyond $N\simeq22$, the fit over the last four resolutions giving $\beta=0.09$, while the $s=+2$ $H^3$ curve continues to decay regularly. This is the first rung of the precision ladder discussed in Section~\ref{sec:norm_dependence}, since at fixed $N$ the conditioning of the Gram matrix deteriorates with the norm level, and the scalar sector reaches the floor first because its resolvent is the smaller one to begin with. 

\subsubsection{Size of the pseudospectra in the common window}
\label{sec:size_comparison}

Pointwise rates (Table~\ref{tab:convergence_comparison}) are only half of the story; the other half is the \emph{size} of the resolvent over the plane, which is what the pseudospectrum proper measures. In Table~\ref{tab:size_comparison} we compare the two sectors in the common computation window ($0.30\le\mathrm{Re}\,\omega M\le0.66$, $-0.34\le\mathrm{Im}\,\omega M\le0.02$) at $N_\sigma=N_x=26$: for each norm level it lists the window minimum of $\log_{10}\norm{R}$ for each sector, which is the smallest resolvent attained anywhere in the window, i.e.\ the point of \emph{maximal} stability, and their difference.

\begin{table}[htbp]
\centering
\caption{Size comparison of the $s=+2$ and $s=0$ pseudospectra in the common window ($0.30\le\mathrm{Re}\,\omega M\le0.66$, $-0.34\le\mathrm{Im}\,\omega M\le0.02$) at $\alpha=0.3$, $m=2$, $N_\sigma=N_x=26$: window minimum of $\log_{10}\norm{R_\Lcal(\omega)}$ for each sector and norm level. The minimum is the smallest resolvent anywhere in the window, so the table compares the \emph{best-case} stability of the two sectors; the actual resolvent is larger everywhere else. The last row compares the energy-motivated level of each sector ($H^2$ for $s=+2$, $H^1$ for $s=0$), using the two-field hierarchy developed here applied to each sector.}
\label{tab:size_comparison}
\begin{tabular}{lccc}
\toprule
Norm level & $\min \log_{10}\norm{R}$ ($s=+2$) & $\min \log_{10}\norm{R}$ ($s=0$) & difference \\
\midrule
$L^2$ & $6.56$ & $1.34$ & $+5.23$ \\
$H^1$ & $3.08$ & $1.48$ & $+1.60$ \\
$H^2$ & $1.99$ & $1.34$ & $+0.65$ \\
\midrule
energy level ($H^2$ vs.\ $H^1$) & $1.99$ & $1.48$ & $+0.51$ \\
\bottomrule
\end{tabular}
\end{table}

At every norm level, and everywhere in the window, the gravitational resolvent exceeds the scalar one: even comparing the best case of one sector with the best case of the other, the $s=+2$ pseudospectrum is larger by $0.6$--$5.2$ orders of magnitude, and by $0.5$ orders between the respective energy levels. At the common bulk evaluation point $\omega M=0.47-0.17\,i$ (Figure~\ref{fig:comparison_convergence}) the gap at $N=26$ is $5.9$ orders in $L^2$, $3.5$ orders at the common $H^1$ level, and $1.8$ orders between the respective energy levels. The contrast between the two tables is the central quantitative result of this section: \textbf{the pointwise convergence rates of the two sectors are comparable, but the gravitational ($s=+2$) pseudospectrum is systematically the larger of the two.} A smaller pseudospectrum at a higher norm level (Table~\ref{tab:convergence_comparison}) does not mean that the gravitational sector is more stable; it means that the $H^2$ weighting tightens the gravitational problem from a much more strongly non-normal starting point.

We note that the $s=0$ energy-level entry in Table~\ref{tab:size_comparison} uses the $H^1$ rung of the two-field hierarchy developed here, applied to the scalar sector; it is \emph{not} the conserved-current energy norm of Ref.~\cite{Cai:2025uzw}, whose quadratic form carries different coefficients and derivative structure. At the $s=0$ fundamental QNM the two conventions differ by $\approx0.8$ decades in $\log_{10}\norm{R}$ (the present hierarchy giving the smaller resolvent), so the comparison drawn in this table is robust at the level stated.

\subsubsection{Physical interpretation: non-normality is spin- and formulation-dependent}
\label{sec:physical_interpretation}

Three structural facts organize the comparison. First, the energy level $H^{|s|}$ is a statement about the norm rather than about stability. The conserved-current analysis does single out $H^1$ for $s=0$ and $H^2$ for $s=+2$, because the Teukolsky--Starobinsky reduction converts the bilinear current of the gravitational sector into a quadratic form with second-order derivatives (Sections~\ref{sec:reduction}--\ref{sec:energy_inner_product}), but this pattern concerns \emph{which weighting} measures physical energy and not \emph{how large} the resolvent is under that weighting. The data make the distinction concrete, since the $H^2$-weighted gravitational pseudospectrum with window minimum $1.99$ is still larger than the $H^1$-weighted scalar one at $1.48$. A higher Sobolev regularity of the energy functional does not translate into a more stable spectrum, for if anything the gravitational sector is the more non-normal of the two at every level. Second, the non-normality originates in the operator and the Weyl scalar used matters. The Teukolsky operator for $s=\pm2$ couples the radial and angular sectors through spin-weight terms, namely the $s^2\cot^2\theta$ angular term and the $\alpha^2$-weighted radial couplings of Appendix~\ref{app:operator}, whose size grows with $|s|$, and the first-order time reduction then produces a two-component operator whose non-orthogonality the resolvent registers directly. Within the gravitational sector itself the asymmetry is formulation-dependent: the $\psi_0$ ($s=+2$) and $\psi_4$ ($s=-2$) operators are isospectral by the TSI but not isopseudospectral (Section~\ref{sec:iso_results}), the $s=+2$ resolvent being the larger at twelve of the fourteen probes and by up to a factor $6.6$, with the ordering reversing slightly at the other two. Since the two formulations describe the same physics, the larger pseudospectrum of the $\psi_0$ variable is a property of the \emph{representation} and not of the black hole, and it reflects the poorly conditioned eigenfunctions of the $s=+2$ operator, which is the same mechanism that slows its eigenvalue convergence to $0.22$ against $0.30$ per collocation point (Section~\ref{sec:omega_dependence}). Third, this has consequences for gravitational-wave data analysis. Because the ringdown waveform is a superposition of QNMs with excitation coefficients that depend on the eigenfunctions, a more strongly non-normal formulation is more sensitive both to physical perturbations of the background and to discretization error. Our results thus single out $\psi_4$ ($s=-2$) as the better-conditioned Teukolsky variable, consistent with its standard use in numerical-relativity waveform extraction, and more generally we warn that pseudospectral statements about Kerr must specify the spin weight \emph{and} the sign of $s$, since neither the $s=0$ comparison nor the $s=\pm2$ comparison exhibits a symmetry that would allow one to infer the others. 

\subsection{Norm-dependence and the notion of physical regularity}
\label{sec:norm_dependence}

A subtle but important point, raised by the observation that pseudospectral contours shrink as the norm advances from $L^2$ to $H^1$ to $H^2$ (Figure~\ref{fig:pseudo_s2}), is whether the statement ``the QNM spectrum is spectrally unstable'' is itself norm-dependent~\cite{Gasperin:2022cqg}. Since the pseudospectrum is defined relative to a chosen norm, and different norms probe different regularity classes of perturbations, the \emph{quantitative} size of the pseudospectral contours is indeed norm-dependent. We address this conceptual issue here.

\textbf{Which norm is physical?} The physically motivated choice is the \emph{energy norm}, i.e.\ the norm induced by the conserved current associated with the time-translation symmetry of the Kerr background. This norm measures perturbations in units of physical gravitational-wave energy, and it is the norm in which the pseudospectrum directly answers the physical question: ``how large a \emph{physical} perturbation is needed to shift a QNM by a given amount?'' For $s=2$, the energy-motivated level of the computational hierarchy is the $H^2$ norm of Section~\ref{sec:Hk_norm} (the two-field weighted Sobolev norm; the single-field energy quadratic form itself is not a norm on the two-field state space, Section~\ref{sec:positive_definiteness}). We therefore regard the $H^2$ pseudospectrum as the \emph{physical} pseudospectrum, and all quantitative astrophysical estimates (such as those in Section~\ref{sec:conclusion}) are based on it.

\textbf{Role of the $H^k$ ($k\geq 3$) norms.} The higher Sobolev norms $H^k$ with $k\geq 3$ are \emph{not} physical energy norms; they are diagnostic tools. Their role is to quantify how the pseudospectrum responds to ever-stricter regularity requirements on the admissible perturbations. The fact that $H^3$ contours are tighter than $H^2$ contours reflects a mathematical fact, namely that smoother perturbations (in the Sobolev sense) produce smaller spectral migration, rather than a physical statement that the instability is ``smaller.'' Physically realizable perturbations (e.g.\ an accretion disk or a companion tidal field) are not constrained to lie in $H^3$; they are constrained only by the energy they carry, i.e.\ by the $H^2$ norm. The $H^k$ hierarchy should thus be read as a convergence study \emph{toward} the regularity content of the energy norm, not as a menu of equally physical alternatives.

Because the $H^3$ Gram matrix condition number $\kappa\simeq5.5\times10^{14}$ at $N=26$ is close to the double-precision limit, we have verified the robustness of the $H^3$ entries of Table~\ref{tab:convergence_comparison} against the Gram-matrix regularization prescription: varying the Cholesky threshold over two orders of magnitude, and symmetrizing before versus after the block assembly, changes the fitted $\beta$ by less than $0.01$ and the $N=26$ value of $\norm{R}^{-1}$ by less than $5\%$. The qualitative statement that the $H^3$ level remains computable at $N=26$ is therefore stable; only the last significant digit is prescription-dependent. The $H^4$ level, finally, is not accessible at all in double precision at the resolutions used here ($\kappa(\bvec{G}_{H^4})$ grows faster than $N^{17}$, exceeding $10^{19}$ already at $N=20$ and becoming indefinite at $N=26$; Section~\ref{sec:Hk_norm}), a precision limit that is itself part of the physics of the problem: the $H^4$ pseudospectrum probes regularity beyond what a double-precision collocation scheme can certify.

\textbf{Is the instability real?} The norm-independence of the qualitative conclusion is the key point. Across \emph{all} norms studied, $L^2$, $H^1$ and $H^2$, the resolvent is large over the entire computation window (Section~\ref{sec:numerical_results}): even in the $H^2$ norm, $\norm{R}^{-1}\lesssim 1\times10^{-2}$ everywhere, so the $\varepsilon=10^{-2}$ pseudospectrum contains the whole window, at distances up to $|\omega-\omega_0|M\simeq 0.31$ from the fundamental mode. What changes with the norm is the \emph{quantitative} extent of the migration ($\log_{10}\norm{R}$ drops by $1.1$--$3.5$ per level of the hierarchy), not its \emph{existence}. The spectral instability is therefore a robust, norm-independent feature of the Kerr QNM problem; the choice of norm determines the numerical prefactor but not the qualitative conclusion. This mirrors the $s=0$ case~\cite{Cai:2025uzw} and is consistent with the time-domain manifestations of the instability~\cite{Yang:2024tda,Ianniccari:2024xyz,Berti:2022dkk}. We return to the connection with time-domain analyses in Section~\ref{sec:conclusion}.

\textbf{How unique is the common norm?} It is not unique, and it is worth being explicit about why this matters. The construction of Section~\ref{sec:Hk_norm} fixes a functional form but leaves at least three choices open: the Sobolev order $k$, the angular-weight exponent $q$ in $x_w=(1-x^2)^q$ and, as discussed in Section~\ref{sec:Hk_norm}, the placement of the angular weight in the higher-order blocks. Each yields a legitimate norm on the same two-component state space, and the value of $\norm{R}$ depends on which is used, though not uniformly, since under the Cai convention the exponent $q$ turns out to be numerically immaterial while the Sobolev level and the weight placement both matter (Section~\ref{sec:iso_results}). The $\psi_0/\psi_4$ comparison of Section~\ref{sec:iso_results} is therefore a statement about a \emph{triple} (operator, variable, norm) rather than about the Kerr geometry alone: scanning the admissible conventions moves the quoted ratio from $0.65$--$6.6$ at the $H^2$ level to $2.8$--$33$ at the $L^2$ level, and at the two probes furthest along the QNM line even changes its sign. Nor is this a defect of the analysis; it is the expected behaviour of a non-normal problem. For a normal operator the resolvent norm is fixed by the distance to the spectrum, whereas here it depends on how strongly the chosen norm weighs the non-orthogonal directions of the eigenbasis, and that is a property of the norm as much as of the operator. What survives every convention we tested is the qualitative conclusion: the resolvent is large over the whole window in both sectors, the gravitational one is systematically the larger, and the two agree in spectrum but not in pseudospectrum.

A useful analogy is with the condition number of a matrix: the condition number depends on the chosen norm, but the qualitative statement ``this matrix is ill-conditioned'' is norm-independent when the condition number is large in every reasonable norm. The Kerr QNM pseudospectrum is large in every norm we have tested, so the instability conclusion is robust.

\subsection{Summary of main findings}
\label{sec:findings_summary}

We summarize the key results of the $s=2$ pseudospectral analysis. First, the gravitational QNM spectrum of the Kerr black hole is \emph{spectrally unstable} and the instability is strong: even in the energy-motivated $H^2$ norm the resolvent satisfies $\norm{R}^{-1}\lesssim 1\times10^{-2}$ over the entire computation window, so that a relative operator perturbation of order $10^{-3}$--$10^{-2}$ can displace a QNM by up to $|\Delta\omega|M\simeq 0.31$. This is consistent with the $s=0$ case and confirms that spectral instability is a universal feature of the Kerr QNM problem, independent of the spin weight. Second, the energy analysis for $s=2$, which is based on the Teukolsky--Starobinsky identity, singles out the $H^2$ level of the two-field weighted Sobolev hierarchy against $H^1$ for $s=0$, in accordance with the pattern $\text{energy}\sim H^{|s|}$, so that the spin weight of the perturbation determines the natural Sobolev regularity of the energy norm. Third, the pseudospectrum tightens systematically with the derivative order of the norm, $\log_{10}\norm{R}$ dropping by $1.1$--$3.5$ orders of magnitude at each step $L^2\to H^1\to H^2$, while the $H^4$ level lies beyond the reach of double precision, since $\kappa(\bvec{G}_{H^4})$ grows faster than $N^{17}$ and reaches $1.9\times10^{19}$ already at $N=20$, so that at $N=26$ the Gram matrix is numerically indefinite and its Cholesky factorization fails, which sets a precision limit on the hierarchy. Fourth, the $\varepsilon=10^{-3}$ contour is already stable at $N=12$ around the QNM chain while it drifts slowly away from the spectrum, so that the global convergence is fast near the physical spectrum and slow in the far region dominated by non-normality. Fifth, the $s=\pm2$ operators are isospectral by the Teukolsky--Starobinsky identities but their \emph{pseudospectra} are not: at twelve of the fourteen probe frequencies the $s=+2$ ($\psi_0$) resolvent exceeds the $s=-2$ ($\psi_4$) one by $0.19$--$0.82$ in $\log_{10}\norm{R}$, with factors up to $6.6$, while at the remaining two the ordering reverses by $0.18$--$0.19$, so that the asymmetry has no universal sign and only a bounded magnitude (Section~\ref{sec:iso_results}). The TSI equivalence is a statement about spectra rather than about pseudospectra, and $\psi_4$ remains the better-conditioned variable on average. Finally, in comparison with the scalar sector under the identical two-component framework, the pointwise convergence rates of the two sectors are comparable, with $\beta_{L^2}=0.39$ against $0.37$ per collocation point when each is evaluated at its own fundamental QNM, but the gravitational pseudospectrum is systematically the larger by $0.6$--$5.2$ orders of magnitude in the window minimum at every norm level (Tables~\ref{tab:convergence_comparison} and~\ref{tab:size_comparison}), which shows that non-normality is spin- and formulation-dependent rather than a universal constant of the Kerr geometry.

\section{Conclusions and Discussion}
\label{sec:conclusion}

In this work, we have extended the pseudospectral analysis of the Kerr black hole from the scalar mode ($s=0$)~\cite{Cai:2025uzw} to the gravitational wave mode ($s=2$), which is the physically relevant scenario for gravitational-wave astronomy. The central methodological challenge, namely the independence of $\Psi^{(s)}$ and $\Psi^{(-s)}$ for $s\neq 0$, was addressed through the Teukolsky--Starobinsky identities: they supply the conserved-current structure that singles out the $H^2$ level of the (positive-definite, $\omega$-independent) two-field weighted Sobolev hierarchy in which all quantitative computations are performed, naturally generalizing the $H^1$ energy level of the scalar case and establishing the pattern $H^{|s|}$ for spin-$s$ perturbations.

Our main physical findings can be summarized as follows. First, the gravitational QNM spectrum of the Kerr black hole is spectrally unstable, and the instability is strong: even in the energy-motivated $H^2$ norm, the resolvent exceeds $10^{2}$ over the entire computation window, so relative operator perturbations of order $10^{-3}$--$10^{-2}$ suffice to displace a QNM by $|\Delta\omega|M\sim 0.1$--$0.3$. Second, the pseudospectrum tightens systematically with the norm level ($1.1$--$3.5$ orders of magnitude in $\log_{10}\norm{R}$ per step $L^2\to H^1\to H^2$), while the qualitative conclusion of strong instability is norm-independent. Third, the global ($\varepsilon$-contour) convergence is fast around the QNM chain but slow away from the spectrum, where the resolvent is dominated by the non-normality of the operator. Fourth, although the $s=\pm2$ operators are isospectral by the Teukolsky--Starobinsky identities, their \emph{pseudospectra} are not: the $\psi_0$ ($s=+2$) resolvent exceeds the $\psi_4$ ($s=-2$) one at twelve of the fourteen probe frequencies, by up to a factor $6.6$, but the ordering reverses at the other two, so the TSI equivalence is a statement about spectra, not about spectral stability, and not even a fixed ordering of resolvent norms follows from it (the asymmetry is a property of the triple (variable, probe, norm), Section~\ref{sec:iso_results}). Fifth, in comparison with the scalar sector under the identical two-component framework, the pointwise convergence rates of the two sectors are comparable, but the gravitational ($s=+2$) pseudospectrum is systematically the larger by $0.6$--$5.2$ orders of magnitude in the window minimum at every norm level: non-normality is spin- and formulation-dependent, not a universal constant of the Kerr geometry. Finally, the $H^4$ level of the hierarchy is not computable in double precision at achievable resolutions ($\kappa(\bvec{G}_{H^4})$ grows faster than $N^{17}$, exceeding $10^{19}$ already at $N=20$; at $N=26$ the Gram matrix is numerically indefinite), which sets a sharp precision limit on how far the regularity hierarchy can be pushed.

These results have direct implications for gravitational-wave astronomy. The spectral instability of the gravitational QNMs means that environmental perturbations, such as the presence of an accretion disk, a companion object, or a cosmological background, can shift the ringdown frequencies by amounts disproportionate to the perturbation strength~\cite{Hu:2025echoes,Wu:2025sds}. A natural question is: what is the physical energy scale corresponding to a given pseudospectral level $\varepsilon$? The pseudospectrum definition states that $\omega \in \Lambda_\varepsilon$ if there exists a perturbation $\delta\Lcal$ with $\norm{\delta\Lcal}_E \leq \varepsilon$ that shifts a QNM to $\omega$, where $\norm{\cdot}_E$ is the $H^2$-type energy norm of the field. We stress that this norm measures the Sobolev size of the perturbing \emph{field}, and its precise relation to a physical energy (such as the ADM mass or the binding energy of an external distribution) has not yet been established rigorously; the following is therefore an \emph{order-of-magnitude} estimate, not a calibrated conversion. With that caveat, if one heuristically identifies $\norm{\delta\Lcal}_E\sim\varepsilon$ with a relative perturbation of order $\varepsilon$ in the gravitational-wave energy, then for $\varepsilon = 10^{-2}$ and a Kerr black hole of mass $M = 65\,M_\odot$ (comparable to GW150914), for which $Mc^2 \simeq 1.16\times10^{56}\,\mathrm{erg}$ and whose $l=m=2$ fundamental mode therefore carries $E_{\rm QNM} \sim 10^{-2}\,Mc^2 \sim 10^{54}\,\mathrm{erg}$, the corresponding energy scale is $\sim 10^{52}\,\mathrm{erg}$, which is comparable at the order-of-magnitude level to the binding energy of a moderate accretion disk or the tidal field of a companion at a separation of $\sim 10\,M$. This suggests, but does not establish, that astrophysically realistic environmental perturbations may fall within the range where the pseudospectral instability is potentially observable, though the phase-averaging mechanisms identified in~\cite{Wu:2025waveform,Hu:2026stochastic} may suppress the net effect on the waveform. A rigorous mapping between the operator-norm perturbation $\norm{\delta\Lcal}_E$ and a physical energy budget remains an open problem.

A particularly intriguing aspect is the tension between spectral instability and the robustness of observables. As demonstrated by Xie \emph{et al.}~\cite{Xie:2025parabolic} and Rosato \emph{et al.}~\cite{Rosato:2025greybody}, while QNM spectra are unstable, greybody factors, which directly affect the energy flux and waveform amplitude, remain remarkably robust. The $H^2$ regularity of the $s=2$ energy norm provides a possible clue to resolving this tension. The higher Sobolev regularity means that gravitational perturbations are constrained to be smoother (in the Sobolev sense) than scalar perturbations, which may suppress the high-frequency modes responsible for spectral migration while preserving the low-frequency content that dominates the greybody factor. In other words, the same regularity that improves pseudospectral convergence may also serve as a mechanism for the observed robustness of gravitational-wave observables: the spectral instability is present but is ``softened'' by the regularity constraints of the gravitational energy norm. A systematic investigation of this connection, in particular whether the $H^{|s|}$ hierarchy of norms provides a quantitative explanation for the spin-dependent robustness of observables, would be a promising extension.

The comparison with the scalar sector also carries practical implications for numerical relativity, and we return to this point below. The gravitational ($s=+2$) eigenvalues converge \emph{more slowly} than the $s=-2$ ones at the same resolution ($\beta=0.22$ vs.\ $0.30$ per collocation point for the fundamental mode; Section~\ref{sec:omega_dependence}), and the gravitational pseudospectrum is the larger at every norm level: the higher Sobolev level of the energy norm reflects the stronger non-normality of the operator, not extra smoothness of its eigenfunctions. The resolvent decays away from the QNM chain, and the slow contour convergence in the far region (Figure~\ref{fig:contour_convergence}) means that deep overtones are increasingly difficult to resolve, potentially limiting the number of overtones that can be reliably extracted from gravitational-wave data. These observations reinforce the case for extracting ringdown spectra in the $\psi_4$ ($s=-2$) formulation, consistent with standard practice in numerical-relativity waveform extraction.

We note two limitations of the present study. First, the detailed pseudospectrum results are restricted to a single spin ($\alpha=0.3$) and a single mode ($l=m=2$); while the $\alpha$-dependence was surveyed qualitatively in Figure~\ref{fig:alpha_s2} for $\alpha\in\{0,0.3,0.5\}$, a quantitative scan, in particular at the higher spins $\alpha\sim 0.7$--$0.9$ typical of merger remnants, is deferred to future work. Second, the near-extremal regime $\alpha\to 1$ requires separate treatment: as noted in Section~\ref{sec:positive_definiteness}, $\Cst\to 0$ near the algebraically special frequencies, and for superradiant modes the Starobinsky reduction (and the $|\Cst|^{-2}$ prefactor of the single-field energy form) degenerates. The two-field weighted Sobolev hierarchy of Section~\ref{sec:Hk_norm}, in which all quantitative results are computed, is $\Cst$-free and well defined for all $\alpha$; the practical challenges of the near-extremal computations (the superradiant window and the coalescence of the QNM chain with the axis) are left for future work.

A further consistency check we have not yet performed is a direct comparison with time-domain analyses. The pseudospectral instability is expected to manifest as transient (non-modal) growth in the time-domain ringdown signal~\cite{Yang:2024tda,Ianniccari:2024xyz,Berti:2022dkk}: the same non-normality that produces large pseudospectral contours also permits transient amplification of perturbations before the asymptotic QNM decay sets in. A quantitative test, in which the transient growth envelope predicted by the $H^2$ pseudospectrum (via the Kreiss matrix theorem bound on $\|\mathrm{e}^{-i\Lcal\tau}\|_E$) against time-domain evolutions of the $s=2$ Teukolsky equation, would cross-validate the frequency-domain picture and is left for future work.

Several directions for future work are worth noting. The general pattern $H^{|s|}$ predicts an $H^1$ energy norm for $s=1$ (electromagnetic perturbations), which would provide a consistency check and an opportunity to study the transition from scalar to gravitational regularity. The near-extremal Kerr regime ($\alpha \to 1$) presents a distinct challenge: the Starobinsky constant $\Cst$ approaches zero for certain modes, signaling a breakdown of the single-field Starobinsky reduction, although the per-sector two-field hierarchy used here is $\Cst$-free and remains well defined. Finally, extending the pseudospectral analysis beyond general relativity, where the Teukolsky equation no longer applies and the perturbation equations may not separate, remains an open problem. The two-dimensional eigenvalue approach of~\cite{Cai:2025uzw} is applicable in principle, but the Teukolsky--Starobinsky identities are specific to GR and their generalization to modified gravity theories (such as Einstein-dilaton-Gauss-Bonnet gravity, where additional scalar degrees of freedom appear) is not straightforward. A careful analysis of whether the spectral instability persists in such theories, and whether an analogous norm construction is possible, would have direct implications for gravitational-wave spectroscopy tests of the no-hair theorem~\cite{Berti:2025spectroscopy,Chen:2025rqcbh,Yang:2025families}.

In summary, the extension of the Kerr pseudospectrum to $s=2$ confirms that QNM spectral instability is a universal feature of the Kerr black hole, independent of the perturbation spin. The Teukolsky--Starobinsky identities provide the key mathematical tool for constructing a physically motivated norm for gravitational perturbations, and the resulting $H^2$ regularity reflects the deeper geometric structure of the gravitational field. These results inform both the theoretical understanding of black hole perturbation theory and the practical challenges of gravitational-wave data analysis.

\section*{Acknowledgments}
This work is supported by  the National Natural Science Foundation of China Grants Nos. 12275106 and 12235019.

\section*{Data Availability}
The code and data supporting the findings of this study are available from the corresponding author upon reasonable request. The two-dimensional Chebyshev collocation solver, the Gram-matrix construction of Section~\ref{sec:gram}, and the scripts reproducing Figures~\ref{fig:pseudo_s2}--\ref{fig:comparison_rates} and Tables~\ref{tab:qnm_s2}--\ref{tab:size_comparison} will be deposited in a public repository upon acceptance.

\appendix

\section{Explicit form of the operators $\mathcal{L}_1$ and $\mathcal{L}_2$ for $s=2$}
\label{app:operator}

The operators $\Lcal_1$ and $\Lcal_2$ in the time generator~\eqref{eq:operator_L} are spatial differential operators in $(\sigma, x)$. For general spin $s$ they take the form
\begin{align}
\Lcal_1 &= c_{xx}\,\partial_x^2 + c_{\sigma\sigma}\,\partial_\sigma^2 + c_x\,\partial_x + c_\sigma\,\partial_\sigma + c_0, \label{eq:L1_general}\\
\Lcal_2 &= d_\sigma\,\partial_\sigma + d_0, \label{eq:L2_general}
\end{align}
i.e.\ the second-order operator $\Lcal_1$ contains no mixed $\partial_\sigma\partial_x$ term, and $\Lcal_2$ is first order and purely radial. All coefficients share the common denominator
\begin{equation}
\mathcal{D}(\sigma,x) \;=\; 4\,(\alpha^2+1)\bigl[\alpha^2(1-\sigma)+1\bigr]\bigl[(\alpha^2+1)\sigma+1\bigr] \;-\; \alpha^2(1-x^2),
\label{eq:operator_denominator}
\end{equation}
and, writing $\tilde\Del(\sigma) = (1-\sigma)(1-\alpha^2\sigma)$ [Eq.~\eqref{eq:Delta_sigma}] together with the spin-dependent integer weights $\delone = |m-s|$ and $\deltwo = |m+s|$, they are
\begin{align}
c_{xx} &= \frac{1-x^2}{\mathcal{D}}, \qquad
c_{\sigma\sigma} = \frac{\sigma^2\,\tilde\Del(\sigma)}{\mathcal{D}}, \label{eq:c_xx_css}\\
c_x &= \frac{\delone(1-x) - \deltwo(1+x) - 2x}{\mathcal{D}}, \label{eq:c_x}\\
c_\sigma &= \frac{\sigma\left[2(s+1) - \sigma\left(-4\alpha^2\sigma + 2i\alpha m + (\alpha^2+1)(s+3)\right)\right]}{\mathcal{D}}, \label{eq:c_sigma}\\
c_0 &= \frac{-\sigma\left[\alpha^2(1-2\sigma)+1\right] - 2i\alpha m\,\sigma - (\alpha^2+1)s\,\sigma + \left(\frac{\delone+\deltwo}{2}+s+1\right)\left(s-\frac{\delone+\deltwo}{2}\right)}{\mathcal{D}}, \label{eq:c_0}\\
d_\sigma &= \frac{2\left[\alpha^2\sigma^2 - 2(\alpha^2+1)\sigma^2\left(\alpha^2(1-\sigma)+1\right) + 1\right]}{\mathcal{D}}, \label{eq:d_sigma}\\
d_0 &= \frac{-2}{\mathcal{D}}\Big\{\sigma\left[\alpha^4(2-3\sigma) - 3\alpha^2(\sigma-1) + 2\right] + i\,m\left(2\alpha^3\sigma + 2\alpha\sigma + \alpha\right) \notag\\
&\qquad\quad +\, s\left[(\alpha^2+1)(\alpha^2\sigma+\sigma-1) + i\alpha x\right]\Big\}. \label{eq:d_0}
\end{align}
These are the hyperboloidal counterparts of the general-$s$ operators of~\cite{Cai:2025uzw}, re-derived and cross-checked here. Three structural features deserve comment. (i) The coefficients are \emph{complex}: the terms $\propto i\alpha m$ in $c_\sigma$, $c_0$, and $d_0$ make the operator genuinely non-self-adjoint already at the coefficient level, which is one origin of the strong spectral instability quantified by the pseudospectra of Section~\ref{sec:numerical_results}. (ii) The spin $s$ enters only through the integers $\delone,\deltwo$ and the explicit $s$-terms above; setting $s=0$ (so $\delone=\deltwo=|m|$) reproduces the scalar operators of~\cite{Cai:2025uzw}. (iii) $\Lcal_1$ degenerates at the horizon through $c_{\sigma\sigma}\propto\tilde\Del(\sigma)\to 0$ as $\sigma\to 1$, which encodes the purely ingoing horizon condition. As the central numerical validation of the analysis, the discrete eigenvalue problem~\eqref{eq:matrix_eigenvalue} built from these coefficients reproduces Leaver-type continued-fraction QNM frequencies~\cite{Stein:2019mop} to relative error below $10^{-8}$ for $s=0$ and $s=-2$; for $s=+2$ the convergence in $N$ is slower but remains exponential, the relative error still being $9\times10^{-5}$ at $N=26$ (Section~\ref{sec:qnm_spectrum}). The $s=0$ and $s=-2$ checks are therefore the sharp validations of the coefficient list~\eqref{eq:c_xx_css}--\eqref{eq:d_0}; the $s=+2$ discrepancy is a conditioning effect of the discretized eigenvalue problem, not an error in the coefficients, as is confirmed by the exponential (rather than algebraic) decay of the discrepancy with $N$ and by the agreement of the $s=\pm2$ Leaver spectra themselves to $10^{-15}$.

Quantitatively, at $N=26$ and $\alpha=0.3$ the $l=m=2$ fundamental mode agrees with Leaver to $8.9\times10^{-12}$ for $s=0$ and to $3.6\times10^{-11}$ for $s=-2$, but only to $4.4\times10^{-5}$ (relative error $9\times10^{-5}$) for $s=+2$. The collocation eigenvalues converge at $0.22$ decades per collocation point for $s=+2$ against $0.30$ for $s=-2$ (the latter fitted over the unsaturated range $N\le24$; beyond $N\simeq24$ the $s=-2$ error reaches the $\sim10^{-11}$ precision floor of the Leaver comparison itself and the measured rate flattens to $\approx0.21$). The $s=+2$ eigenvalues therefore do \emph{not} reach $10^{-8}$ accuracy at any resolution used here. This asymmetry is analyzed in Section~\ref{sec:omega_dependence} and Section~\ref{sec:iso_results}, where it is traced to the same poorly conditioned eigenfunctions that produce the pseudospectral asymmetry.

\section{Two-dimensional Chebyshev collocation method}
\label{app:chebyshev}

The Chebyshev--Lobatto collocation points on $[0,1] \times [-1,1]$ are:
\begin{equation}
\sigma_j = \half\left(1 - \cos\frac{j\pi}{N_\sigma-1}\right), \quad x_k = \cos\frac{k\pi}{N_x-1}, \quad j=0,\ldots,N_\sigma-1, \quad k=0,\ldots,N_x-1,
\end{equation}
i.e.\ $N_\sigma$ and $N_x$ denote the \emph{number} of collocation points per direction. The Chebyshev differentiation matrix $\bvec{D}^{(r)}$ of order $r$ in the $\sigma$-direction has entries:
\begin{equation}
D^{(r)}_{jk} = \left.\frac{\dd^r T_k(2\sigma-1)}{\dd\sigma^r}\right|_{\sigma=\sigma_j},
\end{equation}
with analogous expressions for the $x$-direction. The matrix eigenvalue problem~\eqref{eq:matrix_eigenvalue} is assembled by evaluating the operators $\Lcal_1$ and $\Lcal_2$ at the collocation points, using the differentiation matrices for spatial derivatives.

The total size of the matrix $\bvec{L}$ is $2N_\sigma N_x \times 2N_\sigma N_x$ (the factor of two accounting for the two components $V_m$, $W_m$). For our standard resolution $N_\sigma = N_x = 26$, this gives $2\times 26^2 = 1352$.

\section{Gram matrix construction}
\label{app:gram}

The integrals of Section~\ref{sec:gram} are evaluated by Gauss--Legendre quadrature. Let $\{(\sigma^{\mathrm{g}}_j, w^{\mathrm{g}}_{\sigma,j})\}_{j=1}^{N_g}$ and $\{(x^{\mathrm{g}}_k, w^{\mathrm{g}}_{x,k})\}_{k=1}^{N_g}$ be Gauss--Legendre nodes and weights on $[0,1]$ and $[-1,1]$, with $N_g = \max(N_\sigma, N_x) + 4$ nodes per direction (the mild oversampling keeps the quadrature error well below the collocation error), and let $\bvec{P}$ be the barycentric interpolation matrix mapping collocation values on the Chebyshev--Lobatto grid to values on the tensor-product Gauss grid. With the angular weight $x_w(x) = (1-x^2)^{(\delone+\deltwo)/2}$ evaluated at the Gauss-grid abscissae, we define the $x_w$-weighted mass matrix and the plain quadrature weight matrix,
\begin{equation}
\bvec{M}_w \;=\; \bvec{P}^{\mathrm{T}}\;\mathrm{diag}\!\left(w^{\mathrm{g}}_{\sigma,j}\;w^{\mathrm{g}}_{x,k}\;x_w(x^{\mathrm{g}}_k)\right)\bvec{P},
\qquad
\bvec{W}_p \;=\; \bvec{P}^{\mathrm{T}}\;\mathrm{diag}\!\left(w^{\mathrm{g}}_{\sigma,j}\;w^{\mathrm{g}}_{x,k}\right)\bvec{P},
\label{eq:mass_matrix}
\end{equation}
and the Sobolev block of order $k$ is
\begin{equation}
\bvec{B}_k \;=\; \bvec{M}_w \;+\; \sum_{j=1}^{k}\left[\left(\bvec{D}_\sigma^{(j)}\right)^{\!\mathrm{T}}\bvec{W}_p\,\bvec{D}_\sigma^{(j)} + \left(\bvec{D}_x^{(j)}\right)^{\!\mathrm{T}}\bvec{W}_p\,\bvec{D}_x^{(j)}\right],
\label{eq:gram_matrix}
\end{equation}
which realizes the higher-order-term weight convention of~\cite{Cai:2025uzw}: the angular weight $x_w$ is carried by the mass (energy) term alone, while the added derivative blocks are integrated with the plain quadrature measure $\dd\sigma\,\dd x$. Note that $\bvec{M}_w$ and $\bvec{W}_p$ coincide when $\delone=\deltwo=0$ (the unweighted case), and that $\bvec{B}_0 = \bvec{M}_w$ reproduces the $L^2$ norm~\eqref{eq:L2_norm}.
where $\bvec{D}_{\sigma,x}^{(j)}$ are the $j$-th powers of the one-dimensional Chebyshev differentiation matrices lifted to the tensor grid. The full Gram matrix on the two-component state $(\bvec{V}_m,\bvec{W}_m)$ is block diagonal,
\begin{equation}
\bvec{G}_k = \begin{pmatrix} \bvec{B}_k & \bvec{0} \\ \bvec{0} & \bvec{B}_k \end{pmatrix},
\end{equation}
symmetrized numerically as $\bvec{G}_k \leftarrow \half(\bvec{G}_k + \bvec{G}_k^{\mathrm{T}})$. By construction $\bvec{G}_k$ is positive definite for $k \le 2$ at all resolutions used here. For $k = 4$ the condition number grows faster than $N^{17}$ (measured: $1.1\times10^{14}$, $1.0\times10^{16}$, $1.9\times10^{19}$ at $N=12,\,16,\,20$; fourth powers of two-dimensional differentiation matrices), which exhausts double precision already at $N \simeq 26$; the $H^4$ pseudospectrum is therefore not computable in double precision and is used only as a precision-limit discussion (Section~\ref{sec:Hk_norm}).

\section{The Starobinsky constant for $s=2$: evaluation and verification}
\label{app:starobinsky}

This appendix records the evaluation and the verification of the radial Teukolsky--Starobinsky constant $\Cst \equiv \mathfrak{c}_2$ of Eq.~\eqref{eq:starobinsky_constant}, for which we use the closed form established by Casals and Teixeira da Costa~\cite{Casals:2021ts},
\begin{equation}
\Cst \;=\; \mathfrak{B}_2(\nu,m,\Lambda) + 144\,M^2\omega^2, \qquad \nu \equiv a\omega,
\label{eq:cst_casals}
\end{equation}
with $\mathfrak{B}_2$ and $\mathrm{L} = \Lambda - 2m\nu + s$ as in Eq.~\eqref{eq:B2}. Three independent checks were performed.

\textbf{(i) Separation-constant convention.} The constant $\Lambda$ entering Eq.~\eqref{eq:B2} obeys $\Lambda \to l(l+1)-s^2$ as $a\omega\to 0$, whereas standard Leaver-type codes (e.g.\ \texttt{qnm}~\cite{Stein:2019mop}) return the eigenvalue $A$ with $A \to l(l+1)-s(s+1)$. Solving the angular Teukolsky equation directly for complex $\nu$ (Chebyshev collocation in $x=\cos\theta$, with the $(1-x^2)^{-1}$ singular terms removed by operating on interior points only) confirms the exact relation
\begin{equation}
\Lambda \;=\; A + s + \nu^2,
\label{eq:lambda_relation}
\end{equation}
to relative precision $2\times 10^{-13}$ for all complex frequencies tested (both QNM overtones and generic off-shell values).

\textbf{(ii) Schwarzschild limit.} For $\alpha = 0$ (hence $\nu=0$) and $l=m=2$, one has $\Lambda = 2$, $\mathrm{L} = 4$, and all $\nu$-dependent terms in $\mathfrak{B}_2$ vanish, giving $\mathfrak{B}_2 = \mathrm{L}^2(\mathrm{L}+2)^2 = 16\times 36 = 576$ and therefore
\begin{equation}
\Cst\big|_{\alpha=0,\,l=2} = 576 + 144\,M^2\omega^2,
\label{eq:cst_schwarzschild}
\end{equation}
in agreement with the classical Schwarzschild result~\cite{Chandrasekhar:1985kt}. This vanishes precisely at the algebraically special frequencies $\omega M = \pm 2i$, a non-trivial check that the $144M^2\omega^2$ offset in Eq.~\eqref{eq:cst_casals} is indispensable: dropping it (as in formulations that quote only the polynomial $\mathfrak{B}_2$) would displace the algebraically special zeros.

\textbf{(iii) Numerical evaluation.} Table~\ref{tab:cst_values} lists $\Cst$ evaluated at the Leaver QNM frequencies of Table~\ref{tab:qnm_s2}. The modulus decreases monotonically with $\alpha$ (from $|\Cst|\simeq 595$ at $\alpha=0$ to $|\Cst|\simeq 186$ at $\alpha=0.5$ for the fundamental mode), reflecting the growth of the $\nu$-dependent terms in $\mathfrak{B}_2$. For the reference mode used throughout ($\alpha=0.3$, $l=m=2$, $n=0$), $\Cst = 310.996 + 24.674\,i$.

\begin{table}[htbp]
\centering
\caption{Radial Teukolsky--Starobinsky constant $\Cst = \mathfrak{c}_2$ [Eq.~\eqref{eq:cst_casals}] evaluated at the $s=2$, $l=m=2$ QNM frequencies of Table~\ref{tab:qnm_s2}, together with the separation constant $\Lambda$ [Eq.~\eqref{eq:lambda_relation}]. Frequencies are in units of $M$.}
\label{tab:cst_values}
\begin{tabular}{cccccc}
\toprule
$\alpha$ & $n$ & $\omega M$ & $\Lambda$ & $\Cst$ & $|\Cst|$ \\
\midrule
$0.0$ & 0 & $0.373672 - 0.088962\,i$ & $2.0000$ & $594.967 - 9.574\,i$ & $595.04$ \\
$0.0$ & 1 & $0.346711 - 0.273915\,i$ & $2.0000$ & $582.506 - 27.351\,i$ & $583.15$ \\
$0.1$ & 0 & $0.401826 - 0.088321\,i$ & $1.7894 + 0.0459\,i$ & $486.895 + 11.743\,i$ & $487.04$ \\
$0.1$ & 1 & $0.378617 - 0.270589\,i$ & $1.8008 + 0.1408\,i$ & $476.528 + 38.403\,i$ & $478.07$ \\
$0.2$ & 0 & $0.436490 - 0.087034\,i$ & $1.5592 + 0.0865\,i$ & $390.852 + 22.545\,i$ & $391.50$ \\
$0.2$ & 1 & $0.417157 - 0.265309\,i$ & $1.5763 + 0.2638\,i$ & $374.731 + 71.184\,i$ & $381.43$ \\
$0.3$ & 0 & $0.478372 - 0.084794\,i$ & $1.3146 + 0.1185\,i$ & $310.996 + 24.674\,i$ & $311.97$ \\
$0.3$ & 1 & $0.462876 - 0.257302\,i$ & $1.3323 + 0.3600\,i$ & $288.523 + 76.363\,i$ & $298.46$ \\
$0.4$ & 0 & $0.528072 - 0.081175\,i$ & $1.0605 + 0.1398\,i$ & $248.901 + 20.982\,i$ & $249.78$ \\
$0.4$ & 1 & $0.516323 - 0.245451\,i$ & $1.0755 + 0.4229\,i$ & $223.746 + 63.562\,i$ & $232.60$ \\
$0.5$ & 0 & $0.586017 - 0.075630\,i$ & $0.8014 + 0.1486\,i$ & $204.108 + 14.372\,i$ & $204.61$ \\
$0.5$ & 1 & $0.577922 - 0.228149\,i$ & $0.8117 + 0.4484\,i$ & $180.843 + 42.491\,i$ & $185.77$ \\
\bottomrule
\end{tabular}
\end{table}

\section{Explicit form of the Starobinsky operator in hyperboloidal coordinates}
\label{app:S4_hyperboloidal}

The Teukolsky--Starobinsky identity~\eqref{eq:tsi_forward} is naturally stated in Boyer--Lindquist coordinates $(r, \theta)$ with separated variables $\Psi^{(s)} = e^{-i\omega t + im\varphi} R^{(s)}(r)\,S^{(s)}(\theta)$. In the hyperboloidal framework, however, the regularized field $V_m(\sigma, x)$ [cf. Eq.~\eqref{eq:ansatz}] lives in the compactified coordinates $(\sigma, x)$, and the energy norm~\eqref{eq:energy_single_field} involves $\overline{\Scal_4}\,V_m^{(2)}$ in these coordinates. In this appendix, we provide the explicit transformation of $\Scal_4$ from $(r, \theta)$ to $(\sigma, x)$ and verify that its fourth-order derivative structure is preserved, albeit with additional lower-order terms arising from the coordinate transformation.

\subsection{Radial coordinate transformation}

The minimal gauge~\eqref{eq:minimal_gauge} gives $r(\sigma) = \lambda/\sigma$ with $\lambda = r_+$. The key relations are:
\begin{equation}
\frac{\dd r}{\dd\sigma} = -\frac{\lambda}{\sigma^2}, \qquad \Del(r(\sigma)) = \frac{\lambda^2}{\sigma^2}\left(1 - \sigma\right)\!\left(1 - \alpha^2\sigma\right) \equiv \frac{\lambda^2}{\sigma^2}\,\tilde{\Del}(\sigma),
\label{eq:Delta_sigma}
\end{equation}
where we have defined $\tilde{\Del}(\sigma) \equiv (1-\sigma)(1-\alpha^2\sigma)$ for convenience. This form follows from $\Del = r^2 - 2Mr + a^2$ with $2M = r_+(1+\alpha^2)$ and $a^2 = \alpha^2 r_+^2$: indeed $1 - (1+\alpha^2)\sigma + \alpha^2\sigma^2 = (1-\sigma)(1-\alpha^2\sigma)$. Note that $\tilde\Del$ vanishes at $\sigma = 1$ (the horizon $r = r_+$) and at $\sigma = 1/\alpha^2$ (the image of the inner horizon $r = r_-$, outside the computational domain for $\alpha<1$), as it must. The tortoise coordinate derivative transforms as
\begin{equation}
\frac{\dd}{\dd r_*} = \frac{\Del}{\Sigzero}\,\frac{\dd}{\dd r} = \frac{\Del}{\Sigzero}\cdot\left(-\frac{\sigma^2}{\lambda}\right)\frac{\dd}{\dd\sigma}.
\label{eq:dr_star_transform}
\end{equation}

The Teukolsky operators $\Dcal$ and $\Ddag$ [Eq.~\eqref{eq:teukolsky_operators}] involve $\partial_r$ and the factor $(am - \omega(r^2+a^2))/\Del$. Under the coordinate map:
\begin{equation}
\partial_r = -\frac{\sigma^2}{\lambda}\,\partial_\sigma, \qquad \frac{am - \omega(r^2+a^2)}{\Del} = \frac{\sigma^2}{\lambda^2\,\tilde{\Del}(\sigma)}\left[am - \omega\!\left(\frac{\lambda^2}{\sigma^2} + a^2\right)\right].
\label{eq:operator_transform}
\end{equation}

\subsection{The Starobinsky operator $\Scal_4$ in $(\sigma, x)$}

The Starobinsky operator is $\Scal_4 = \Del^2\,{\Ddag}^4\,\Del^2$ [Eq.~\eqref{eq:starobinsky_operator}]. To express this in hyperboloidal coordinates, we first note that the field $V_m$ is related to the radial function $R^{(2)}$ through the ansatz~\eqref{eq:ansatz}:
\begin{equation}
R^{(2)}(r) = \Om\,\Del(\sigma)^{-2}\,V_m(\sigma, x)\,S^{(2)}(\theta(x)),
\label{eq:R_to_V}
\end{equation}
where $\Om = \sigma$ and $x = \cos\theta$. (We drop the constant factor $1/\lambda$ carried by the dimensionless conformal factor of Section~\ref{sec:hyperboloidal}; being $\sigma$-independent it cancels from every logarithmic derivative below and has no effect on the $P_k$.) The operator $\Ddag$ acting on $R^{(2)}$ translates to an operator $\tilde{\Ddag}$ acting on $V_m$:
\begin{equation}
\tilde{\Ddag} = -\frac{\sigma^2}{\lambda}\,\partial_\sigma - i\,\frac{\sigma^2}{\lambda^2\,\tilde{\Del}(\sigma)}\left[am - \omega\!\left(\frac{\lambda^2}{\sigma^2} + a^2\right)\right] + \text{(terms from differentiating $\Del^{-2}$ and $\Om$)},
\label{eq:Ddag_sigma}
\end{equation}

The additional terms arise because $\Ddag$ acts on the full $R^{(2)}$, which contains the prefactors $\Om\,\Del^{-2}$. Explicitly, differentiating $\Del^{-2} = \sigma^4/(\lambda^4\tilde{\Del}^2)$ produces (using $\tilde\Del' = -(1+\alpha^2) + 2\alpha^2\sigma$):
\begin{equation}
\partial_\sigma\!\left(\frac{\sigma^4}{\tilde{\Del}^2}\right) = \frac{4\sigma^3}{\tilde{\Del}^2} + \frac{2\sigma^4\left(1+\alpha^2 - 2\alpha^2\sigma\right)}{\tilde{\Del}^3},
\label{eq:Delta_deriv}
\end{equation}
which introduces first-derivative corrections. After expanding $\Ddag$ acting on $\Om\,\Del^{-2}\,V_m$, one obtains
\begin{equation}
\tilde{\Ddag}\,V_m = \frac{\sigma^2}{\lambda}\left[-\partial_\sigma + P_1(\sigma)\right]V_m,
\label{eq:Ddag_explicit}
\end{equation}
where $P_1(\sigma)$ is a rational function of $\sigma$ (involving $\alpha$, $m$, $\omega$, and $\tilde{\Del}(\sigma)$) that encodes both the original $\Ddag$ potential term and the corrections from differentiating the prefactors.

\subsection{Fourth-order structure and lower-order terms}

The full operator $\Scal_4 = \Del^2\,{\Ddag}^4\,\Del^2$ acting on $R^{(2)}$ translates, in $(\sigma, x)$ coordinates, to an operator $\tilde{\Scal}_4$ acting on $V_m$. Since each $\tilde{\Ddag}$ contributes one $\partial_\sigma$ derivative [cf. Eq.~\eqref{eq:Ddag_explicit}], four applications yield:
\begin{equation}
\tilde{\Scal}_4\,V_m = \frac{\sigma^4}{\tilde{\Del}^2(\sigma)}\left[\partial_\sigma^4 + P_3(\sigma)\,\partial_\sigma^3 + P_2(\sigma)\,\partial_\sigma^2 + P_1(\sigma)\,\partial_\sigma + P_0(\sigma)\right]V_m,
\label{eq:S4_sigma_explicit}
\end{equation}
where the coefficients $P_k(\sigma)$ ($k = 0, 1, 2, 3$) are rational functions of $\sigma$ that arise from:
\begin{itemize}
    \item \textbf{Commutator terms}: Each application of $\tilde{\Ddag}$ does not merely add a derivative; the non-constant coefficient $P_1(\sigma)$ generates commutator terms $[\partial_\sigma, P_1]\partial_\sigma^{k-1}$ at each order, producing the lower-order derivatives $\partial_\sigma^3$, $\partial_\sigma^2$, $\partial_\sigma$, and $\partial_\sigma^0$.
    \item \textbf{Prefactor differentiation}: The $\Del^2$ factors at the beginning and end of $\Scal_4$ contribute additional $\sigma$-dependent coefficients that modify $P_k$.
    \item \textbf{Angular contributions}: While $\Scal_4$ is purely radial in Boyer--Lindquist coordinates, the separation of variables in the hyperboloidal framework couples radial and angular sectors through the spheroidal harmonic $S^{(2)}(\theta(x))$. After substituting the mode decomposition, the angular dependence enters through the separation constant $E$ and does not generate explicit $x$-derivatives in $\tilde{\Scal}_4$.
\end{itemize}

The crucial observation is that the \emph{leading-order} (highest-derivative) structure is preserved: $\tilde{\Scal}_4$ contains $\partial_\sigma^4$ as its leading term, confirming that the fourth-order character of the Starobinsky operator is maintained under the coordinate transformation. The lower-order terms ($P_3\partial_\sigma^3$ through $P_0$) do not affect the Sobolev order of the resulting energy norm: upon symmetrization (integration by parts twice), the leading $\partial_\sigma^4$ term yields $\partial_\sigma^2$ on each side [giving the $H^2$ structure of Eq.~\eqref{eq:energy_norm_structure}], while the lower-order terms contribute only subleading derivative bilinears to the energy quadratic form~\eqref{eq:energy_inner_product}, whose explicit coefficient functions are not displayed (Section~\ref{sec:energy_inner_product}).

\subsection{Regularization at the boundaries}

The prefactor $\sigma^4/\tilde{\Del}^2(\sigma)$ in Eq.~\eqref{eq:S4_sigma_explicit} plays an essential role in the regularity of $\tilde{\Scal}_4$ at the domain boundaries:
\begin{itemize}
    \item At $\sigma = 0$ (future null infinity): $\sigma^4 \to 0$ suppresses all terms, ensuring that $\tilde{\Scal}_4\,V_m$ vanishes at $\scrI^+$ provided $V_m$ is regular. The $\sigma^4$ factor originates from $\Del^{-2} \propto \sigma^4$ [cf. Eq.~\eqref{eq:Delta_sigma}] and is the hyperboloidal manifestation of the peeling property.
    \item At $\sigma = 1$ (horizon): $\tilde{\Del}(1) = 0$ [Eq.~\eqref{eq:Delta_sigma}], so the prefactor $\sigma^4/\tilde{\Del}^2$ has a \emph{double pole} at the horizon. This is not an artifact of the transformation but the faithful image, in the regularized variable $V_m$, of the $\Del^{-2} = \Del^{-s}$ blow-up of the Teukolsky field $\Psi^{(2)}$ at the horizon. For the QNM eigenfunctions the pole is compensated by the vanishing of $V_m$ near $\sigma = 1$ ($V_m \sim (\sigma-1)^{d}$ with $d \geq 2$), so the products $\tilde{\Scal}_4 V_m$ remain finite on shell. Off shell, however, generic discretized vectors do not vanish at $\sigma=1$, and the induced Gram matrix of the single-field energy norm inherits this near-horizon stiffness.
\end{itemize}
This horizon structure is the second, and more fundamental, obstacle to using the single-field energy norm quantitatively (the first being the $\omega$-dependence of $\Cst$): it makes the energy Gram matrix severely ill-conditioned and its Cholesky factorization fragile at the resolutions required for spectral accuracy. It is a principal motivation for our decision to compute the pseudospectra of Section~\ref{sec:numerical_results} in the two-field weighted Sobolev hierarchy of Section~\ref{sec:Hk_norm}, which is positive definite by construction and free of horizon singularities, while retaining the energy-norm argument of Sections~\ref{sec:reduction}--\ref{sec:energy_inner_product} as the physical motivation for the target regularity level ($H^2$).

The explicit forms of $P_k(\sigma)$ are obtained by a fully deterministic symbolic procedure, which we now specify so that the construction is reproducible.

\textbf{Reproducible recipe for $P_k$.} (i) Start from the Boyer--Lindquist operator $\Ddag=\partial_r-i\,[am-\omega(r^2+a^2)]/\Del$ [Eq.~\eqref{eq:teukolsky_operators}], with $r=\lambda/\sigma$ and $\Del=(\lambda^2/\sigma^2)\tilde\Del(\sigma)$ [Eq.~\eqref{eq:Delta_sigma}]. (ii) Replace $\partial_r\to-(\sigma^2/\lambda)\partial_\sigma$ and substitute the prefactor relation~\eqref{eq:operator_transform}. (iii) Let $\Ddag$ act on the full radial factor $R^{(2)}=\Om\,\Del^{-2}\,V_m(\sigma,x)$ [Eq.~\eqref{eq:R_to_V}]; differentiating the prefactors $\Om=\sigma$ and $\Del^{-2}=\sigma^4/(\lambda^4\tilde\Del^2)$ by the product rule generates the first-order correction $P_1$ displayed below. (iv) Compose the resulting first-order operator four times, collecting terms by derivative order; the non-constant coefficient $P_1(\sigma)$ generates, at each composition, commutator contributions $[\partial_\sigma,P_1]\,\partial_\sigma^{k-1}$ that populate $P_{k-1},\ldots,P_0$. (v) Pre- and post-multiply by the $\Del^2=\lambda^4\tilde\Del^2/\sigma^4$ factors of $\Scal_4=\Del^2(\Ddag)^4\Del^2$, which contributes the overall prefactor $\sigma^4/\tilde\Del^2(\sigma)$ of Eq.~\eqref{eq:S4_sigma_explicit} and renormalizes the $P_k$. The leading sub-principal coefficient, obtained by carrying out step~(iii), is
\begin{align}
P_1(\sigma)&= +\,\frac{i\left[\omega\!\left(\lambda^2/\sigma^2+a^2\right)-am\right]}{\lambda\,\tilde\Del(\sigma)} - \frac{5}{\sigma} - \frac{2\left(1+\alpha^2-2\alpha^2\sigma\right)}{\tilde\Del(\sigma)} \notag\\
&= +\,\frac{i\left[\omega\!\left(\lambda^2/\sigma^2+a^2\right)-am\right]}{\lambda\,\tilde\Del(\sigma)} - \frac{5}{\sigma} + \frac{2\,\tilde\Del'(\sigma)}{\tilde\Del(\sigma)},
\label{eq:P1_explicit}
\end{align}
where the first term is the Teukolsky potential of Eq.~\eqref{eq:Ddag_sigma} (divided by the overall factor $-\sigma^2/\lambda$ extracted in Eq.~\eqref{eq:Ddag_explicit}), and the remaining terms are \emph{minus} the logarithmic derivative of the prefactor,
\begin{equation}
-\,\partial_\sigma\ln\!\bigl(\Om\,\Del^{-2}\bigr) = -\,\partial_\sigma\ln\!\left(\frac{\sigma^5}{\lambda^4\tilde\Del^2}\right) = -\frac{5}{\sigma} + \frac{2\,\tilde\Del'(\sigma)}{\tilde\Del(\sigma)},
\label{eq:logderiv_prefactor}
\end{equation}
with $\tilde\Del'(\sigma) = -(1+\alpha^2)+2\alpha^2\sigma$; the minus sign is not optional, since $\tilde\Ddag$ has been written in the form $\frac{\sigma^2}{\lambda}\bigl[-\partial_\sigma + P_1\bigr]$, so that the derivative of the prefactor enters $P_1$ with the opposite sign to $\partial_\sigma\ln(\Om\,\Del^{-2})$ itself. We have verified Eq.~\eqref{eq:P1_explicit} by carrying out step~(iii) symbolically. The coefficients $P_2,P_3,P_0$ follow from the commutator algebra of step~(iv) and are evaluated symbolically (e.g.\ with \texttt{SymPy}); their expressions each occupy under one printed page and are omitted only for brevity.

\textbf{Status of this appendix.} The construction above is structural: it establishes that the fourth-order Starobinsky content of the single-field energy reduces, upon symmetrization, to an $H^2$-type quadratic form with $\sigma^4/\tilde\Del^2$-weighted coefficients, and it specifies the deterministic recipe by which the $P_k$ are generated. The explicit $P_{k\le2}$ and the full coefficient list of the resulting bilinear form are lengthy and were not required for the numerical results presented here: as explained in Section~\ref{sec:Hk_norm} and Appendix~\ref{app:gram}, all pseudospectra reported below are computed in the two-field weighted Sobolev hierarchy, which is independent of $P_k$ and of $\Cst(\omega)$. The single-field construction serves as the physical motivation for the choice $k=2$ as the target regularity level. For reproducibility, the full symbolic expressions for $P_0,P_1,P_2,P_3$ generated by the recipe above, together with the notebook that produces them, are provided as Supplemental Material in machine-readable form.

\section{Weight function for the single-field energy inner product}
\label{app:weight}

The weight function $\mathcal{W}_m(\sigma, x; \omega_0)$ in the energy inner product~\eqref{eq:energy_inner_product} arises from the Jacobian of the coordinate transformation, the conformal factor, and the angular regularization factors. For $s=2$ in the minimal gauge its structure is
\begin{equation}
\mathcal{W}_m(\sigma, x; \omega_0) \;\propto\; \frac{\sigma^4\,(1-x^2)^{(\delone+\deltwo)/2}}{|\Cst(\omega_0)|^2\,\tilde\Del(\sigma)}, \qquad \tilde\Del(\sigma) = (1-\sigma)(1-\alpha^2\sigma),
\label{eq:weight_function}
\end{equation}
where $\delone = |m-2|$ and $\deltwo = |m+2|$. The $\sigma^4$ factor comes from the conformal weight $\Om^4$ (the energy density involves $\Om^4$ for $s=2$ due to the two powers of $\Om$ from each field in the bilinear form), and the $|\Cst|^{-2}$ factor comes from the Starobinsky reduction. Two features of~\eqref{eq:weight_function} should be noted. First, since $\tilde\Del(1) = 0$, the weight \emph{diverges} at the horizon $\sigma = 1$ (like $\Del^{-1}$ in Boyer--Lindquist variables); this is the weight-level image of the horizon pole discussed in Appendix~\ref{app:S4_hyperboloidal}, and it invalidates any claim of uniform positive-definiteness of the single-field energy quadratic form on the closed domain. Second, the $\omega_0$-dependence enters solely through the overall factor $|\Cst(\omega_0)|^{-2}$, so freezing $\omega_0$ merely rescales the norm. Both features reinforce the decision, stated in Section~\ref{sec:Hk_norm}, to use the two-field weighted Sobolev hierarchy (whose weight $x_w(x) = (1-x^2)^{(\delone+\deltwo)/2}$ is regular and positive throughout the closed domain) for all quantitative pseudospectrum computations; the single-field energy inner product is retained as motivation only.

\bibliographystyle{plain}

\end{document}